\documentclass{jfm_arxiv}
\usepackage{graphicx}
\usepackage[svgnames, dvipsnames]{xcolor}

\usepackage[figuresright]{rotating}
\graphicspath{ {./} }
\DeclareGraphicsExtensions{{.pdf}}

\definecolor{inner}{HTML}{4C136B}
\definecolor{log}{HTML}{FF9C08}
\definecolor{outer}{HTML}{14615E}

\newcommand{\au}{{\overline u}}
\newcommand{\av}{{\overline v}}
\newcommand{\aw}{{\overline w}}
\newcommand{\xb}{\boldsymbol{x}}
\newcommand{\ub}{\boldsymbol{q}}

\newcommand{\mletter}[3]{\raisebox{#1}[0pt][0pt]{\makebox[0pt]{\hspace{#2}#3}}}

\def\spacce#1{\hskip #1pt}
\def\drawline#1#2{\raise2.5pt\vbox{\hrule width #1pt height #2pt}}
\newcommand{\solid}{\drawline{24}{.5}\nobreak}

\newcommand{\solidcircle}{{\Large$\bullet$}\nobreak}
\newcommand{\linesolidcircle}{\hbox%
{\drawline{8}{.5}\spacce{2}\solidcircle\drawline{8}{.5}}\nobreak}

\newcommand{\dd}{{\, \rm{d}}}
\newcommand{\dr}{{\rm{d}}}
\newcommand{\Dr}{{\rm{D}}}

\def\beq{\begin{equation}}
\def\eeq{\end{equation}}
\newcommand{\la}{\label}
\newcommand{\ii}{{\rm i}}

\let\Im\relax
\DeclareMathOperator{\Im}{Im}
\let\Re\relax
\DeclareMathOperator{\Re}{Re}

\renewcommand{\r}[1]{(\ref{#1})}

\newcommand{\lambdaf}{\Lambda}
\newcommand{\Sl}{S_{\!\lambdaf}}

\colorlet{refcolor1}{Black}
\colorlet{refcolor2}{Black}
\colorlet{refcolor3}{Black}

\shorttitle{Momentum transfer by linearised eddies in channel flows}
\shortauthor{M. P. Encinar and J. Jim\'enez}

\title{Momentum transfer by linearised eddies in turbulent channel flows}

\author{Miguel P. Encinar\corresp{\email{mencinar@torroja.dmt.upm.es}}
 \and Javier Jim\'enez}

\affiliation{School of Aeronautics, Universidad Polit\'ecnica de Madrid, 28040, Madrid, Spain}

\begin{document}

\maketitle

\begin{abstract}
    The presence and structure of an Orr-like inviscid mechanism are studied in fully developed, large-scale turbulent channel flow. Orr-like `bursts' are defined by the relation between the amplitude and local tilting angle of the wall-normal velocity perturbations, and extracted by means of wavelet-based filters. They span the shear-dominated region of the flow, and their sizes and lifespans are proportional to the distance from the wall in the logarithmic layer, forming a self-similar eddy hierarchy consistent with Townsend's attached-eddy model. Except for their amplitude, which has to be determined nonlinearly, linearised transient growth represents their evolution reasonably well. Conditional analysis, based on wavelet-filtered and low-pass-filtered velocity fields, reveals that bursts of opposite sign pair side-by-side to form tilted quasi-streamwise rollers, which align along the streaks of the streamwise velocity with the right sign to reinforce them, and that they preferentially cluster along preexisting streak inhomogeneities. On the other hand, temporal analysis shows that consecutive rollers do not form simultaneously, suggesting that they incrementally trigger each other. This picture is similar to that of the streak-vortex cycle of the buffer layer, and the properties of the bursts suggest that they are different manifestations of the well-known attached Q$_2$-Q$_4$ events of the Reynolds stress. 
\end{abstract}

\section{Introduction}\label{sec:intro}

Nonlinearity is an essential characteristic of turbulent phenomena \citep{ten:lum:72}.
Linear models of statistically steady turbulent flow retain neither its chaotic behaviour
\citep{rue:71} nor its self-similar multiscale organisation \citep{kol:41b}. In addition,
linear models cannot predict the intensity of turbulent perturbations. Nevertheless, some
features of shear flows, such as the structure of the velocity perturbations at the
energy-injection scale, can be described reasonably well by the linearised Navier--Stokes
equations. In fact, since the source of turbulent kinetic energy in a shear flow is the mean
shear, it is unavoidable that at least part of the energy production mechanism should be
linear, in the sense of involving interactions of the perturbations with the mean flow,
rather than of the perturbations among themselves.

The best-known example is the Kelvin--Helmholtz instability of mean profiles with inflection
points, which controls many of the properties of the large-scale structures in fully
turbulent free-shear flows \citep{bro:ros:74}. On the contrary, the mean profiles of
wall-bounded flows lack an inflection point, and are linearly stable \citep{rey:tie:67}.
Still, attempts to derive properties of wall-bounded turbulence by assuming that the mean
profile is marginally stable to perturbations have a long history \citep{mal:56}, and it has
been known for some time \citep{but:92} that modally stable linearised perturbations can
transiently grow by drawing energy from the mean shear. Many linear models reproduce some of the
large-scale features of channel flow well from linearised dynamics
\citep{far:93, ala:jim:06, hwa:cos:10}.

One such mechanism was proposed by \citet{orr:1907}. The cross-shear velocity component
(wall-normal component in wall-bounded flows) is amplified over times of the order of the
inverse of the shear when backwards-leaning perturbations are tilted forward by the mean
velocity profile. The perturbations grow until they are normal to the shear and
are damped past that point, because continuity requires that the change in vertical scale
due to the tilting is balanced by the velocity amplitude. Since perturbations
eventually decay, there is no net production of kinetic energy for the wall-normal velocity,
but, for three-dimensional perturbations whose wavefronts are oblique to the mean stream,
some of the energy is transferred to the other two velocity components. Moreover, since
continuity does not interact with the wall-normal variation of the wall-parallel velocities,
the effect of this `lift up' remains after the wall-normal velocity has vanished
\citep{jim:pof:13}, forming the characteristic streamwise-velocity streaks that populate
wall-bounded flows \citep{kli:rey:sch:run:67}. Because the growth of the wall-normal
velocity is much shorter lived than the resulting streaks, we use the term `burst' to
describe it. The term was originally coined for the ejections of low-speed streaks close
to the wall in boundary layers \citep{kim:kli:rey:71}. It was abandoned for a while due to
controversies about the nature of the ejections, but eventually reclaimed, in the sense
introduced above, to denote the intermittent behaviour of wall-bounded flows
\citep{kaw:kid:01, jim:kaw:sim:nag:shi:05, flo:jim:2010}.

The connection between linearised Orr amplification and bursting was studied by
\citet{jim:pof:13}, who showed that the length and time scales predicted by the linear model
were in agreement with the bursting of the logarithmic layer in minimal channels. This work
was extended by \citet{jim:pof:15}, who tracked the temporal evolution of the largest
Fourier modes in channels designed to be minimal for the logarithmic layer. The strongest
wall-normal velocity events were found to closely follow the predictions of the linearised
equations when the mode was initially `coherent' across the wall-normal direction
(i.e. could be represented by a wavetrain). {\color{refcolor2} In this paper we use the term wavetrain to refer to a periodic (infinite) travelling wave, in contrast to a localised travelling wavepacket}. A key aspect of these results is the relation
between the `tilting' angle of the perturbations and their amplitude, which is predicted by
the linearised Orr--Sommerfeld equations, and serves as an identification criterion for
Orr-like bursts.

Several problems remain. The works just cited treat Orr bursts as individual Fourier
modes. This is no problem for linearised theory over spatially homogeneous directions, since
modes can eventually be added to form more complicated structures, but it becomes
questionable when bursts are to be identified in (nonlinear) extended flow fields. The
bursts observed in experiments and simulations are localised in space as well as in time,
and cannot be described as infinite wavetrains \citep{adr:2007}. Extensions of the classical
quadrant analysis \citep{wal:eck:bro:72, wil:lu:72} to three-dimensional tangential Reynolds-stress
structures \citep{loz:flo:jim:2012, don:loz:sek:jim:17}, and later to their temporal evolution
\citep{loz:jim:2014}, show that much of the momentum transfer in channels and in other shear
flows is carried by transient eddies with geometric aspect ratios of order unity, rather
than by infinite wavetrains. The reason why \citet{jim:pof:15} looked for Orr-like events in a
minimal channel was precisely to be able to use Fourier methods while maintaining the aspect
ratios of experimental observations, and the analysis failed when applied to wavelengths
much shorter than the size of the simulation box. It is not obvious whether the results from
uniform wavetrains can be generalised to a population of localised bursts of different sizes
in a large simulation box, and many of the tools used for the former, such as the wavetrain
inclination angle, need to be redefined for the latter.

Another important question is to what extent Orr bursting explains the momentum transfer in
full-sized channels. For example, it was shown by \citet{jim:pof:15} that the largest-scale
field of the wall-normal velocity in a minimal channel can be described as a linearised Orr
burst over 65\% of the time, but it is unclear which is the equivalent fraction in large
boxes. For example, the tangential Reynolds stress ($-uv$) in minimal boxes is
estimated to burst 20\% of the time \citep{jim:18}, but the stress structures identified by
\cite{loz:flo:jim:2012} only cover 8\% of the wall.

In the present work, we introduce a band-pass-filtering method that allows us to extend
the analysis of bursting beyond minimal channels, detecting structures of arbitrary size in full-sized
simulations. Using large boxes is particularly important when studying the characteristic time
and length scales of the turbulent structures, because the largest scales analysed in minimal
computational boxes are also the most prone to suffer simulation artefacts. We study {\color{refcolor3} a wide range of}
scales across different Reynolds numbers, and simulation boxes, uncovering a range of
self-similar Orr-like events in the logarithmic region, with sizes of the order of their
distance from the wall. {\color{refcolor3} The identified wall-normal velocity eddies are self-similar with respect to their three spatial dimensions and time.}

On the other hand, it is important to understand the limitations of our approach. The main
one has to do with scale. The reason why Fourier modes are useful in analysing the
linearised equations is that harmonic functions are eigenfunctions of uniform translations
in space and time. Any linear mechanism without an explicit dependence on the homogeneous
coordinates cannot change the wall-parallel length scales, including the Fourier
wavenumbers. The range of scales mentioned in the previous paragraph cannot be generated by
the linearised equations, and remains an empirical input parameter. The effect of the
band-pass filters used in this paper is also to isolate a narrow range of scales, which
remains constant throughout the observed evolutions. In this respect, both linearisation and
the band-pass analysis exclude the multiscale aspect of turbulence dynamics.
To remedy this, we also analyse the effect of low-pass filters that allow us to include the
longer streaks of the streamwise velocity, and some multiscale effects.

The remainder of this paper is organised as follows. The linear theory for the Orr
mechanism in turbulent channel flows is briefly reviewed in \S\ref{sec:linear}. Section
\S\ref{sec:ne} covers the domain definition and the database that will be used in the paper,
and the rest of \S\ref{sec:methods} describes the methodology used to detect and characterise Orr
events. Results are presented in \S\ref{sec:bandpass} and \S\ref{sec:lowpass}. Section
\S\ref{sec:bandpass} discusses the statistics and spatio-temporal properties of
band-pass-filtered flows, which isolate the bursts, and \S\ref{sec:lowpass} extends the
results to the low-pass-filtered velocity. Conclusions are gathered in \S\ref{sec:conclusions}.

\section{The linear model}\label{sec:linear}

Let $\xb = (x, y, z)$ be the streamwise, cross-shear and spanwise directions, 
$\check{\ub} = (\check{u}, \check{v}, \check{w})$ being the corresponding velocity components. We
restrict ourselves to shear flows with mean velocity profiles $\boldsymbol{U} = (U(y), 0,
0)$ with no inflection points, where the average is taken over the two homogeneous
directions $(x, z)$ and time. The kinematic pressure and viscosity are $p$ and $\nu$,
respectively, defined as the quotients between the dynamic quantities and the constant fluid
density, which does not appear explicitly in the equations. The Navier--Stokes equations can be
written as
\begin{gather}
\partial_t \ub = -(\boldsymbol{U} + \ub)\boldsymbol{\cdot}\nabla(\boldsymbol{U} + \ub) - \nabla(P + p) + \nu\nabla^2(\boldsymbol{U} + \ub),\label{eq:nvs}\\
\nabla\boldsymbol{\cdot}\ub = 0;
\end{gather}
where $\check{\ub} = \boldsymbol{U} + \ub$ is the standard Reynolds decomposition, and $P$
is the mean pressure. In general, upper-case letters denote ensemble averages and lower-case ones are fluctuations. Primes are reserved for root-mean-square (rms) intensities. The linearised version of \eqref{eq:nvs} is
\begin{equation}
\partial_t \ub_L = - \boldsymbol{U}\boldsymbol{\cdot}\nabla{\ub_L} - \ub_L\boldsymbol{\cdot}\nabla\boldsymbol{U} - \nabla p_L + \nu\nabla^2\ub_L, \label{eq:nslin}
\end{equation}
where $p_L$ is the fast pressure \citep{kim:89}, although we suppress the subscript
{\it`L'} when referring to linearised variables for the remainder of the section. The
linearised equations can be manipulated into the Orr--Sommerfeld--Squire system
\citep{orr:1907, squ:33}:
\begin{align}
  \partial_t \phi &= - U\partial_x \phi + \dr_y S\partial_x v + \nu \nabla^2\phi,\label{eq:orr}\\
  \partial_t \omega_y &= - U\partial_x \omega_y - S\partial_z v + \nu \nabla^2\omega_y;\label{eq:sqr}
\end{align}
where $\phi = \nabla^2 v$, $\boldsymbol{\omega} = (\omega_x, \omega_y, \omega_z)$ is the
vorticity, and $S = \dr_y U$ is the mean shear. Although \eqref{eq:orr} is autonomous in
$v$, continuity and \r{eq:sqr} connect it with the wall-parallel velocities. Continuity and
the linear pressure do not appear in \r{eq:orr}, which is a vorticity equation, but their
effect is hidden in the definition of $\phi$. For example, the term $\dr_yS\partial_x
v$, which is responsible for the Kelvin--Helmholtz instability of free shear-flows
\citep{bro:ros:74}, vanishes identically in a homogeneous shear, $U = Sy$, and \r{eq:orr}
reduces to a simple advection--diffusion equation. Since such flows share many of their
bursting properties with boundary layers and channels, it was argued by \cite{jim:pof:13}
that the amplification of the bursts in \eqref{eq:orr} has to be mediated by the pressure
through the Laplacian operator in $\phi$. In fact, for a homogeneous shear in the limit
$\nu\to 0$, and a pure Fourier mode of streamwise wavenumber $k_x$, and spanwise wavenumber
$k_z = 0$, the solutions to \eqref{eq:orr} are `Orr' bursts of the form
\begin{equation}
v = \cos^2(\Psi)\exp{\left[\ii k_x(x- Syt)\right]},\label{eq:psijim}
\end{equation}
where $\Psi = \arctan(St)$ is the inclination angle of the
wavefront with respect to the $y$ axis. The inclination angle is a reinterpretation of
time, and \r{eq:psijim} relates the amplitude of the $v$ perturbations to the inclination of
their wavefronts, being maximum when $\Psi = 0$ and vanishing towards $\Psi = \pm \pi/2$. 
The burst starts from a weak backwards-leaning perturbation, which amplifies as it is
tilted forward by the shear, is strongest when it is normal to the shear, and is damped again
after this moment. The whole evolution unfolds during times of the order of the shear time,
$S^{-1}$, and the linearisation is only valid if this time is shorter than other evolution
times in the equation.

For non-zero viscosity, there is a perturbation Reynolds number, $\Rey = S/k_x^2\nu$, and \r{eq:psijim} decays in times of $O(S^{-1}\Rey^{1/3})$ \citep{jim:pof:13}, which are
typically long for large structures. Nonlinearity is represented by the local eddy
turnover time, $\|\ub\|^2/\varepsilon$, where $\|\ub\|^2/2$ is the
turbulent kinetic energy and $\varepsilon$ is the energy dissipation rate. Its ratio to the shear
time is the Corrsin shear parameter, $S^* = S\|\ub\|^2/\varepsilon$. In
wall-bounded flows, $S^* \approx 10$ in the range $50~\nu/u_\tau < y < 0.7~h$, where
$u_\tau$ is the friction velocity \citep{jim:pof:13}, and \r{eq:nslin} is therefore valid at
least for the energy-containing structures in this range.

\subsection{Optimal transient growth\label{ssec:otg}}

\begin{figure}
  \centering
  \newcommand{\sca}{0.855}

  {\hspace{50pt}\includegraphics[scale=\sca]{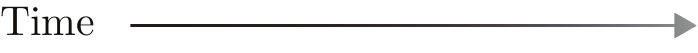}\hfill}

  \includegraphics[scale=\sca]{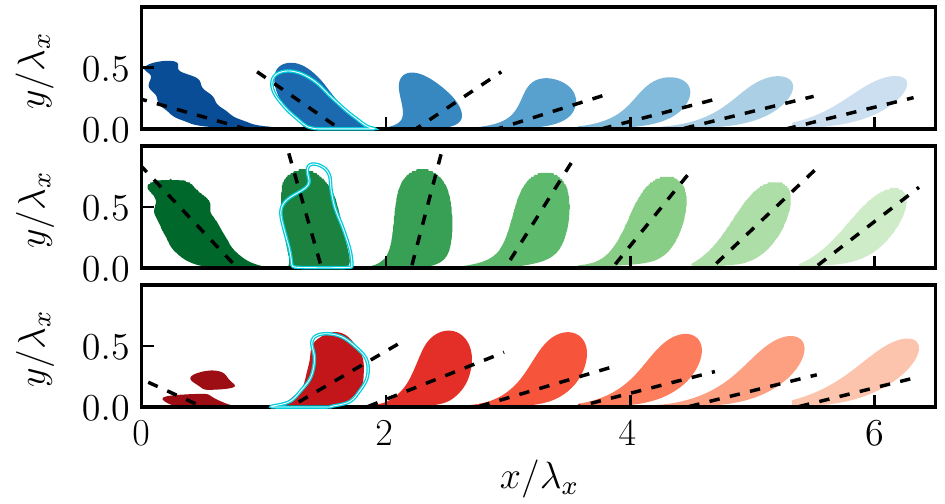}\mletter{4.6cm}{-16.1cm}{(a)}%
  \hspace{5pt}\includegraphics[scale=\sca]{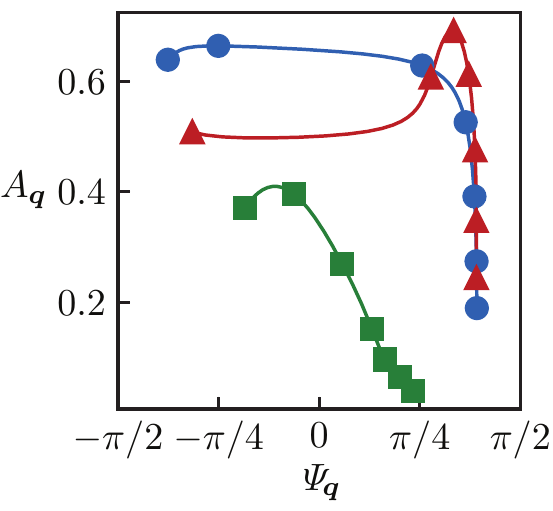}\mletter{4.6cm}{-9cm}{(b)}
  \caption{Evolution of the amplitude and inclination angles of an optimal transient growth mode in a channel flow. $\lambda_x=\lambda_z$.
  (a) Snapshots of the velocity components (top to bottom, $u$, $v$, $w$), spaced by $St \approx 0.52$, from left to right. The dashed lines show the inclination angle computed as in \cite{jim:pof:15}. The contours levels are $u = 4, v =1, w=3$ in uniform arbitrary units.
  (b) Amplitude as a function of the inclination angle. The markers are, {\fullcirc}, $u$; {\fullsquare}, $v$; {\fulltriangle}, $w$; and correspond to the snapshots in {\sl (a)}}\label{fig:lineal}
\end{figure}

For more complex mean velocity profiles, Orr-like solutions can be computed numerically. We
use the algorithm in \cite{sch:hen:01} to find solutions that optimally amplify the total
kinetic energy, including a $y$-dependent eddy viscosity to account for the nonlinear
turbulent dissipation, and a velocity profile and kinematic viscosity corresponding to a
turbulent channel at $\Rey_\tau = u_\tau h/\nu = 2000$, as in \cite{ala:jim:06} and
\cite{puj:gar:09}. {\color{refcolor1} We use the approximation to the mean eddy viscosity $\nu_T$ of \citet{ces:58}:
\begin{equation}
  \nu_T = \frac{\nu}{2}\left\{1 + \frac{K^2\Rey_\tau}{9}\left[2\eta-\eta^2\right]\left[3-4\eta+\eta^2\right]^2\left[1-\exp{\left(\frac{-\eta\Rey_\tau}{A}\right)}\right]^2\right\}^{1/2} + \frac{1}{2},
\end{equation}
where $\eta = y/h$, and the parameters $A = 25.4$ and $K =0.426$ are tuned to match the profile of \citet{hoy:jim:06}. The mean velocity profile is obtained by integrating $\left(1 - \eta \right) {u_\tau}^2/{\nu_T}$. Using the same profiles,} \citet{ala:jim:06} found that the maximum energy amplification occurs for
$\lambda_x\gg \lambda_z$, but \cite{jim:18} remarked that these wavenumbers are associated
with the slow amplification of the streamwise-velocity streaks through \r{eq:sqr}, and to
very long evolution times that may be limited by the nonlinear or viscous effects mentioned
above. We are here interested in situations such as \r{eq:psijim}, where the energy
amplification is associated with the transient bursting of $v$ over a few shear times. These
were shown by \cite{jim:18} to be strongest at $\lambda_x\approx \lambda_z$.

Figure \ref{fig:lineal}(a) shows as shaded contours several snapshots of the most amplified
transient mode with $\lambda_x = \lambda_z \approx 0.7h$, plotted at intervals of
$S(y_g)\rmDelta t \approx 0.23$, where $S(y_g)$ is the shear at the centre of gravity of
the $v^2$ profile at maximum amplification. It was shown in \cite{jim:pof:13} that the
dimensions of the most amplified inviscid $v$-modes in the logarithmic layer are proportional to
$\lambda_0=2\pi/(k_x^2+k_z^2)^{1/2}$, and that their centre of gravity is at $y_g\approx
0.22 \lambda_0$, or $y_g\approx 0.16 \lambda_x$ for equilateral wavevectors. The same is
true in figure \ref{fig:lineal}(a), even if this case differs from \cite{jim:pof:13} in
using an eddy viscosity. The temporal evolution of the perturbations in \cite{jim:pof:13} is
controlled by $\lambda_0S(y_g)t/\lambda_x$. In the range of wavelengths $250 \nu/u_\tau
\lesssim \lambda_x \lesssim 4h$, corresponding to $40 \nu/u_\tau \lesssim y_g \lesssim 0.64h$, we can
assume that $S(y_g) \propto u_\tau/y_g\propto u_\tau /\lambda_0$, and that the scaling of
the time is $\lambda_0S(y_g)t/\lambda_x\propto u_\tau t/\lambda_x$. The result is that both
the dimensions and the temporal evolution of the amplified structures scale with
$\lambda_x$. The relation between the inclination angle $\Psi$ and the amplitude of $v$ was shown
in \cite{jim:pof:13} to be fairly independent of the wavenumbers, and even of the type of
flow. These results are extended here to the three velocity components, and to the
eddy-viscosity equations, and we indeed find that the optimal modes in the neighbourhood of
$\lambda_x \approx \lambda_z$ scale well with $\lambda_x$ in the range of wavenumbers
mentioned above. For example, the shape and inclination angles of the solution in figure
\ref{fig:lineal}(a) are remarkably similar to those in figure 19 of \cite{jim:18}, where the
wavelengths are much larger, $\lambda_x = 3.9h, \lambda_z = 5.55h$, and the solid contour
lines superimposed on the second snapshot in figure \ref{fig:lineal}(a), which show the
moment of maximum amplitude of a mode with $\lambda_x = \lambda_z \approx 0.35h$, also agree
very well with those at the longer wavelengths. The self-similar behaviour ceases to hold at
heights corresponding to the buffer layer, where the shear is no longer inversely
proportional to $y$.

The dashed lines in figure \ref{fig:lineal}(a) are the vertically averaged inclination
angles at each time, computed using the formula for single modes in \cite{jim:pof:15}; for a
more general one see (\ref{eq:Psi}) below. They represent well the instantaneous structure
of $v$, but they are biased towards the lower half of the structures for $u$ and $w$ because
the averaging in \r{eq:Psi} is weighted with the energy, which tends to be concentrated near
the wall for these two components. This is also where the shear is highest, leading to
flatter inclinations. Figure \ref{fig:lineal}(b) shows the amplitude--inclination relation
for the three velocity components in figure \ref{fig:lineal}(a). {\color{refcolor2} The amplitude and inclination angles of the wavetrain are defined as in \citep{jim:pof:15}. Both definitions are extended in (\ref{eq:A})--(\ref{eq:Psi}), later in this paper, to more general wavepackets}. The trajectory for $v$ is
reminiscent of \eqref{eq:psijim}, but the wall-parallel components tilt forwards much faster,
as shown by the larger separation of their first few symbols, which are uniformly spaced in
time. Inspection of figure \ref{fig:lineal}(a) shows that this is also due to the weighting
of the angle close to the wall, where the higher shear results in faster characteristic
times. The upper part of the $u$ and $w$ structures does not tilt much faster than that of
$v$, while the root of the latter tilts as fast as those of the other two components. The
last half of the angle--amplitude evolution is also different for the three velocity
components. The spanwise velocity has an amplitude peak at approximately the same time as
$v$, which is not present for $u$. At the end of the evolution, $v$ decays at a rate
dictated by the shear, while $u$ and $w$ only decay under the effect of the eddy viscosity. For
inviscid perturbations in a uniform shear, the three velocity components have well-differentiated
behaviours. During the burst, $v$ grows and decays in $t=O(S^{-1})$, as in
\r{eq:psijim}, but $u$ and $w$ never decay \citep{jim:pof:13}. In all cases, continuity
requires that the two wall-parallel have to be proportional to each other once $v$ has died.
An important message of figure \ref{fig:lineal} is that optimum growth solutions should only
be taken as indicative. The properties enumerated in the last few sentences are common to
any initial condition, while optimum growth is meaningless in an inviscid flow. The
behaviour of $v$, whose equation \r{eq:orr} is autonomous, is essentially independent of the
initial conditions, which mostly define the origin of time, but $u$ and $w$ satisfy the
forced equation \r{eq:sqr}, whose evolution depends on $v$ and on the initial conditions.
The optimum transient growth criterion selects the initial condition with a highest local
maximum, but the $v$ burst and the $v$-less final decay are robust properties of most
initial conditions involving backwards-leaning perturbations. Note also that the energy
growth in figure \ref{fig:lineal}(b) is not particularly large, and that the same is true of
all the cases mapped in figure 29 of \cite{jim:18}.

\section{Methods}\label{sec:methods}

The rest of the paper seeks to relate the behaviour of fully nonlinear channel flow
simulations to the linearised dynamics sketched above.
\subsection{The numerical datasets}\label{sec:ne}

\begin{table}
  \begin{center}
    \def~{\hphantom{0}}
    \begin{tabular}{lccccccccr@{}lr}
      Case  & $\Rey_\tau$ & $L_x/h$ & $L_z/h$ & $N_x$ & $N_y$ & $N_z$ & $D_y$ & $N_f$ & \multicolumn{2}{c}{\color{refcolor3} $T_{\mkern-1mu h}$} & \multicolumn{1}{c}{Reference}\\[3pt]
      S1000 & 932 & $2\pi$ & $\pi$ & $512$ & $385$ & $512$ & CH & $1800$ & 20& & {\small \cite{loz:jim:2014}}\\
      L1000 & 934 & $8\pi$ & $3\pi$ & $2048$ & $385$ & $1536$ & CH & $72$ & 8&.5 & {\small \cite{ala:jim:zan:mos:04}}\\
      S2000 & 2009 & $2\pi$ & $\pi$ & $1024$ & $633$ & $1024$ & FD & $590$ & 11& & {\small \cite{loz:jim:2014}}\\
      F2000 & 2000 & $8\pi$ & $3\pi$ & $512$ & $512$ & $512$ & FD & $1146$ & 14&& {\small \cite{encinar18}}\\
      L2000 & 2003 & $8\pi$ & $3\pi$ & $4096$ & $633$ & $3072$ & FD & $30$ & 10&.3 & {\small \cite{hoy:jim:06}}\\
      S4000 & 4164 & $2\pi$ & $\pi$ & $2048$ & $1081$ & $2048$ & FD & $40$ & 10& & {\small \cite{loz:jim:2014}}\\
    \end{tabular}
    \caption{Parameters of the simulations. $L_{x, z}$ are the streamwise and spanwise periods of
    the numerical box. $N_{x, y, z}$ are the number of collocation grid points in $y$, or Fourier modes 
    in $x, z$, except for F2000, which was computed at full resolution but stored
    at the reduced one given in the table. $D_y$ represents the discretisation used for the
    wall-normal direction: `CB' refers to Chebyshev polynomials and `FD' to compact finite
    differences. $N_f$ is the number of fields used to compute statistics. {\color{refcolor3} $T_h$ is the time spanned by the approximately equidistant $N_f$ snapshots in eddy turnover times.} The snapshots of
    S1000 and F2000 are close enough in time to compute temporal derivatives.}
    \label{tab:sim}
  \end{center}
\end{table}

We use direct simulations of canonical incompressible turbulent channel flows whose
half-height is $h$. Normalisation with $\nu$ and $u_\tau$ is represented by a `$+$'
superscript, and our primary Reynolds number is $Re_\tau=h^+$. The simulations use periodic
boundary conditions in the wall-parallel directions, with periods $L_x$ and $L_z$, and are
summarised in table \ref{tab:sim}. The Reynolds numbers range from $\Rey_\tau \approx 900
\text{ to } 4000$, and the simulations include both medium-sized $(L_x = 2\pi h, L_z = \pi h)$
and large $(L_x = 8\pi h, L_z = 3\pi h)$ computational boxes. The equations of motion are
written as evolution equations for the $(x, z)$ averages of the streamwise and spanwise
velocities, and for $\phi$ and $\omega_y$, as in \r{eq:orr}--\r{eq:sqr} without
linearisation \citep{kim:moi:mos:87}. The spatial discretisation is Fourier spectral in the
two periodic directions, dealiased using the pseudo-spectral 3/2 rule, but the
discretisation along $y$ varies among simulations, as indicated in table \ref{tab:sim}. Some
cases use Chebyshev polynomials collocated at the Gauss--Lobatto--Chebyshev nodes. Others use
spectral-like compact finite differences \citep{lel:92, flo:jim:06}, up to 12th-order
consistent, in grids whose spacing is adjusted to keep the resolution approximately constant
in terms of the local isotropic Kolmogorov scale $\eta = (\nu^3/\varepsilon)^{1/4}$. Full
details can be found in the original publications. {\color{refcolor2} F2000 has the peculiarity of being run as a direct numerical simulation, but stored at a coarser `large-eddy simulation' resolution \citep{encinar18}}. The data stored are
snapshots of the three velocity components, which in most cases  are approximately statistically
independent and can only be used to compile instantaneous statistics. However, the snapshots in
S1000 and F2000 are stored closely enough in time to provide temporal derivatives
and histories without recomputing the flow evolution.

\subsection{Filtering}\label{sec:filterdef}

As discussed in \S \ref{sec:linear}, the relation between the instantaneous tilting
angle of the wall-normal velocity perturbations and their amplitude can be used as a
diagnostic property for Orr bursts. \cite{jim:pof:15} defined the inclination
angle of the wavefronts of a pure Fourier mode in terms of the wall-normal derivative of
its complex phase, but this definition, as well as that of the amplitude, needs to be generalised for
spatially localised objects. Consider the low-pass-filtered field
\begin{equation}
\widetilde{\ub}(x, y, z, t; \lambdaf) = \int_\mathcal{B} \Gamma(x'-x, z'-z, \lambdaf) \ub(x', y, z', t){\dd}x'{\dd}z', \label{eq:lpass}
\end{equation}
where $\mathcal{B}$ stands for the full computational box{\color{refcolor3}, $\varLambda$ serves as the filter width} and
\begin{equation}
  \Gamma(x, z;\lambdaf_{0L},\lambdaf_{0z} ) =
  \gamma(x,\lambdaf_{0L})\,\gamma(z,\lambdaf_{0z}),\label{eq:env}
\end{equation}
with
\begin{equation}
  \gamma(\xi, \lambdaf_0) = \frac{1}{\lambdaf_0\sqrt{2\pi}}
\exp\left[-\tfrac{1}{2}(\xi/\lambdaf_0)^2\right];
\label{eq:filgauss}
\end{equation}
and the band-pass-filtered one
\begin{equation}
\overline{\ub}(x, y, z, t; \lambdaf) = \int_\mathcal{B} G(x'-x, z'-z, \lambdaf) \ub(x', y, z'){\dd}x'{\dd}z', \label{eq:bpass}
\end{equation}
with
\begin{equation}
  G(x, z;\lambdaf,\lambdaf_{0x},\lambdaf_{0z} ) =
  \Gamma(x, z; \lambdaf_{0x},\lambdaf_{0z}) \exp (2\pi\ii x/\lambdaf).\label{eq:fil}
\end{equation}
The spectral transfer functions of the two kernels are
\begin{eqnarray}
|\widehat{\Gamma}|^2 &=& \exp\left[ -(k_z\lambdaf_{0z})^2-(k_x\lambdaf_{0L})^2\right],
\label{eq:tfGam}\\
|\widehat{G}|^2 &=& 
\exp\left( -(k_z\lambdaf_{0z})^2-[(k_x-2\pi/\lambdaf)\lambdaf_{0x}]^2\right),
\label{eq:tfG}
\end{eqnarray}
where the carat stands for the $(x, z)$ Fourier transform. Both filters have spectral width $\rmDelta
k=O(\lambdaf_0^{-1})$, but the band-pass filter $G$ is centred at the wavenumber $k_x=2\pi/\lambdaf$ (and
$k_z=0$), while the low-pass filter $\Gamma$ is centred at $k_x=k_z=0$. Thus, if $\lambdaf$ and the
$\lambdaf_0$'s are chosen to be of the same order, $\Gamma$ smooths the flow by damping
everything shorter and narrower than $O(\lambdaf_0)$, while $G$ selects only structures
which may be wide, but whose length is of the order of $\lambdaf$. The latter is useful to isolate
localised bursts, while the former also retains the larger structures of the flow around
them.

\begin{figure}
  \centering
  \raisebox{20pt}{\includegraphics[width=.45\textwidth]{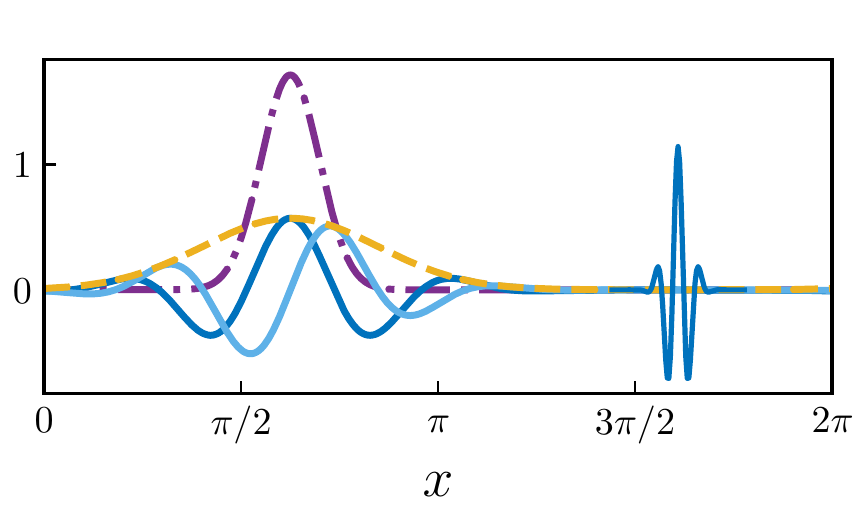}}\mletter{4.15cm}{-11.2cm}{(a)}%
\hspace*{2mm}
  \includegraphics[width=.45\textwidth]{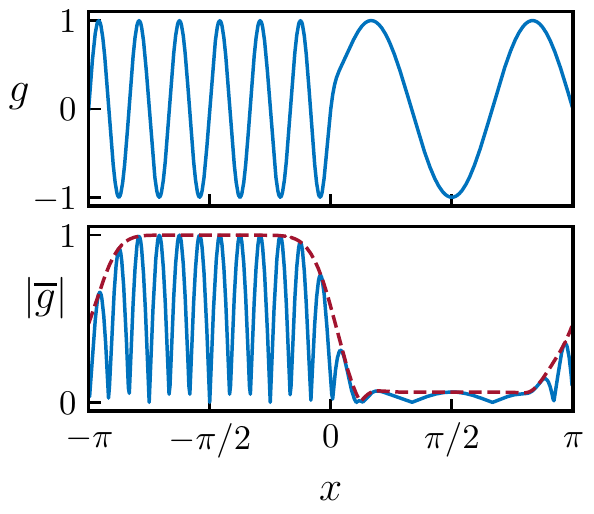}%
  \mletter{4.8cm}{-11.8cm}{(b)}\mletter{2.5cm}{-11.8cm}{(c)}%
\caption{(a) The wider filter on the left is the band-pass kernel $G(x,0)$ for $\lambdaf
= {4\pi}/{9}$. The solid darker line is $\Re(G)$, and the lighter one is $\Im(G)$. The two
envelopes are: \dashed, $\gamma(x,\Lambda_{0L})$; \chain, $\gamma(x,\Lambda_{0z})$. The narrower filter on the right
corresponds to $\lambdaf = {\pi}/{18}$, although the amplitude has been divided by four for
the purpose of plotting. (b) Sample signal, $g = \sin(12x) \text{ if } x < 0$, and $\sin(3x)$
otherwise. (c) Its amplitude after band-pass filtering with $\lambdaf = \pi/6$. The solid
line in (c) is the amplitude $|\Re(G)|$, and the dashed one corresponds to $|G|$.
}
\label{fig:mono}
\end{figure}

\begin{figure}
  \centering
\includegraphics[width=.5\textwidth]{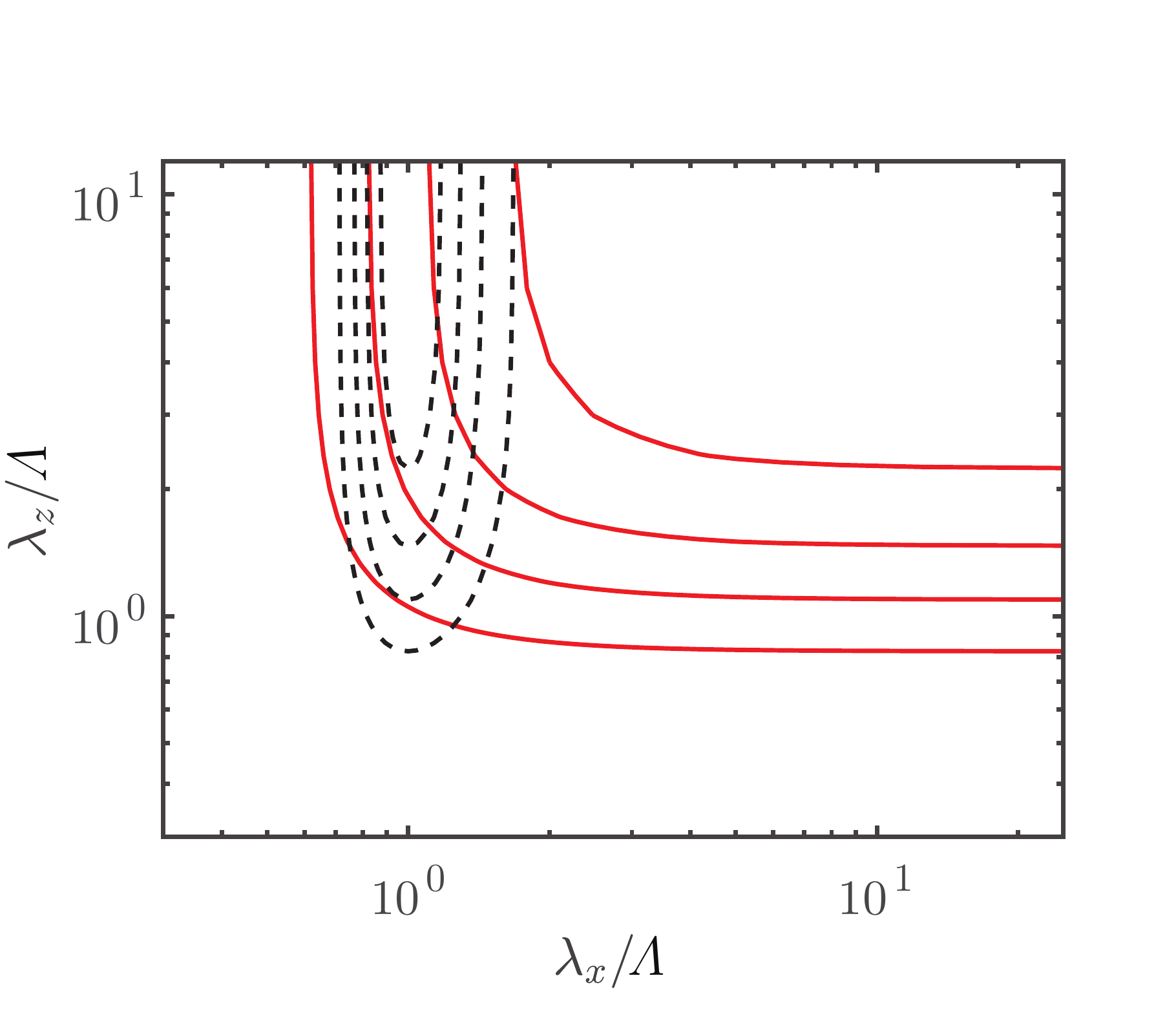}\mletter{4.7cm}{-13.2cm}{(a)}%
\hspace*{-2mm}%
\raisebox{4pt}{\includegraphics[width=.5\textwidth]{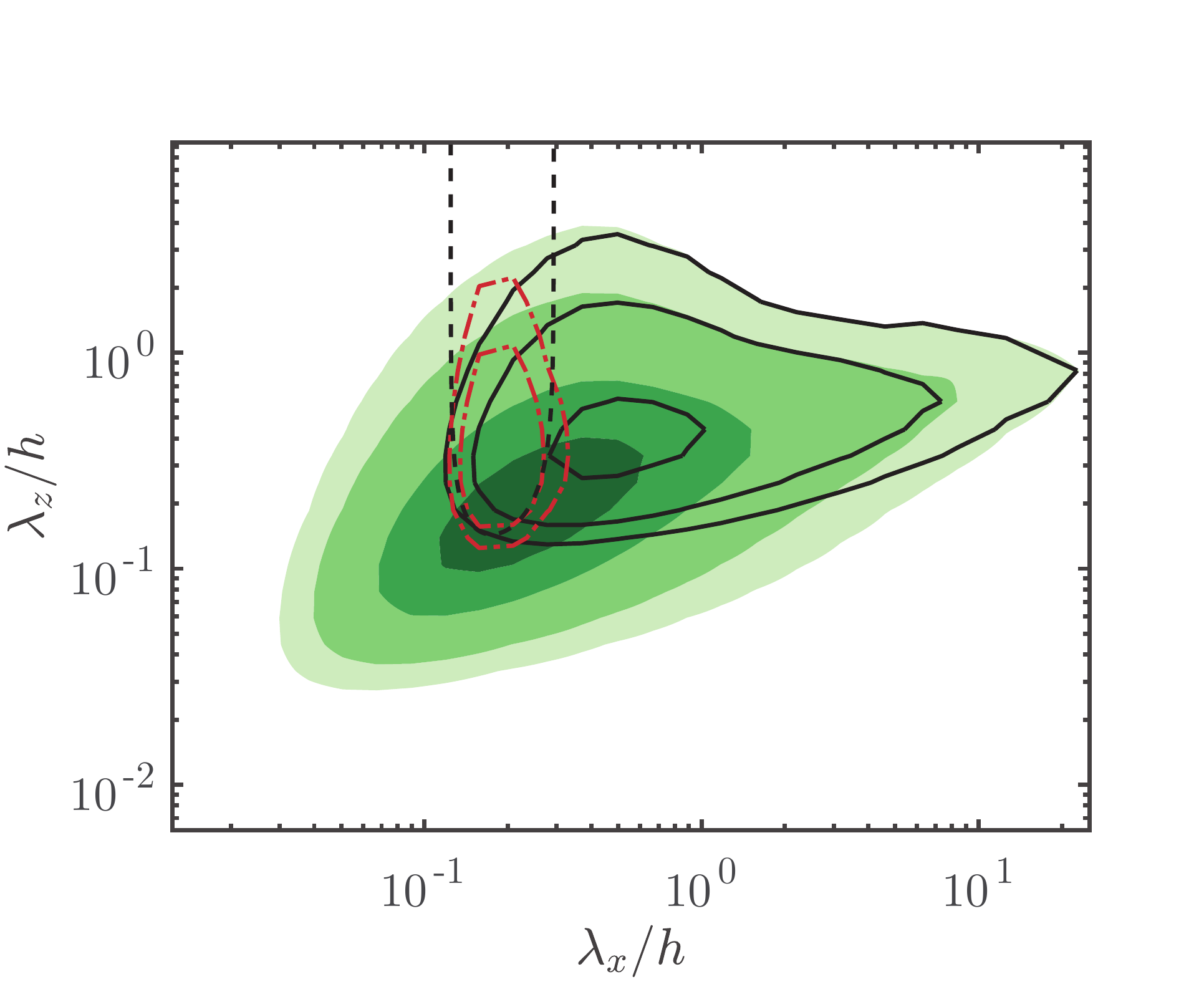}}\mletter{4.7cm}{-13.1cm}{(b)}%
\caption{(a) Transfer functions of the: \dashed, band-pass filter $G$; \solid, low-pass
filters $\Gamma$. The contours are $(0.2:0.2:0.8)$. (b) Sample filtered premultiplied spectra of 
$v$ in L2000, at $y/h=0.15$ and $\Lambda/h=0.175$. The dashed contour is
$|\widehat{G}|^2=0.2$. The shaded contours
are the full $k_x k_z E_{vv}$; \solid, low-pass filtered; \chain, band-pass filtered. The spectral
contours are $k_x k_z E_{vv}/u_\tau^2 =(0.014, 0.029, 0.071, 0.114)$.}
\label{fig:filters}
\end{figure}

The low-pass filter $\Gamma$ presents few conceptual problems, and its results are
described in \S\ref{sec:lowpass}, but the band-pass filter \r{eq:bpass} generates a complex
field which is not easily interpreted as a flow, and requires some discussion. Its
kernel $G$ is a monochromatic complex wave in the streamwise direction whose amplitude is
modulated by the Gaussian \r{eq:filgauss} in the streamwise and spanwise directions. It is
essentially a complex-valued continuous Morlet or Gabor wavelet \citep{farge:annrev}, whose
real and imaginary parts are represented in figure \ref{fig:mono}(a) as functions of $x$.
The number of oscillations contained within the Gaussian envelope is approximately
$2\lambdaf_{0x}/\lambdaf$. To keep the filters self-similar we choose
$\lambdaf_{0x}=\lambdaf/2$ and $\lambdaf_{0z}=\lambdaf/6$, implying an aspect ratio 
$\lambdaf_{0x}:\lambdaf_{0z}\approx$ $3:1$, which captures the strongest wall-normal
velocity perturbations (see figure \ref{fig:filters}b). Changing this aspect ratio to $2:1$
or 4:1 did not qualitatively change the results.

The main advantage of filtering a real-valued function with a complex wavelet is that the
real and imaginary parts of the resulting field preserve phase (positional) information,
while its absolute value hides the oscillations of the wavelet \citep{sre:85, farge:annrev}. A
visualisation for synthetic data is presented in figure \ref{fig:mono}(b).

In fact, the band-passed field, $\overline{a}$, of a generic velocity component can be
interpreted as the coefficient of a local Fourier expansion that optimally approximates the
flow within the Gaussian envelope \eqref{eq:env}. It minimises the weighted $L_2$ error
\begin{equation}
J[\overline{a}](x) = \int_\mathcal{B} \Gamma \left(x'-x, z'- z\right)\left| a(x', y, z') -
\overline{a}^\dagger (x, y, z) \exp\left[2\pi\ii (x'-x)/\lambdaf\right] \right|^2\dd x' \dd z',
\end{equation}
where the dagger denotes complex conjugation and $a$ is either $u$, $v$ or $w$. Note that $a^\dagger$ in this
integral is evaluated at the centre of the filtering kernel, $x$, rather than at the
integration variable, $x'$, and that the Fourier wavetrain, $\exp\left[2\pi\ii
(x'-x)/\lambdaf\right]$, is centred in each case at the filter position. This implies that
$a^\dagger$ can be interpreted as a field of coefficients of `local' Fourier
wavetrains, whose absolute value is $\left|a(x, y, z)\right|$, and whose argument
is $-\arctan(\Im(a)/\Re(a))$. Thus, although $\overline{\ub}$
cannot be used as a filtered velocity field, its absolute value is the
local wavetrain amplitude \citep{sre:85}, and the wall-normal derivative of its argument is a local
inclination angle, as in \cite{jim:pof:15}.

It is convenient to choose the short-wave limits of the low-pass and band-pass filters to be
of the same order, so that they isolate related structures. The (1/e) limit of \r{eq:tfG}{\color{refcolor2}, i.e. the argument that makes the filter gain equal to $(1/e)$,  } is
$k_x=2\pi/\lambdaf + 1/\lambdaf_{0x}$, while that of \r{eq:tfGam} is $k_x=1/\lambdaf_{0L}$.
Equating them results in $\lambdaf_{0L} = \lambdaf/(2\pi +\lambdaf/\lambdaf_{0x})$. For our
previous choice of $\lambdaf_{0x}=\lambdaf/2$, this implies $\lambdaf_{0L} = \lambdaf/8$. The
transfer function of the two filters is represented in figure \ref{fig:filters}(a). The
cospectrum of any two filtered variables can be computed from the unfiltered cospectrum as
$E_{\overline{a}\,\overline{c}}= E_{ac} |\widehat{G}|^2$, with an equivalent formula for
$E_{\widetilde{a}\widetilde{c}}$. An example of the effect of the band- and low-pass filters
on the spectrum of a representative plane of $v$ is shown in figure \ref{fig:filters}(b).

Note that the computation of the filtered fields can be done economically because the
streamwise and spanwise directions are periodic, and fast Fourier transforms can be used to
compute the convolutions {\color{refcolor3} \citep{can:hus:qua:zan:88}}.

\subsection{The band-pass-filtered pseudo-spectrum}\label{ssec:ops}

\begin{figure}
  \centering
  \vspace*{2ex}
  \includegraphics[width=\textwidth]{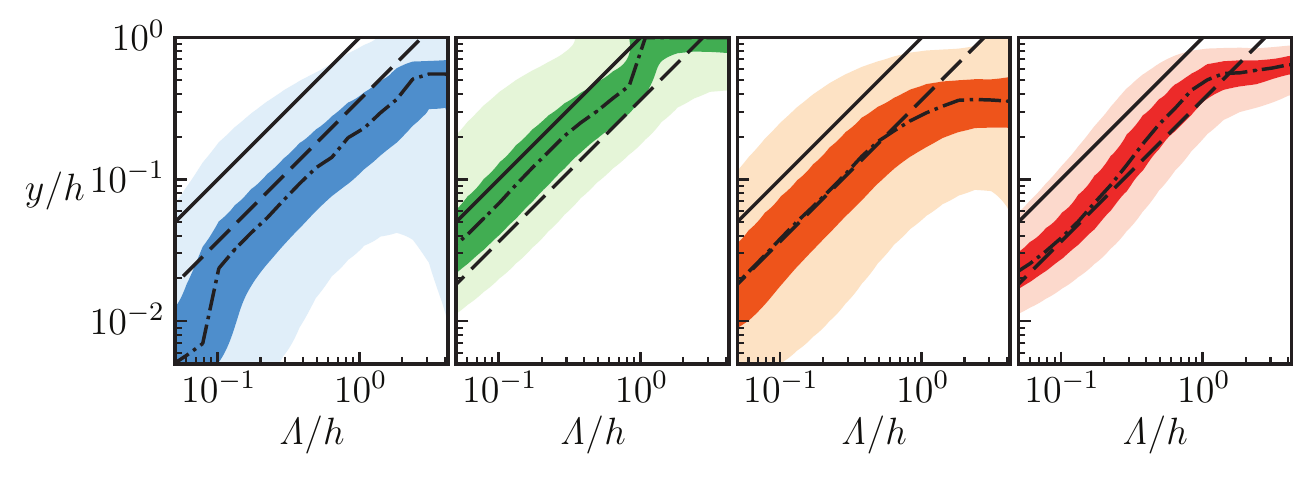}\mletter{4.7cm}{-20.7cm}{(a)}\mletter{4.7cm}{-15cm}{(b)}\mletter{4.7cm}{-9.3cm}{(c)}\mletter{4.7cm}{-3.5cm}{(d)}
\caption{Pseudo-spectra of the band-pass filtered velocity components and of their
tangential Reynolds stress, as a function of the filter wavelength and of the distance from
the wall. $Re_\tau = 5200$ \citep{lee:mos:15}. The contours are 90\% and 50\% of the maximum {\color{refcolor2} as a function of $y$, }
for each filter wavelength. The solid and dashed lines bracket the most energetic part of
$\mathcal E_{vv}$ in (b). The upper (solid) limit is $y = \Lambda$. The lower (dashed) one
is $y = 4\Lambda/11$. The chain-dotted line is the wall-normal location of the maximum for
each filter width. (a) $\mathcal E_{uu}$. (b) $\mathcal E_{vv}$. (c) $\mathcal E_{ww}$. (d)
$-\mathcal E_{uv}$.}
\label{fig:speck}
\end{figure}

The perturbation energy of the filtered flow fields can be directly computed from their spectrum:
\begin{equation}
\mathcal E_{ac}(y; \lambdaf) = \sum_{\forall k_x, k_z} E_{\overline{a}\,\overline{c}}(k_x, y, k_z; \lambdaf) = \sum_{\forall k_x, k_z} |\widehat{G}|^2 (k_x, k_z;\lambdaf) E_{ac}(k_x, y, k_z),
\end{equation}
which is represented in figure \ref{fig:speck} for the three velocity components and for
the tangential Reynolds stress, as a function of $y$ and of the filter wavelength $\lambdaf$.
Because the filter wavelength represents the size of the scales retained by the filter, we
can think of $\mathcal E_{ac}(y; \lambdaf)$ as a `pseudo-spectrum' with pseudo-wavelength
$\lambdaf$. Figure~\ref{fig:speck} shows that the wall-normal location of the maximum
energy is proportional to the filter width for a range of heights corresponding to an
extended logarithmic layer, $y \in (100\nu/u_\tau, 0.4\, h)$. This is also approximately the region where a
self-similar hierarchy of wall-attached velocity structures with sizes proportional to their
distance from the wall is found to exist \citep{Tow:76}, and we will refer to it as our
definition of the logarithmic layer from now on. {\color{refcolor2} Note that, although the filter is self-similar in the streamwise and spanwise directions, it contains no information about the sizes in $y$, and thus the uncovered self-similarity in $y$ is not a property of the filter, but of the spectra of the velocity components.} {\color{refcolor3} Owing to the removal of the large scales by the filter, the wall-parallel velocities also show long self-similar ranges, albeit centred at larger scales. This was shown also to be the case in \citet{abe:ant:18}, where a minimal streamwise unit was used to remove the large scales of $w$, uncovering a self-similar behaviour for this component.}

Figure \ref{fig:speck} allows us to determine which distances from the wall should be
used to characterise eddies of different scales. The band-pass filter \eqref{eq:fil}
produces a flow field that is locally averaged within each wall-parallel plane, but eddies
and bursts also have a vertical dimension, which figure \ref{fig:speck} shows to be proportional to
their horizontal ones \citep{loz:flo:jim:2012}. It was shown by \cite{jim:pof:13} that bursts
may be characterised by a vertically averaged amplitude, 
\begin{equation}
A^2_{\overline a} = \frac{1}{y_1 - y_0}{\int_{y_0}^{y_1} \left|\overline{a}\right|^2 {\dd}y},\label{eq:A}
\end{equation}
and by a mean inclination angle,
\begin{equation}
\Psi_{\overline a} = \arctan\dfrac{\lambdaf\int_{y_0}^{y_1} \Im(\overline{a}^\dagger\partial_y \overline{a})
 {\dd}y}{2\pi\int_{y_0}^{y_1} \left|\overline{a}\right|^2 {\dd}y},\label{eq:Psi}
\end{equation}
where $a$ stands for either $u$, $v$ or $w$, and the integrals extend over the
intense part of the eddy. The average in \eqref{eq:Psi} is weighted with the
square of the amplitude because strong perturbations typically coexist in
the filtering window with weaker ones, whose inclination angle is not necessarily
small, nor relevant.

The limits $y_0$ and $y_1$ are chosen to bracket the intense band of $\mathcal E_{vv}$ in
figure \ref{fig:speck}(b), and the same limits are used for all variables. The upper
limit is very close to the Corrsin spectral scale, $y_1= \Lambda$, above which the shear is
too weak to interact with eddies of size $\lambdaf$ \citep{jim:18}. The lower limit,
$y_0=4\lambdaf/11$, also scales with $\lambdaf$, and represents the point below which
impermeability damps $v$. Because these limits are adjusted for the wall-normal velocity, while
the $u$ or $w$ structures are larger than those of $v$, figure \ref{fig:speck}(a, c) shows
that they are too far from the wall to capture more than a fraction of the energy of the wall-parallel
components. {\color{refcolor2} Approximately energy of the  band-passed spanwise component is captured within the band, closer to the lower limit, potentially associated with the short scales of $w$. Longer features of $w$ fall outside the band, and probably come from larger structures farther away from the wall \citep{ala:jim:zan:mos:04}.} On the other hand, the tangential Reynolds stress in figure \ref{fig:speck}(d)
is captured well, supporting the classical classification into `active' and `inactive'
{\color{refcolor2} eddies of} \cite{Tow:76}{\color{refcolor2}, as a substantial amount of longer $u$ and $w$ perturbations carry little tangential Reynolds stress}. 

From now on, we will use the quantities in \r{eq:A}--\r{eq:Psi} to represent the local
amplitude and inclination angle of our flow fields. Their main practical advantage is that they reduce
the dimensionality of our dataset, because each filter has an associated three-dimensional
data space (the two wall-parallel positions and time), instead of the full four-dimensional
one. In addition, they act as a filter in $y$, excluding structures outside the band
$(y_0,\, y_1)$, or with very different vertical dimensions from the band thickness. They are centred at
$[x, (y_0 + y_1)/2, z]$, and the vertical average is only meaningful whenever a velocity
perturbation of size comparable to $\lambdaf$ happens to be within the vertical integration window.
Otherwise, the average reverts to the unconditional mean. For example, when several wavetrains
with different inclinations are stacked along the wall-normal direction, the integral of
their phase derivatives produces an average inclination angle which is similar to the global
ensemble-averaged inclination.

\subsection{Filter parameters}\label{ssec:fp}

Table \ref{tab:fil} gathers the information for the filters used in the paper. We use six wavelengths,
\begin{equation}
\lambdaf_i/h = \frac{8\pi}{9}2^{-i},\quad i = 1,\ldots 6, 
\end{equation}
with the corresponding filtered fields denoted by $\overline{\ub}_i=\overline{\ub}(x, y, z, t;
\lambdaf_i)$ for the band-pass filters discussed in \S\ref{sec:bandpass}, and
$\widetilde{\ub}_i=\widetilde{\ub}(x, y, z, t; \lambdaf_i)$ for the low-pass ones in
\S\ref{sec:lowpass}. They cover the energy spectrum with logarithmically equispaced bands of
wavelengths, and span the range that can be expected to be relevant for the logarithmic
layer. The long-wavelength limit of the widest filter is kept constant in outer units across
Reynolds numbers, $\lambdaf_1/h = {4\pi}/{9}$. It follows from figure \ref{fig:speck} that
this filter corresponds to $y/h\in(0.5,\,1.4)$, which is well above the expected
self-similar region, but it is retained here to compare it to the experiments in
\cite{jim:pof:15}, who analysed the first streamwise mode of a minimal channel with $L_x =
{\pi}/{2}$. It allows us to test whether the minimal domain used in that work affected
the results. The short-wavelength limit, $\lambdaf^+\approx 180$, scales in wall units,
resulting in an increasing number of filter bands as the Reynolds number increases. In the
highest Reynolds number case S4000, this filter corresponds to $\lambdaf_6$. Since both our
Reynolds numbers and our filter widths are approximately spaced by powers of two, there are
matching filters for all the simulations, both in outer and in inner units, allowing us to
compare scaling criteria.

\begin{table}
  \begin{center}
    \def~{\hphantom{0}}
    \begin{tabular}{c c c c c c c c c c}
      Name & $\lambdaf/h$ & $y_0/h$ & $y_1/h$ & S1000 & L1000 & S2000 & F2000 & L2000 & S4000\\
      $\lambdaf_1$ & $1.4$ & 1.0 & 0.51 &  $\circ$ & $\circ$ & $\circ$ & $\circ$ & $\circ$ & $\circ$\\
      $\lambdaf_2$ & $0.7$ & 0.7 & 0.25 & $\circ$ & $\circ$ & $\circ$ & $\circ$ &  & $\circ$\\
      $\lambdaf_3$ & $0.35$ & 0.35 & 0.13 & $\circ$ & $\circ$ & $\circ$ & $\circ$ &  & $\circ$\\
      $\lambdaf_4$ & $0.175$ & 0.175 & 0.063 & $\circ$ & $\circ$ & $\circ$ & $\circ$ &  & $\circ$\\
      $\lambdaf_5$ & $0.087$ &  0.087 & 0.032 &  & & $\circ$ &  &  & $\circ$\\
      $\lambdaf_6$ & $0.044$ &  0.044 &0.016 &  &  & &  & & $\circ$\\
    \end{tabular}
    \caption{Properties of the filters used. $\lambdaf$ is the filter width defined in
    \eqref{eq:fil}, and $y_0$, $y_1$ are the integration limits in \eqref{eq:A}--\eqref{eq:Psi}. A
    circle below a simulation indicates that that filter was computed for it.}
    \label{tab:fil}
  \end{center}
\end{table}

To estimate how much energy is retained by our band-pass filtering and vertical
integration operations, we compute the ratio between the energy of the filtered and
unfiltered wall-normal velocity field within a given band of wall distances, as a function of the filter wavelength. It ranges from $10\%$ to $20\%$, with the higher limit corresponding to the narrower filters.
These values are similar to that for the single most energetic mode
in \citet{jim:pof:15}. In our case, we can study multiple sizes covering a broad range of
scales, but the quantification of the total filtered energy is not straightforward because
the spectral and wall-normal bands of the filters overlap. One way of getting the equivalent to a Parserval's
theorem for localised basis functions would be to project the velocity onto an orthonormal wavelet basis
instead of using continuous wavelets \citep{men:91}, but at the cost of limiting the spatial
locations at which we could have information of a given scale, making it unsuitable for
smoothly tracking the  temporal evolution of the structures. Another approach is to
rescale the kernels so that they recover the total energy when integrated over scale space
\citep{leu:swa:dav:12}, but this distorts the amplitudes, and only works for a particular
spectrum. We have chosen to make our transfer functions unity at their nominal wavelength.
The energy contained in the overlap of neighbouring filters is then about 2\%--5\%
(non-neighbouring bands are irrelevant because they do not overlap vertically). Accounting for
overlaps, our filters approximately retain 20\% of the total energy of the wall-normal velocity in
$60~\nu/u_\tau < y < h$ at $Re_\tau = 2000$, with minor differences among Reynolds numbers. {\color{refcolor2} This energy ratio concerns the band-pass filter that is designed to isolate the bursts. The less isolating low-pass filter retains 50\%($\lambdaf_1$)--70\%($\lambdaf_4$) of the turbulent kinetic energy, and we will see through the paper that most of the conclusions obtained for the band-passed structures carry over to the low-passed ones.}

\section{Band-pass-filtered amplitude and inclination fields}\label{sec:bandpass}

\begin{figure}
  \centering
  \includegraphics[width=\textwidth]{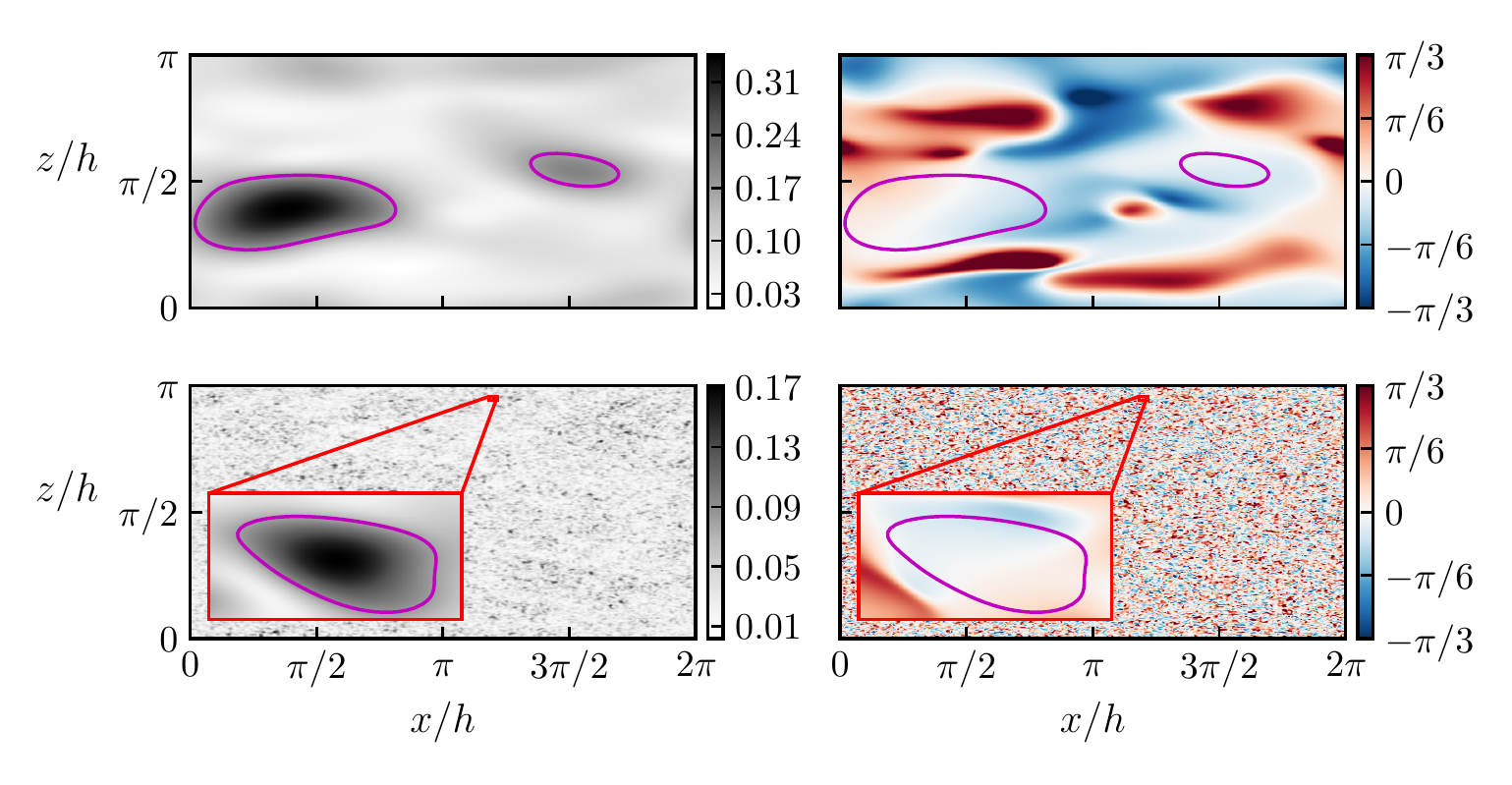}
  \caption{Snapshots of the amplitude, $A^+_{\av}$(left), and inclination angle, $\Psi_{\av}$(right), of the band-pass-filtered wall-normal velocity of S4000. Top corresponds to $\overline{v}_1$ and bottom to $\overline{v}_6$. While the top snapshot has been carefully picked to contain a burst at this scale, the bottom one is randomly chosen, because all snapshots contain at least one burst. The inset amplifies a region of $\overline{v}_6$ with a burst. The amplification factor, 32, corresponds to the wavelength ratio between both filters. The contours are $A^+_\av = 0.5\max{\{A^+_\av\}}$.\label{fig:planes}}

\end{figure}

Figure \ref{fig:planes} shows snapshots of the amplitude and inclination angle of the band-pass-filtered
wall-normal velocity, as defined in \eqref{eq:A}--\eqref{eq:Psi}. Figure
\ref{fig:planes}(a, b) uses the widest of our six filters, $\overline{v}_1\,(\lambdaf=1.4
h)$, while figure \ref{fig:planes}(c, d) uses the narrowest one, $\overline{v}_6\,
(\lambdaf=0.044 h)$, which is 32 times narrower. To facilitate comparison, the insets in
figures \ref{fig:planes}(c, d) are magnified by the ratio of the two filter widths, resulting
in structures of similar size to those in figures \ref{fig:planes}(a, b). This supports
our remark in \S\ref{sec:filterdef} that the band-pass filter isolates structures of size
proportional to $\lambdaf$. The amplitude fields in figures \ref{fig:planes}(a, c) are
smooth, with distinct intense regions which are candidates for Orr events at the moment of
peak amplitude. They are highlighted as line contours in both the amplitude and inclination
fields, and it is visually clear from figures \ref{fig:planes}(b, d) that the regions of high
intensity are associated with vertical inclinations.

\begin{figure}
  \centering
  \newcommand{\sca}{0.91}
    \includegraphics[scale=\sca]{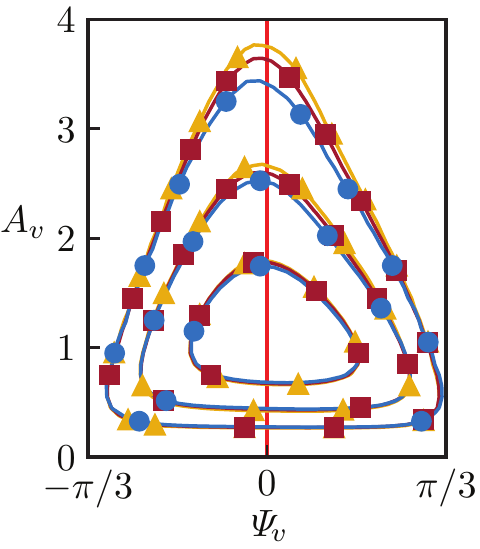}\mletter{4.5cm}{-6.6cm}{(a)}\mletter{2.8cm}{-8.6cm}{\rotatebox[origin=c]{90}{\setlength{\fboxsep}{0pt}\colorbox{white}{\rule{0pt}{1.5em}${A_{\overline v}}/{A_{\overline vM}}$}}}\hfill%
    \includegraphics[scale=\sca]{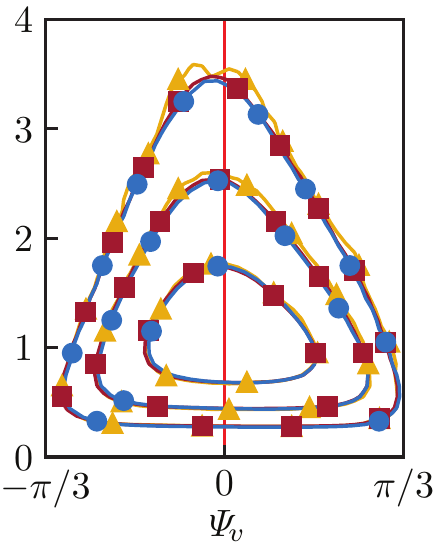}\mletter{4.5cm}{-6.6cm}{(b)}\hfill%
    \includegraphics[scale=\sca]{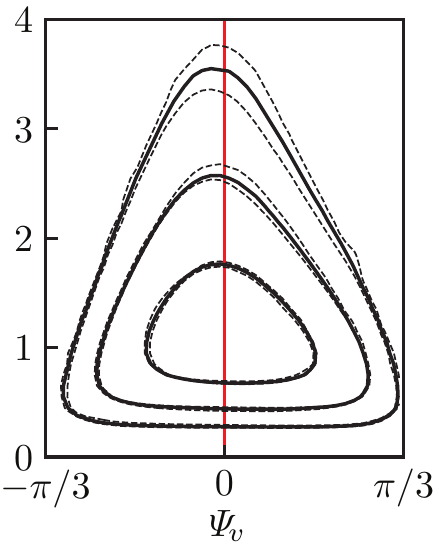}\mletter{4.5cm}{-6.6cm}{(c)}
    \vspace{10pt}\\
    \includegraphics[scale=\sca]{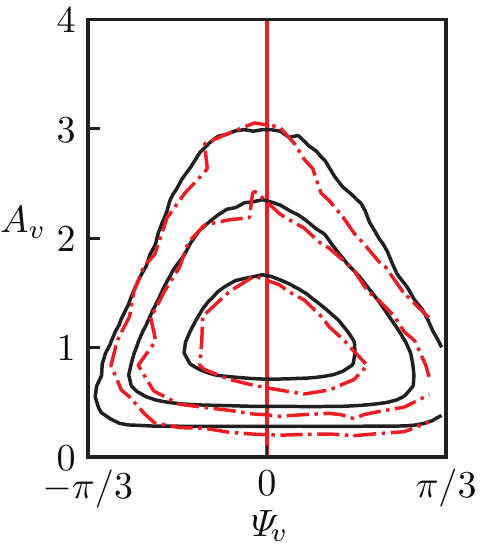}\mletter{4.5cm}{-6.6cm}{(d)}\mletter{2.8cm}{-8.6cm}{\rotatebox[origin=c]{90}{\setlength{\fboxsep}{0pt}\colorbox{white}{\rule{0pt}{1.5em}${A_{\overline v}}/{A_{\overline vM}}$}}}\hfill%
    \includegraphics[scale=\sca]{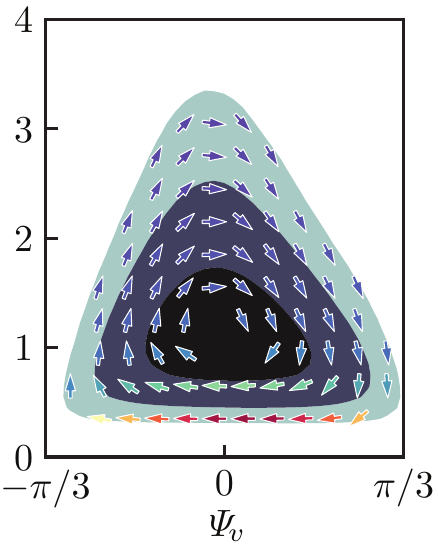}\mletter{4.5cm}{-6.6cm}{(e)}\hfill%
    \includegraphics[scale=\sca]{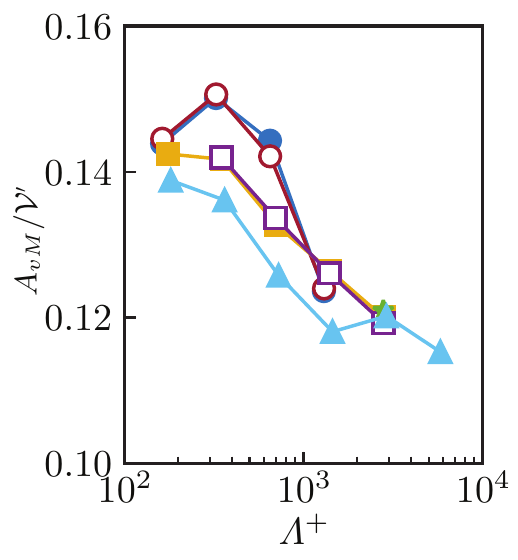}\mletter{4.5cm}{-6.6cm}{(f)}
    \caption{(a--e) Joint p.d.f.s of the amplitude, $A_{\av}$, and inclination angle, $\Psi_{\av}$, of the filtered wall-normal velocity. The amplitudes are normalised with their mode, given  in (f). Contours contain $50, 90, 99\%$ of the probability mass.
    (a) Joint p.d.f. of $\overline{v}_3$ for S1000, S2000 and S4000.
    (b) Joint p.d.f. of $\overline{v}_3, \overline{v}_4, \overline{v}_5$ for S1000, S2000 and
    S4000, respectively.
    (c) Averaged joint p.d.f. for the filters that fall in the logarithmic region, $\overline{v}_2$
    to $\overline{v}_6$, for the available cases, as in table \ref{tab:fil}. The solid lines are
    the averaged p.d.f.. The dashed lines along each contour bracket the minimal area that contains
    all of the individual joint p.d.f.s.
    (d) The solid lines are the average of all the joint p.d.f.s of $\overline{v}_1$. The
    chain-dotted red line is as in figure 3 in \cite{jim:pof:15}. 
    (e) Joint p.d.f. of $\overline{v}_3$ for F2000. The arrows are the conditional mean velocities
    (CMV), normalised to unit length. The colours are the standard deviation of the CMVs
    normalised with their mean, ranging from 0.4 (dark blue) to 6.5 (red).
    (f) Modal values used to normalise the amplitudes.
    Symbols: \fullcirc, S1000; \opencirc, L1000; \fullsquare, S2000; \opensquare, F2000;
    \fullstar, L2000; \fulltriangle, S4000.
    }%
  \label{fig:jpdf1}
\end{figure}

The joint probability density function (p.d.f.) of $A_\av$ and $\Psi_\av$ is presented in figure
\ref{fig:jpdf1}(a--e), with the amplitude normalised by the modal value, $A_{\av M}$, at which
its p.d.f. is maximum. As in \cite{jim:hoy:2008}, the p.d.f.s collapse better in this
normalisation than with the mean. While figure \ref{fig:planes} emphasises the
self-similarity of the filtering operation, figure \ref{fig:jpdf1}(a--c) tests the
self-similarity of the filtered flow, which should hold both in inner and in outer units
throughout the logarithmic region. Figures \ref{fig:jpdf1}(a) and \ref{fig:jpdf1}(b) each
includes p.d.f.s at three Reynolds numbers, but those in figure \ref{fig:jpdf1}(a) have the
same filter width in outer units $(\lambdaf/h=0.35)$, and those in figure
\ref{fig:jpdf1}(b) have the same width in inner units $(\lambdaf^+\approx 350)$. In both
cases, the p.d.f.s agree well. Figure \ref{fig:jpdf1}(c) collects the p.d.f.s for all the Reynolds
numbers and all the filters whose vertical domain is contained within the logarithmic region
(see table \ref{tab:fil}). The solid contours are the average of all the p.d.f.s, and the
dashed ones bracket the narrowest band that contains the corresponding contours of all the
cases. It is clear from these results that the similarity of the p.d.f.s is satisfied extremely
well.

Figure \ref{fig:jpdf1}(d) displays the p.d.f. of the widest filter $\overline{v}_1$, averaged
over all the flows in table \ref{tab:fil}. This filter is too wide to collapse with the ones
in the logarithmic layer, but all the p.d.f.s used in figure \ref{fig:jpdf1}(d) also agree well
among themselves (not shown). The averaged p.d.f. is compared in the figure with the results in
\cite{jim:pof:15}, who analysed a single Fourier mode of wavelength comparable to
$\lambdaf_1$. The good agreement validates the approximate equivalence between the new
methodology and the monochromatic analysis in \cite{jim:pof:15}.

The collapse of the p.d.f.s in figure \ref{fig:jpdf1} also supports that the behaviour of the
$v$-bursts is relatively independent of the numerical box. Figure \ref{fig:jpdf1}(a--e)
contains p.d.f.s from a variety of numerical boxes, which agree well among themselves, and the
p.d.f. from \cite{jim:pof:15} in figure \ref{fig:jpdf1}(d) uses a very small box, $L_x/h =
\pi/2, L_z/h = \pi/4$, which only represents well the structures below $y/h\approx 0.25$,
but which is too small for the larger ones farther from the wall.

The p.d.f.s in figure \ref{fig:jpdf1} have two distinct regions. Their core, which includes
most of the probability mass, contains vertically oriented structures with amplitudes of the order of
the modal value $A_{\av M}$. That the typical inclination of $v$ in channels is vertical
had previously been shown using proper orthogonal decomposition by \cite{moi:mos:1989},
and using autocorrelation functions by \cite{sil:jim:14}.

The upper part of the outermost isocontour of the p.d.f.s contains large amplitudes and
inclinations, and is approximately triangular. The extreme values and low probabilities in this region
suggest that these points represent individual structures, while the triangular shape
implies a definite statistical relation between the inclination angle and the intensity, as
graphically suggested by figure \ref{fig:planes}. Weak regions are inclined either forwards
or backwards, and strong ones are approximately vertical or slightly tilted backwards, as in
the transient linearised bursts discussed in \S\ref{ssec:otg}.

Moreover, the upper part of the p.d.f. is traversed from left to right, as expected of
shear-dominated structures. This was already shown to be the case for monochromatic
wavetrains in \citet{jim:pof:15}, but is confirmed here for localised structures in large
boxes. In the time-resolved cases, S1000 and F2000, we can define conditional mean
velocities (CMVs) in the $(\Psi_\av, A_\av)$ parameter space as
 \begin{equation}
\bigg\langle\left[ \frac{{\Dr} \Psi_\av}{{\Dr} t}, \frac{{\Dr} A_\av}{{\Dr} t}\right]
\bigg\rangle_{(\Psi_\av, A_\av)},
\label{eq:CMV}
 \end{equation}
where $\langle\,\rangle_{(\Psi_\av, A_\av)}$ denotes conditional averaging at $(\Psi_\av, A_\av)$,
and $\Dr /\Dr t = \partial_t + C_{\!\lambdaf}\partial_x$, is a semi-Lagrangian approximation to the total
derivative that incorporates an advection velocity, $C_{\!\lambdaf}$, 
estimated by linear regression of the time-dependent $x$ location of the maximum of the
$x$--$t$ two-point two-time autocorrelation of $A_\av$, using at each instant a centred
seven-point time stencil. In this approximation, the advection velocity is assumed to be unique for the
whole $(x, z)$ plane, independently of the number of bursts present at each moment, and depending only on the filter width. This is
justified because the structures of the logarithmic layer were shown not to be dispersive by
\cite{loz:jim:2014} and by \cite{jim:18}, and because the uniform scale of the band-pass-filtered
variables would probably guarantee a uniform advection velocity even in they were. The rest
of the temporal derivatives in \r{eq:CMV} use five-point centred finite differences. Note
that, although the advection velocity of individual structures is known to be approximately
equal to the mean flow velocity, and is therefore a function of $y$, the velocity used here
for the vertically integrated bursts is independent of the wall distance.

The result is presented in figure \ref{fig:jpdf1}(e) for F2000, where the CMVs are plotted
in the $(\Psi_\av, A_\av)$ plane as arrows pointing towards the next most probable state. The
arrows are coloured by the conditional standard deviation of the CMV, which is high in the
core and lower edge of the p.d.f., and low in its upper edge, as in \citet{jim:pof:15}. This
suggests again that the mean velocities in the upper part of the p.d.f. are representative of
individual coherent events, although it is unclear at this point whether the full periphery 
can be considered to represent a single burst, or whether it is formed by tangents of different 
burst trajectories at different locations. 

While the coherent part of the p.d.f. is traversed from left to right, its lower part is
traversed from right to left, closing the cycle. This is inconsistent with linear models,
but the standard deviation of the CMVs in this region is large, suggesting that it is populated
by structures that cannot be characterised solely by their position in the $(\Psi_\av,
A_\av)$ plane, and whose evolution cannot be modelled by quasi-linear dynamics.  

Finally, figure \ref{fig:jpdf1}(f) shows the modal values used to normalise $A_\av$ in the
rest of figure \ref{fig:jpdf1}. Different box sizes yield the same modal value, which is
fairly constant when compared with the wall-normal velocity fluctuation averaged within the
same bands:
\begin{equation}
    (\mathcal{V}^\prime)^2 = \frac{1}{y_1-y_0}\int_{y_0}^{y_1} (v^\prime)^2 \dd y.
\end{equation}
This constancy is a consequence of the self-similar definition of the filter, whose transfer
function has constant logarithmic width in $\lambda_x$, independently of $y$ and $\lambdaf$.
Since figure \ref{fig:speck} shows that the wavelengths of the spectrum of the wall-normal
velocity also scale self-similarly with $y$, the energy selected by a truly self-similar
filter should be approximately constant, but, because our band-pass filters are defined as
low-pass in $\lambda_z$ (see \S\ref{sec:filterdef}), the transfer function of the narrower filters
spans a wider range of $\lambda_z$. The extra energy in these wavelengths explains
the slightly larger modal values of the narrow filters in figure \ref{fig:jpdf1}(f).

\subsection{Wall-parallel velocities}\la{ssec:parallel}

\begin{figure}
  \centering
  \newcommand{\sca}{0.88}
  
  \includegraphics[scale=\sca]{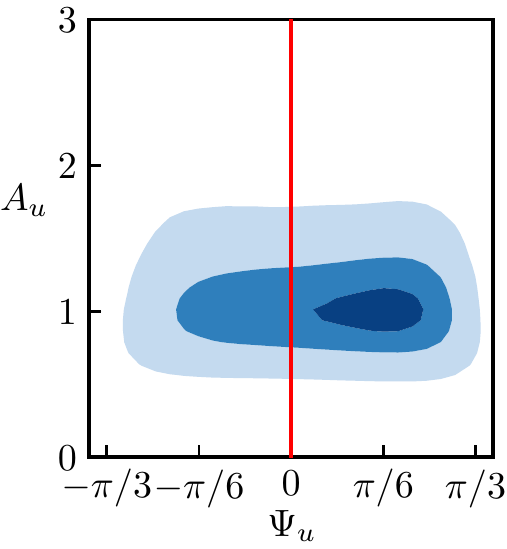}\mletter{4.3cm}{-6.6cm}{(a)}\mletter{2.8cm}{-8.8cm}{\rotatebox[origin=c]{90}{\setlength{\fboxsep}{0pt}\colorbox{white}{\rule{0pt}{1.5em}${A_{\overline u}}/{A_{\overline uM}}$}}}\hspace{1.4cm}%
  \includegraphics[scale=\sca]{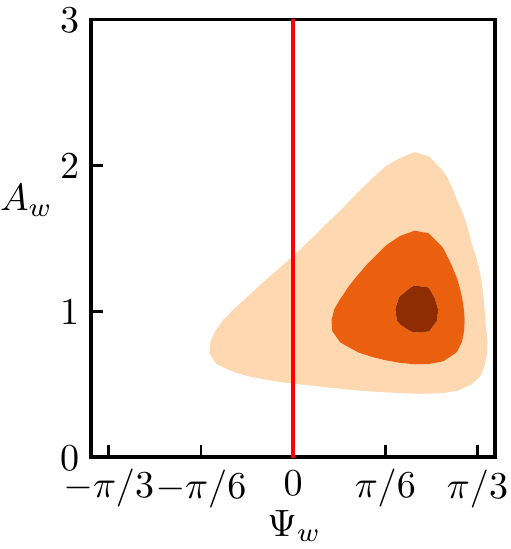}\mletter{4.3cm}{-6.6cm}{(b)}\mletter{2.8cm}{-8.8cm}{\rotatebox[origin=c]{90}{\setlength{\fboxsep}{0pt}\colorbox{white}{\rule{0pt}{1.5em}${A_{\overline w}}/{A_{\overline wM}}$}}}\hspace*{18pt}%

  \includegraphics[scale=\sca]{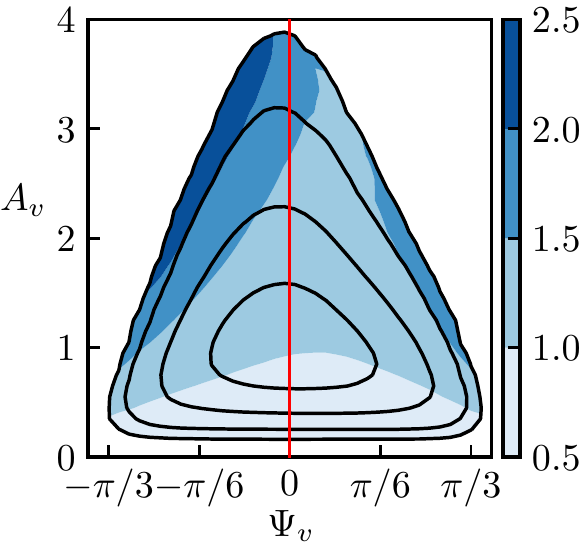}\mletter{4.3cm}{-7.8cm}{(c)}\mletter{2.8cm}{-10.2cm}{\rotatebox[origin=c]{90}{\setlength{\fboxsep}{0pt}\colorbox{white}{\rule{0pt}{1.5em}${A_{\overline v}}/{A_{\overline vM}}$}}}\hspace{20pt}
  \includegraphics[scale=\sca]{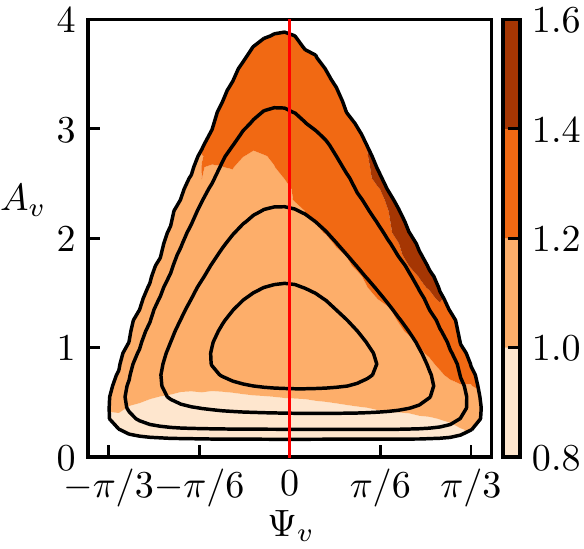}\mletter{4.3cm}{-7.8cm}{(d)}\mletter{2.8cm}{-10.2cm}{\rotatebox[origin=c]{90}{\setlength{\fboxsep}{0pt}\colorbox{white}{\rule{0pt}{1.5em}${A_{\overline v}}/{A_{\overline vM}}$}}}

  \includegraphics[scale=\sca]{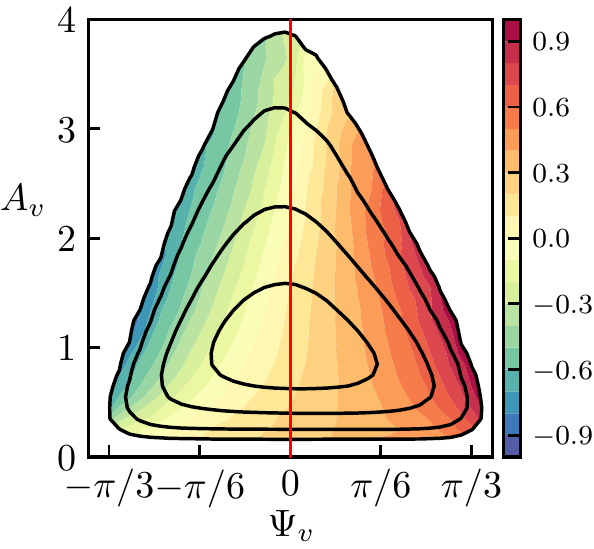}\mletter{4.3cm}{-7.8cm}{(e)}\hspace{20pt}\mletter{2.8cm}{-11.7cm}{\rotatebox[origin=c]{90}{\setlength{\fboxsep}{0pt}\colorbox{white}{\rule{0pt}{1.5em}${A_{\overline v}}/{A_{\overline vM}}$}}}
  \includegraphics[scale=\sca]{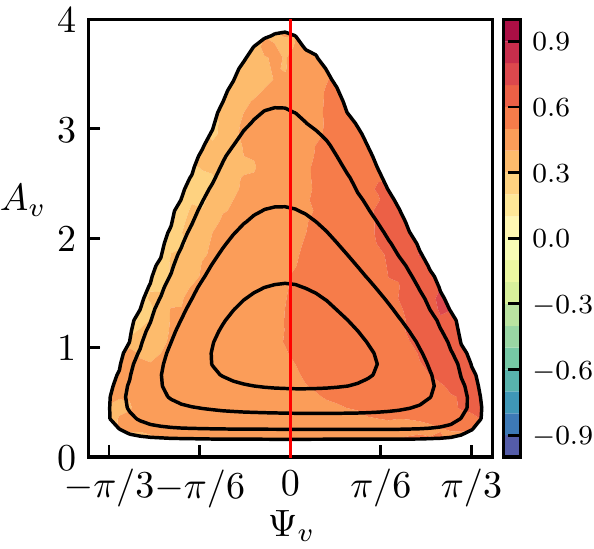}\mletter{4.3cm}{-7.8cm}{(f)}\mletter{2.8cm}{-10.3cm}{\rotatebox[origin=c]{90}{\setlength{\fboxsep}{0pt}\colorbox{white}{\rule{0pt}{1.5em}${A_{\overline v}}/{A_{\overline vM}}$}}}

\caption{Joint p.d.f.s of $A$ and $\Psi$ involving $\overline{u}_3$ and $\overline{w}_3$ for
S1000. All amplitudes are normalised with their modal values. (a) $\Psi_\au$ and
$A_\au/A_{\au M}$, contour levels are $90\%$, $50\%$ and $10\%$ of the probability mass. (b)
$\Psi_w$ and $A_\aw/A_{\aw M}$, contours as in \emph{(a)}. (c--f) Black contours are the
joint p.d.f. of $A_{\av}/A_{\av M}$ and $\Psi_{\av}$, contours containing $99.9\%$, $99\%$,
$90\%$ and $50\%$ of the probability mass. Shaded contours are the mean of several
quantities conditioned to the value of $A_{\av}$ and $\Psi_{\av}$. (c) Conditional average
of $A_\au/A_{\au M}$. (d) $A_\aw/A_{\aw M}$. (e) $\Psi_\au$. (f) $\Psi_\aw$.
\label{fig:jpdfmeans}}
\end{figure}

We next compute the amplitude and inclination angles of the band-pass-filtered wall-parallel
velocities. Figure \ref{fig:jpdfmeans}(a, b) shows the joint p.d.f.s of $(\Psi,\, A)$ for
$\overline u$ and $\overline w$, respectively. Their structures are preferentially tilted
towards the direction of the shear, in contrast to the wall-normal velocity structures,
which are almost equally distributed between forward and backward inclinations. This is more
pronounced for the spanwise velocity, which is rarely tilted backwards (10\% probability),
and subtler for the streamwise component, which is tilted backwards 40\% of the time. As
with the wall-normal velocity, there is a statistical correlation between the amplitude and
inclination of the spanwise velocity, but the same is not true for the streamwise velocity,
for which the two quantities are essentially unrelated. Scrambling $A_\au$ with respect to
$\Psi_\au$ leaves their joint p.d.f. unchanged (not shown). The behaviour of the inclination
angles and amplitudes of $\overline u$ and $\overline w$ in the upper part of their joint
p.d.f.s is qualitatively similar to the linearised trajectories in figure \ref{fig:lineal}(b),
although the inclination angles differ quantitatively. For example, the spanwise velocity
shares with the transient-growth model an amplitude `peak' towards positive inclinations,
but the tilting angle of the linearised peak is always larger than that of the direct numerical simulation by
0.1--0.2 radians. Using unconditional two-point correlations of the wall-normal and spanwise
velocity components, \cite{jim:18} argued that the cross-stream velocities contributing to
the correlation are part of a quasi-streamwise `roller'. Some of the features of those
autocorrelation functions are shared by the intense core of the joint p.d.f.s in figures
\ref{fig:jpdf1} and \ref{fig:jpdfmeans}. For example, the autocorrelation of the
wall-normal velocity in the logarithmic layer is approximately vertical ($\Psi_\av \approx
0$), and the spanwise velocity is tilted forward by $\Psi_\aw \approx \pi/6$, in agreement
with the amplitude peaks of the joint p.d.f.s of $\overline v$ and $\overline w$. Because the
correlation function is dominated by strong events, these similarities are not surprising,
but they confirm that the band-pass filter retains some of the structure of the intense
events of the velocity.

To explore the relation between the different variables during the bursting cycle, we
compute averages conditioned to ($\Psi_\av, A_\av$). The conditional mean amplitudes of
$\overline{u}$ and $\overline{w}$ are presented in figures \ref{fig:jpdfmeans}(c, d), and
their conditional inclinations are shown in figure \ref{fig:jpdfmeans}(e, f). If we assume
that the CMVs in figure \ref{fig:jpdf1}(e) reflect the temporal evolution of individual
strong bursts, we can describe the burst by the evolution of the conditional mean values of
the different quantities as we move from left to right along the upper edge of the p.d.f. of
($\Psi_\av, A_\av$). The burst starts from intense $\overline{u}$, with weaker
$\overline{v}$ and $\overline{w}$. At this stage $\overline{u}$ and $\overline{v}$ are
tilted upstream, but $\overline{w}$ is too weak for its inclination to be defined. The
ambient shear then tilts everything forward, and $\overline{v}$ and $\overline{w}$ are
amplified while the amplitude of $\overline{u}$ decreases. Beyond the point where
$\overline{v}$ is vertical, the amplitude of the wall-normal velocity decreases again, but
$\overline{w}$ reaches its maximum value. The interpretation of these observations will be
deferred to \S\ref{ssec:cm}, after we have examined the conditional temporal behaviour of
individual bursts.

\subsection{Conditional mean evolutions \label{ssec:cm}}

\begin{table}
  \begin{center}
    \def~{\hphantom{0}}
    \begin{tabular}{r c c c c c}
       Number of bursts & Dataset & $\lambdaf_1$ & $\lambdaf_2$ & $\lambdaf_3$ & $\lambdaf_4$\\
       In total & S1000 & $70$ & $228$ & $1301$ & $9777$\\
       In total & F2000 & $863$ & $5098$ & $34802$ & $274265$\\
       Per time-area in $\Sl/\lambdaf_i^2$ & S1000 & $0.267$ & $0.106$ & $0.085$ & $0.091$\\
       Per time-area in $\Sl/\lambdaf_i^2$ & F2000 & $0.16$ & $0.114$ & $0.106$ & $0.111$\\
    \end{tabular}
    \caption{Number of bursts found in different band-pass filtered time series. See table \ref{tab:sim} for details of the simulations and section \S\ref{ssec:cm} for the details of the time-area normalisation.}
    \label{tab:nburst2d}
  \end{center}
\end{table}

\begin{figure}
  \centering
  \def\sca{0.81}
  \includegraphics[scale=0.91]{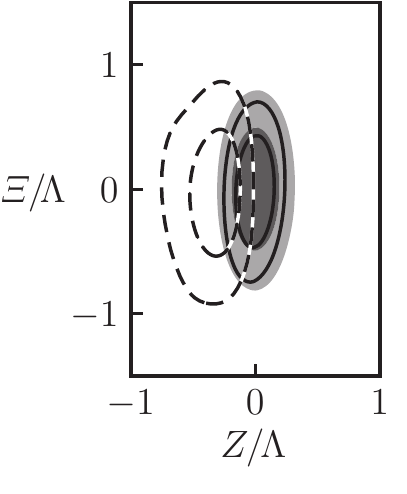}\mletter{4.25cm}{-6.9cm}{(a)}
  \includegraphics[scale=\sca]{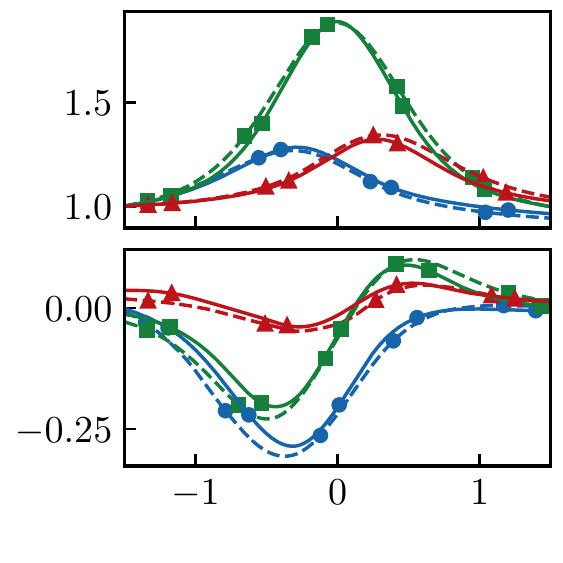}\mletter{4.25cm}{-9.2cm}{(b)}\mletter{2.25cm}{-9.2cm}{(c)}\mletter{0.15cm}{-3.7cm}{$\Sl T$}\mletter{1.5cm}{-8.9cm}{$\psi_{\overline \ub}$}\mletter{3.35cm}{-8.9cm}{$\dfrac{A_{\overline \ub}}{\langle A_{\overline \ub}\rangle}$}\hspace{5mm}%
  \includegraphics[scale=\sca]{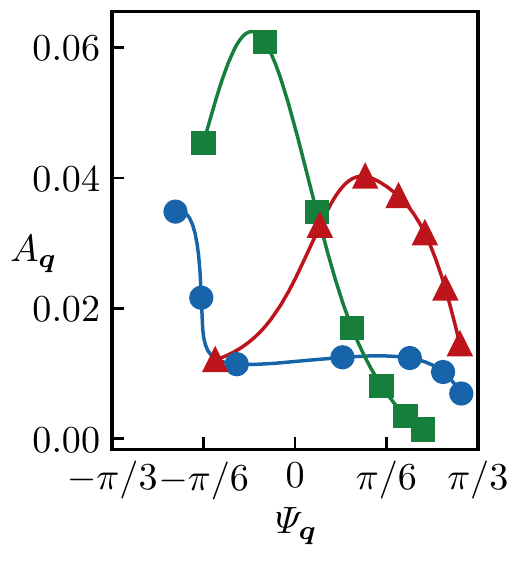}\mletter{4.25cm}{-8.6cm}{(d)} \\
  \includegraphics[scale=1.0]{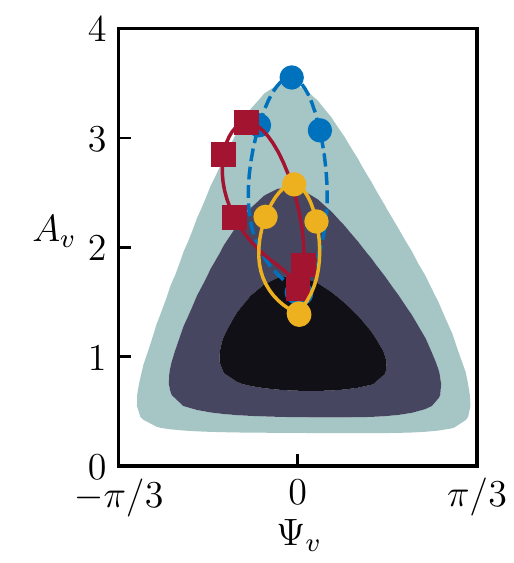}\mletter{4.75cm}{-9.3cm}{(e)}\mletter{3.0cm}{-9.5cm}{\rotatebox[origin=c]{90}{\setlength{\fboxsep}{0pt}\colorbox{white}{\rule[-2ex]{0pt}{2.0em}\large${A_{\overline v}}/{A_{\overline vM}}$}}}\hspace{12pt}
  \includegraphics[scale=.97]{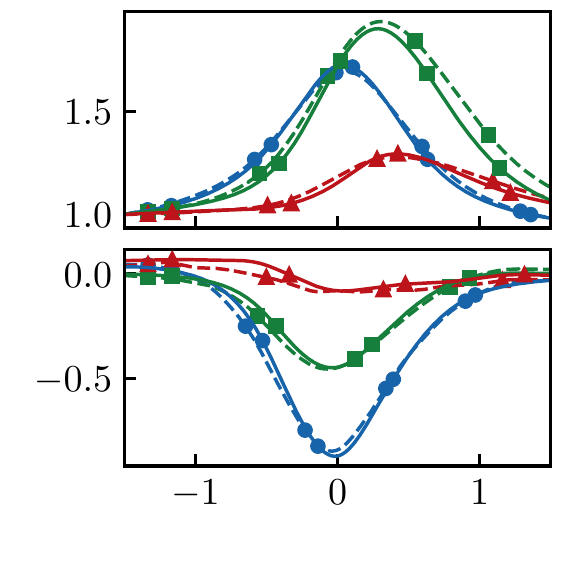}\mletter{4.95cm}{-10.8cm}{(f)}\mletter{2.72cm}{-10.8cm}{(g)}\mletter{0.18cm}{-4.43cm}{$\Sl T$}\mletter{2.2cm}{-10.66cm}{$\psi_{\overline \ub}$}\mletter{4.01cm}{-10.66cm}{$\dfrac{A_{\overline \ub}}{\langle A_{\overline \ub}\rangle}$}
\caption{(a) Snapshot of the conditional bursts of $\overline{\ub}_2$ at $\Sl T=0$, the peaking time of $\overline{v}_2$. The contours, normalised with the maximum of each component are: shaded, $A_\av = (0.6, 0.8)$; solid $A_\au = (0.8, 0.9)$; dashed $A_\aw = (0.8, 0.9)$.
(b--e) Different aspects of the evolution of $A$ and $\Psi$, conditioned to the presence of a burst. F2000.
(b) Temporal evolution of the maximum amplitude, as a function of time. \full, conditional $\overline{u}_2$; \dashed, conditional $\overline{u}_3$. Symbols are \fullcirc, $\overline{u}$; \fullsquare, $\overline{v}$; \fulltriangle, $\overline{w}$.
(c) As \emph{(b)}, for the deviations of the inclination angle respect to their mean, at the
position of the maxima of each component at every instant.
(d) Transient amplification of a linearised Orr burst. Conditions are as in figure
\ref{fig:lineal}(b), but the energies are computed over the range of $y$ corresponding
to the $\lambdaf_2$ filter.
(e) The shaded contours are the joint p.d.f. of $(\Psi_\av, A_\av)$ for $\lambdaf_2$.
\protect\mbox{---{\Large$\bullet$}---}, Evolution in (b--c), conditioned to \r{eq:cond_orr};
\protect\mbox{-\ -\ {\Large$\bullet$}\ -\ -}, conditioned to the higher threshold \r{eq:cond_orrh};
\protect\mbox{---{\,\vrule height5pt depth0pt width5pt}---} evolution in (f--g), conditioned to \eqref{eq:cond_noorr}. The five markers are $\Sl T = [-1.5,-0.3, 0, 0.3, 1.5]$. (f, g) As in {(b, c)}, but conditioned to a burst precursor.\label{fig:cond2dv}}
\end{figure}

While in the previous section we inferred a hypothetical burst evolution from the
statistical relations among the different quantities along the upper edge of the joint p.d.f.
of $(\Psi_\av, A_\av)$, we now turn our attention to the direct study of their conditional
temporal evolution by examining structures that pass through the upper `tip' of the joint
p.d.f.,
\begin{equation}
  A_{\av} > 2 A_{\av M}, \qquad \left|\Psi_{\av}\right| < 0.15,\label{eq:cond_orr}
\end{equation}
at some stage of their lives. We know from figure \ref{fig:jpdf1}(e) that the standard
deviation of the CMVs passing through these very intense events is low, and that the
statistics along the upper edge of the p.d.f. are approximately the same as those of linearised
Orr bursts. This suggests that they represent the evolution of individual bursts, at least
as long as the amplitude threshold is chosen high enough to separate bursts from
each other. The restriction in \r{eq:cond_orr} to a small inclination angle is not strictly necessary, since it follows from figure \ref{fig:jpdf1} that intense wall-normal
velocity structures are almost always vertical, but including it allows us to relax somewhat
the requirement for large amplitudes, and to identify bursts that would be too weak
otherwise. The analysis is most easily performed in a convective frame of reference, $(\xi =
x - C_{\!\lambdaf}t,\, z, t)$ where $C_{\!\lambdaf}$ is the convection velocity used in \eqref{eq:CMV}. We
identify the position and time, $(\xi_c, z_c, t_c)$, of the maximum amplitude for each
region as representative of the `peaking' point, $(\xi_c, z_c, t_c) = \text{argmax }{A_\av(\xi, z, t)}$, of the burst, and use it as the centre for
our conditionally averaged evolution $\langle \cdot\rangle_B$, defined as
\begin{equation}
  {\langle a\rangle_B}(\Xi, Z, T)=\frac{1}{N}\sum_{c=1,..., N} a(\xi_{c}+\Xi, z_{c}+Z, t_{c}+T),\label{eq:cond2d}
\end{equation}
where $N$ is the number of identified bursts for each filter (see table \ref{tab:nburst2d}),
and $a$ is either $A_{\au,\av,\aw}$ or $\Psi_{\au,\av,\aw}$. To avoid `false positives', we
discard small bursts for which $\mathcal{S} < 2.25\Lambda^2$, where $\mathcal{S}$ is the
temporally averaged area enclosed in the $(x, z)$ plane by the threshold in \r{eq:cond_orr}.
Independently of the chosen threshold, approximately 5\% of the identified bursts are
temporally clustered in quick succession, reminiscent of the `packets' of vortical
structures reported by \cite{adr:mei:tom:00}, but most cases cannot be easily grouped into
such packets.

Bursts defined in this way are spatio-temporal objects in $(\xi, z, t)$. To normalise their
density per unit duration--area, we assume that two bursts cannot occupy the same volume in
space--time, and that they scale spatially with $\Lambda$, and temporally with the inverse of
the average shear across their integration band \r{eq:A},
\begin{equation}
\Sl = (U(y_1(\lambdaf)) - U(y_0(\lambdaf)))/(y_1(\lambdaf) - y_0(\lambdaf)).
\la{eq:meanS}
\end{equation}
Their expected duration-size would then be proportional to $\Lambda^2/\Sl$, and the number of
detected bursts should collapse when their total duration-area is normalised in these units.
Table \ref{tab:nburst2d} shows that this is approximately true, with a density close to
$10\%$ except for the largest filter, which is too large for the scaling to hold. This
density is of the same order as the volume fraction of the `Q' structures studied in
\cite{loz:flo:jim:2012}.

Figure \ref{fig:cond2dv}(a) shows a snapshot of the conditional burst of $\overline\ub_2$ in
F2000 at peaking time. Other filters collapse well with $\lambdaf$, with average dimensions
$(\ell_\Xi/\lambdaf, \ell_Z/\lambdaf) \approx (2,1)$. The filtered spanwise velocity can be seen
to be offset to one side of the streamwise and wall-normal components, while the latter fall on top
of each other. If the average had been computed strictly as in \eqref{eq:cond2d}, the
conditional mean evolution would have two symmetric lobes of $A_\aw$ centred on
the burst, but this symmetry is statistical, and does not imply the symmetry of individual
events. The equations of motion and boundary conditions of the channel are invariant to
reflections with respect to $(x, y)$ planes, so that any solution $(u(x, y, z), v(x, y, z), w(x, y,
z))$ implies that $(u(x, y, -z), v(x, y, -z), -w(x, y, -z))$ is also a solution.
Individual Orr events are equally likely to be chiral-left or chiral-right, and the reflection symmetry
can be applied to each snapshot to obtain an average that is more representative of individual
events \citep{stretch90, loz:flo:jim:2012}. There is no unique way of deciding how to do this
consistently for complete histories, and the criterion that we found to produce the least
symmetric conditional mean was to always locate to the left of the burst the strongest
maximum of $A_\aw$ as it passes through the spatio-temporal neighbourhood of its peak:
\begin{equation}
|\Xi|\le 0.2\lambdaf,\, |Z|\le \lambdaf,\,  \Sl |T| \le 0.2.
\end{equation}
The conditional histories computed in this way only contain one region of strong $w$, as in
figure \ref{fig:cond2dv}(a), suggesting that most of the underlying events have only one
strong spanwise velocity region, corresponding to a single streamwise roller. Double rollers
are found more seldom. Visual inspection of a representative sample of cases shows that
approximately 5\% of the bursts are almost symmetric, with two $w$ lobes.

Figure \ref{fig:cond2dv}(b) shows the conditional evolution of the maximum amplitude of the
three filtered velocity components as a function of time. Far from $\Sl T=0$, the three
components tend to their unconditional mean, which is used to normalise the plot. The
wall-normal velocity, which is used to condition the temporal evolution, is amplified the
most, but the streamwise and spanwise velocities are also amplified, peaking at $\Sl T \approx
\pm 0.3$, respectively before and after the peak of $\langle A_\av \rangle_B$, as also
suggested by figure \ref{fig:jpdfmeans}(d, e). The evolution of the deviation of the
inclination angles from their unconditional means,
\begin{equation}
  \langle\psi_{\overline a}\rangle_B = \langle\Psi_{\overline a}\rangle_B - \langle\Psi_{\overline a}\rangle,
\end{equation}
is presented in figure \ref{fig:cond2dv}(c), which plots the inclination at the location of
the maximum amplitude of each component. The evolution of the inclination angles of the streamwise and
wall-normal velocity is from negative to positive, and tends to the unconditional mean
beyond $|\Sl T| \approx 1$. The changes in $\Psi_\aw$ are less pronounced, with values closer to
its unconditional mean across the whole evolution.

A reference to figure \ref{fig:lineal} shows that this behaviour is qualitatively similar to
linearised equilateral Orr bursts, but there are some interesting differences. The
inclination angles of $\overline{u}$ and $\overline{v}$ are proportional to each other in
the upper part of the p.d.f. of $(\Psi_\av, A_\av)$, where the presumed quasi-linear cycle takes
place, but the inclination angle of $\overline{w}$ is not. While $\Psi_\au$ evolves from
negative to positive in both figures \ref{fig:lineal} and \ref{fig:jpdfmeans}(e), the
inclination $\Psi_\aw$ is never negative in figures \ref{fig:jpdfmeans}(f) and
\ref{fig:cond2dv}(c). Moreover the conditional amplification of $\overline{u}$, as given by
the ratio between the highest and lowest conditional averages in figure
\ref{fig:jpdfmeans}(c), is almost three times higher than that for $\overline{w}$ in
figure \ref{fig:jpdfmeans}(d), while the opposite seems to be true in figure
\ref{fig:lineal}(b), and the maximum amplitudes of the two variables in figure
\ref{fig:cond2dv}(b) are approximately equal. The reason turns out to be the different
spatial location of the various velocity components. We saw in figure \ref{fig:cond2dv}(a)
that, while the structures of $\overline{u}$ and $\overline{v}$ are spatially collocated in
$(x, z)$, the $\overline{w}$-eddy is offset to one side, resulting in a weaker footprint in
the conditional joint p.d.f.s. Even more important is the vertical offset. The amplitudes in figure
\ref{fig:lineal}(b) are integrated over the whole channel, but those in figure
\ref{fig:jpdfmeans} are band-pass filtered. This includes the wall-parallel filtering by
wavelength, which has no effect on the monochromatic wavetrain in figure \ref{fig:lineal},
but also the restriction in \r{eq:A}--\r{eq:Psi} to a band of wall distances, which changes
the balance among the different components as they drift vertically in and out of the
filtering band. This effect is visually clear from figure \ref{fig:lineal}(a), and figure
\ref{fig:cond2dv}(d) shows that it can be reproduced by restricting the amplitudes in figure
\ref{fig:lineal}(b) to the same band of wall distances as in figure \ref{fig:jpdfmeans}. It
was already mentioned when discussing figure \ref{fig:lineal} that the definition of the
angles is biased by the different amplitude distribution of the three velocity components,
and it is now clear that so is the definition of the band-pass amplitudes. Thus, while
figures \ref{fig:jpdfmeans}(c, d) and \ref{fig:cond2dv}(b--d) might give the impression that
the earlier part of the burst is dominated by the association of $u$ and $v$ (such as in a
spanwise-oriented roller), while the later part is dominated by a streamwise
roller (of  $v$ and $w$), this {\color{refcolor2} interpretation could also be an artefact of how the flow is filtered. It
should be remembered that linear models cannot change the wall-parallel wavevector of a
single mode, nor its wall-parallel orientation, and changes in them have to be attributed to nonlinearity, or to the different amplification of different wavenumbers within a wavepacket, which probably would not apply to the relatively narrow wavebands here. This suggests that the interpretation of the phenomenon as a `rotating' roller outlined above should be disregarded. Nevertheless, figures \ref{fig:cond2dv}(d) and \ref{fig:lineal}(b) are similar, even though the relative amplitudes of the velocity components are different in both figures due to the energy contained outside the $y$-band limits. Moreover, \citet{jim:pof:13} showed that the cross-correlation of $v$ and $w$, integrated across the channel half-height, has a delayed peak when it is conditioned to a $v$-burst, albeit with slightly different delay between components. These evidences corroborate the causal relation between the burst and the increase in the magnitude of $w$.  Other interpretations are also possible. For example, $A_{\overline{w}}$ could also be indicative of streak breakdown, which has been observed in the buffer-layer streaks \citep{kli:rey:sch:run:67, jim:moi:91, ham:kim:wal:95} or in transient growth modes in the transition of boundary layers in the presence of a streaky background \citep{bra:hen:02}.}

Figure \ref{fig:cond2dv}(b) includes results for two different filters, which scale well
when normalised with the mean shear \r{eq:meanS}, and imply lifetimes of the order of
$\Sl T\approx 1$, measured at one-half the maximum amplification. Using slightly different
definitions of mean shear and lifetime, \cite{loz:jim:2014} and \cite{jim:pof:15} report the
somewhat longer value $\Sl T\approx 5$. In fact, the evolution of the inclination angle in
figure \ref{fig:cond2dv}(c) suggests that the conditional evolution only represents
individual bursts when they are relatively near the conditioning time, $\Sl |T|\lesssim 0.5$, beyond
which both the angle and the amplitude tend to their unconditional values.

This is better seen in figure \ref{fig:cond2dv}(f) which displays the $(\Psi_\av, A_\av)$
trajectory of figure \ref{fig:cond2dv}(b, c) on top of the joint $( \Psi_{\av}, A_{\av})$
p.d.f.. The trajectory follows the direction of the CMVs near the conditioning point, but soon
drops into the high-probability region at the core of the p.d.f.. Note that the trajectory
crosses the lowest possible point consistent with \r{eq:cond_orr}. Most eddies are in the
core of the joint p.d.f., and any conditional mean is dominated by the points closest to the core. In
a similar way, once the effect of the condition is lost, trajectories naturally fall towards
the probability maximum. Thus, if condition \r{eq:cond_orr} is substituted by the
higher-amplitude one,
\begin{equation}
  A_{\av} > 3 A_{\av M}, \qquad \left|\Psi_{\av}\right| < 0.15,\label{eq:cond_orrh}
\end{equation}
the conditional trajectory changes to the qualitatively similar, but higher-intensity, trajectory
plotted as dashed in figure \ref{fig:cond2dv}(f).

\begin{figure}
  \includegraphics[]{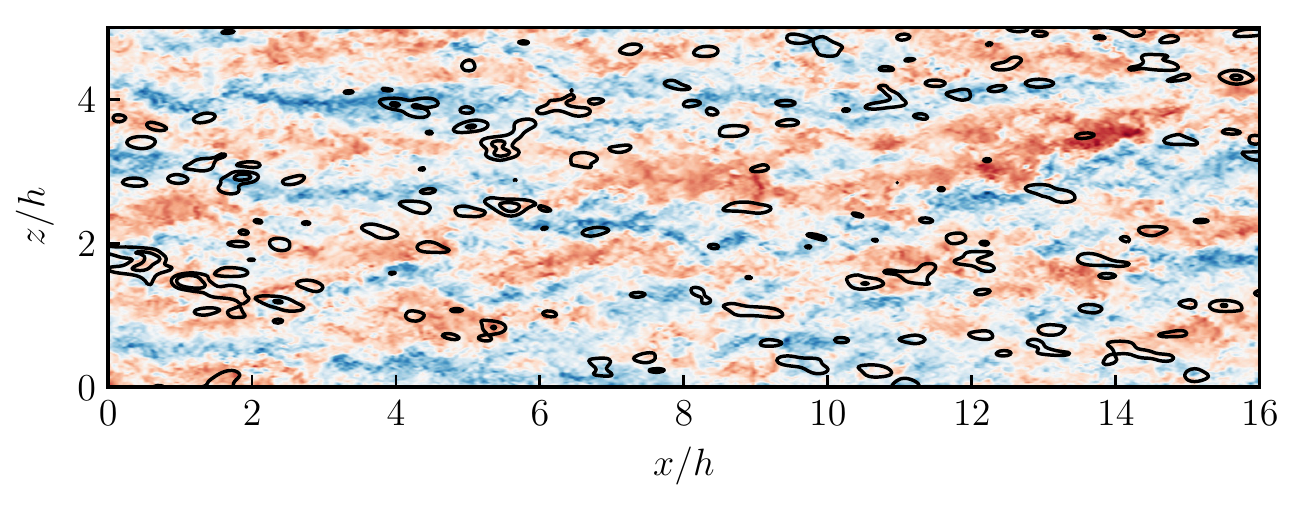}
  {\color{refcolor2} 
  \caption{Wall-parallel plane of the streamwise velocity of F2000. The shading is $u^+$ at $y/h\approx 0.2$, ranging from $u^+ \approx -5$ (blue) to $u^+ \approx 5$ (red). The line contours are $A_{\au_2}$ for the same snapshot, representing 50 and 75\% of the absolute maximum of $A_{\au_2}$ at the plane.\label{fig:uvsau}}}
\end{figure}

Perhaps the most interesting feature of the conditional evolution is the peak of
$\overline{u}$ preceding the burst, also represented by the high values of the conditional
amplitude of $\overline{u}$ over the left part of the joint p.d.f. of $( \Psi_\av, A_\av,)$ in
figure \ref{fig:jpdfmeans}(c, e). Note that a strong $A_\au$ should not be interpreted as a
especially strong streak of the streamwise velocity, because the band-pass filter is
designed to capture the energy of $v$, and the structures of $u$ are between six and ten
times longer than those of $v$ \citep{jim:18}. The band-pass-filtered amplitude of a uniform
infinite streak is zero, and we should consider $A_\au$ as {\color{refcolor2} a measure of the inhomogeneity of the streak, e.g. due to break-up or to `meandering'}. Streak inhomogeneities have often been
identified as important features of the logarithmic layer cycle \citep{flo:jim:2010, gio:sun:hwa:17}. {\color{refcolor2} The difference between $u$ and $A_\au$ is illustrated in figure \ref{fig:uvsau}, which shows both quantities in a wall-parallel plane of F2000. The streamwise velocity perturbations are shaded at $y/h\approx 0.2$, and the line contours are strong regions of $A_{\overline{u}_2}$. It is clear from the figure that both flow features are markedly different, and should not be confused.}

The previous discussion shows that a burst of $\overline{v}$ is preceded by
backwards-leaning perturbations of $\overline{u}$. The converse, that backwards-leaning
perturbations of $\overline{u}$ act as precursors of the bursts of $\overline{v}$, is tested
in figure \ref{fig:cond2dv}(f, g) by the conditional evolution of bursts conditioned to
\begin{equation}
  A_\au > 1.8A_{\au M},\quad \Psi_\au < -0.4, \la{eq:cond_noorr}
\end{equation}
which is on the left edge of the joint $(\Psi_\av, A_\av)$ p.d.f., but relatively far from its
top. As shown in the figure, the conditional burst develops as predicted, with qualitatively
similar temporal relations and delays among the different components as in figure
\ref{fig:cond2dv}(b, c). It is particularly striking that, even if figure
\ref{fig:cond2dv}(e, f) is conditioned on $\overline{u}$ rather than on $\overline{v}$, the
amplification of the latter is even stronger than that of the former, suggesting that at
least the left part of the joint p.d.f. represents well-organised bursts. A similar conclusion
was reached by \cite{jim:pof:15}, who showed that bursts could be linearly `predicted' from
conditions up to about half their lifetime before the peak amplification of $v$. The mean
trajectory in $(\Psi_\av, A_\av)$ space of the evolutions initialised within
\r{eq:cond_noorr} is plotted with squares in figure \ref{fig:cond2dv}(e). Attempts to use
the later peak of $A_\aw$ to `postdict' the burst of $\overline v$ were not successful. This
may be interpreted as that strong, forward-leaning perturbations of $\overline w$ are not
generated solely by Orr bursts. {\color{refcolor1} If this kind of $w$-perturbations can be generated by some other dynamics, denoted in this paragraph as `B', it can only be assured that Orr bursts lead to forward-leaning $w$ perturbations, as well as do `B' events; but identifying forward-leaning $w$ perturbations is not enough to determine if they come from Orr bursts, or from `B' dynamics.}

\section{Low-pass-filtered velocity fields}\label{sec:lowpass}

\begin{figure}
  \centering
  \newlength{\mywidth}
  \setlength{\mywidth}{0.38\textwidth}

  \hspace*{10pt}\includegraphics[width=\mywidth]{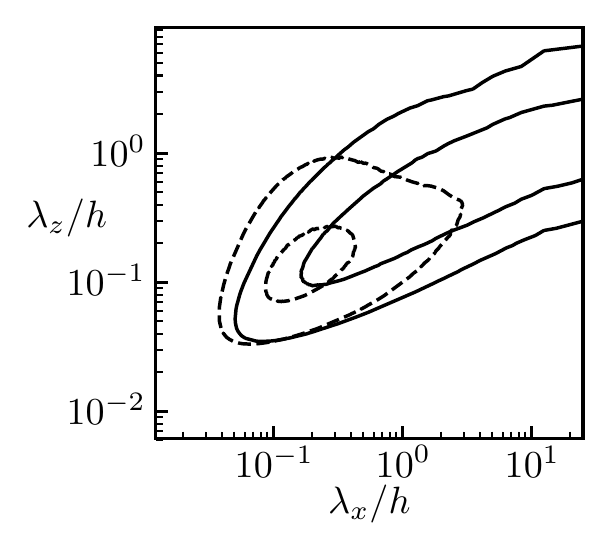}\mletter{3.8cm}{-9.8cm}{(a)}\hspace{20pt}
  \raisebox{5.3pt}{\includegraphics[width=\mywidth]{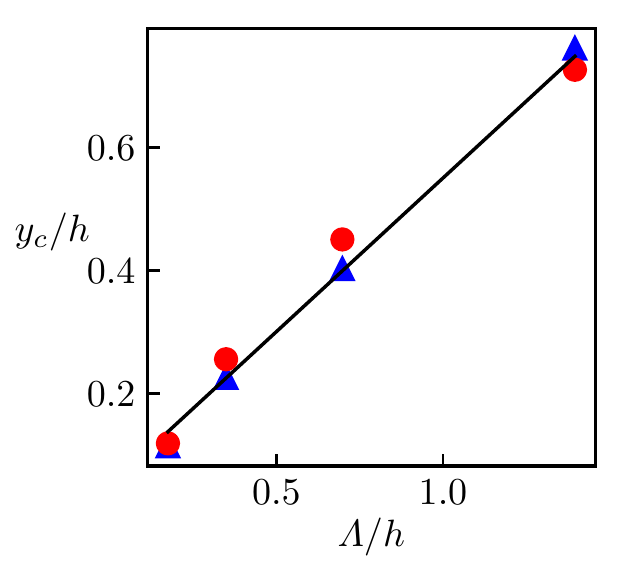}\mletter{3.8cm}{-9.8cm}{(b)}}

  \caption{
  (a) Spectra of: \solid, the streamwise velocity; \dashed, the wall-normal velocity,
  integrated over $y/h \in [0.2, 0.4]$. Contours contain 50\% and 10\% of the integrated spectral mass.
  (b) Distance from the wall of the point of highest intensity of the conditional burst. \textcolor{red}{\fullcirc}
  ejections; \textcolor{blue}{\fulltri} sweeps. The diagonal line is $y_c/h = 0.5\lambdaf/h
  + 0.05$. 
}\label{fig:cond4d_spe_y}
\end{figure}

We mentioned in \S\ref{sec:filterdef} that, while the band-pass-filtered field
\r{eq:bpass} is useful in isolating the amplitude and inclination of eddies of a given size,
it is not easily interpreted as a velocity. As a consequence, the conditional results in the
previous section give only limited information about the flow structure. In particular, the
local sign of the velocities is missing, and so is the relation of individual features with
the flow around them. Both are important. Sweeps (negative wall-normal velocity events) and
ejections (positive wall-normal velocity events) are known to have different characteristics
\citep{lu:wil:73, wal:eck:bro:72, loz:flo:jim:2012}, and this asymmetry is strongest for the
wall-attached structures responsible for most of the tangential Reynolds stress
\citep{don:loz:sek:jim:17}. Similarly, bursts of the wall-normal velocity are known to be
associated with streamwise-velocity streaks \citep{loz:flo:jim:2012, don:loz:sek:jim:17}, but
the two velocity components have very different sizes. Figure \ref{fig:cond4d_spe_y}(a)
shows that the spectrum of $u$ in the logarithmic region is much longer than that of $v$,
and it is difficult to study the interaction of the two variables if they are band-passed to
a single scale.

Both deficiencies are substantially remedied by the low-pass-filtered velocity \r{eq:lpass},
which is easily interpreted as a smoothed flow field, and retains the largest features of
the flow, including the long streaks of $u$. Unfortunately, these low-passed fields are not
approximate wavetrains or wavepackets, and we lose the information about the local inclination angle
used in \r{eq:cond_orr} as part of the condition to identify bursts. However, we saw when
discussing that condition that the inclination angle was only intended to relax the
intensity identification threshold, since figure \ref{fig:jpdf1} shows that high
amplitudes are unlikely to be anything but vertical. As a consequence, we study in this
section conditional flow histories conditioned only on the intensity of the events, and
expect their inclination, if any, to emerge as a consequence of that conditioning. Since this
requires both a large computational box to contain the largest flow structures, and
temporal information, the rest of the section only uses the temporally resolved
simulation F2000.

\begin{table}
  \begin{center}
    \def~{\hphantom{0}}
    \begin{tabular}{r c c c c}
      & $\lambdaf_1$ & $\lambdaf_2$ & $\lambdaf_3$ & $\lambdaf_4$\\
      Number of sweeps & 637 & 2898 & 17079 & 147686\\
      Number of ejections & 797 & 3282 & 18159 & 160207\\
      Number of pairs & 410 & 1717 & 9995 & 82009\\
      Sweeps per time-area in $\Sl/\lambdaf^2$ &  0.118 & 0.065 & 0.052 & 0.06\\
      Ejections per time-area in $\Sl/\lambdaf^2$ &  0.147 & 0.073 & 0.055 & 0.065
    \end{tabular}
    \caption{Number of bursts found in the low-pass-filtered temporal series of F2000. The details of the time-area normalisation are in \S\ref{ssec:cm}.}
    \label{tab:nswejc}
  \end{center}
\end{table}

We create filtered fields using the low-pass filters described in \S\ref{sec:filterdef} and,
to obtain structures that are approximately equivalent to the band-passed ones in
\S\ref{ssec:cm}, integrate the resulting velocity over the same $y$ bands. The resulting
two-dimensional fields recall the amplitudes studied in \S\ref{ssec:cm}, but contain intense
regions of negative as well as of positive velocity. Therefore, the process of isolating
intense regions of the wall-normal velocity now consists of two independent thresholding
operations, $\widetilde{v}_i>2\widetilde{v}_i^\prime$ for ejections, and
$\widetilde{v}_i<-2\widetilde{v}_i^\prime$ for sweeps, where $\widetilde{v}_i^\prime$ is the
root-mean-square intensity of the filtered time series. This results in two sets of intense
events, which are treated independently, and the rest of the section includes averages
conditioned to one or to the other. As a consequence, the presence of a sweep in a flow
conditioned to ejections should be considered a feature of the flow, not of the
conditioning, and vice versa.

As in \S\ref{ssec:cm}, the purpose of the threshold used to define the sweeps and ejections
is mostly to separate individual events, and its value is not critical. We also discard
bursts that are too small, and, in addition, merge into a single object bursts of the same
sign whose centres, $(\xi_c, z_c, t_c)$, defined as the location of their maximum intensity,
are too close to each other. The centre of the resulting object is taken to be the centre of the
stronger of the two eddies being merged. We saw in \S\ref{ssec:cm} that the size of the
bursts is $\ell_\Xi/\lambdaf \sim \ell_Z/\lambdaf \sim Tu_\tau/\lambdaf \sim O(1)$, and,
after some experimentation, choose as the merging criterion that
\begin{equation}
\sqrt{(\rmDelta t_cu_\tau)^2 + (\rmDelta \xi_c)^2 + (\rmDelta z_c)^2} < 0.15\lambdaf.  \label{eq:close}
\end{equation}
The number of centres determined in this way is given in table \ref{tab:nswejc} for each
filter size. Centring each burst on its defining extremum, the conditional structures are
computed as in \eqref{eq:cond2d}, using the four-dimensional filtered histories without
integrating them in $y$, to retain the wall-normal burst structure:
\begin{equation}
    {\langle a\rangle_L}(\Xi, y, Z, T)=\frac{1}{N}\sum_{c=1,..., N} a(\xi_{c}+\Xi, y, z_{c}+Z, t_{c}+T).\label{eq:cond4d}
\end{equation}
For each burst, we use a centred four-dimensional box spanning the channel half-height,
$y=(0, h)$, the wall-parallel box with dimensions $\Xi/h\! =\! \pm 4\pi,\, Z/\lambdaf\! =\!
\pm 3$, and a time interval consistent with the burst lifetimes discussed in
\S\ref{ssec:cm}, $\Sl T\! \approx\! \pm 5$. The conditional mean evolutions for the different
filters are found to collapse spatially when normalised with the height of the most intense
point of the conditional evolution, $y_c$, which can be interpreted as the distance from the
wall of the centre of the conditional burst at the moment of its highest intensity. Figure
\ref{fig:cond4d_spe_y}(b) shows that $y_c$ increases linearly with the filter width as
$y_c^+ = \lambdaf^+/2 + 100$, corresponding to bursts whose central height is proportional
to $\lambdaf/2$, with an offset of $100\nu/u_\tau$ representing the buffer layer. Because of
this offset, the spatial structure of the bursts collapses better with $y_c$ than with
$\lambdaf$, and this scaling will be used in the rest of the section.

\begin{figure}
  \centering
  \setlength{\mywidth}{0.38\textwidth}

  \includegraphics[width=\mywidth]{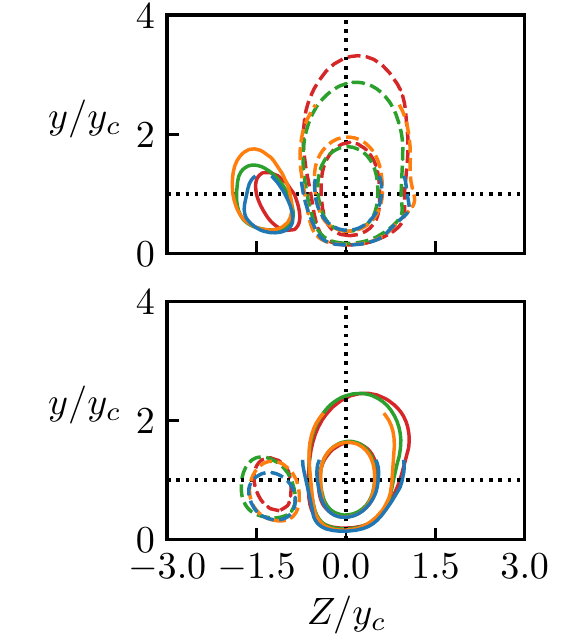}\mletter{5.4cm}{-9.8cm}{(a)}\mletter{5.4cm}{-1.9cm}{\raggedleft sweep}\mletter{2.76cm}{-2.15cm}{\raggedleft ejection}\hspace{17pt}
  \includegraphics[width=\mywidth]{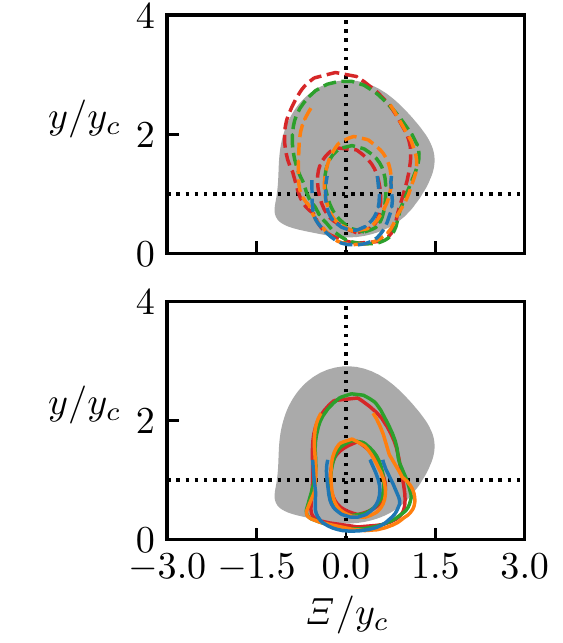}\mletter{5.4cm}{-9.8cm}{(b)}\mletter{5.4cm}{-1.9cm}{\raggedleft sweep}\mletter{2.76cm}{-2.15cm}{\raggedleft ejection}\\[10pt]

  \phantom{.}\hspace*{8pt}\includegraphics[width=0.9\mywidth]{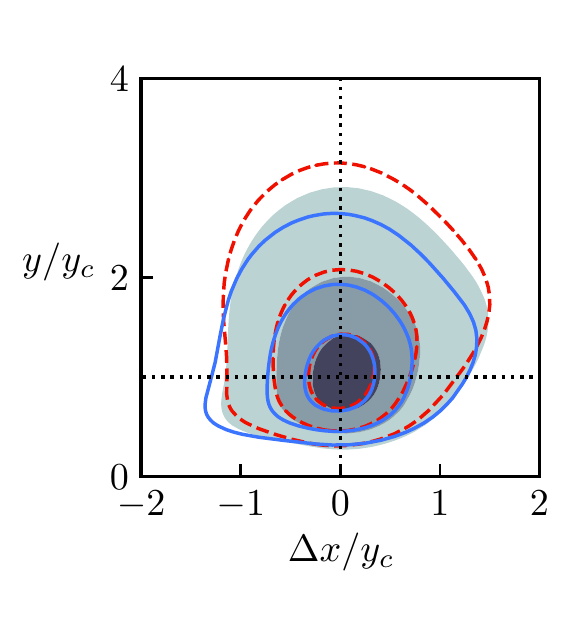}\mletter{4.4cm}{-9.0cm}{(c)}\hspace{22pt}
  \includegraphics[width=\mywidth]{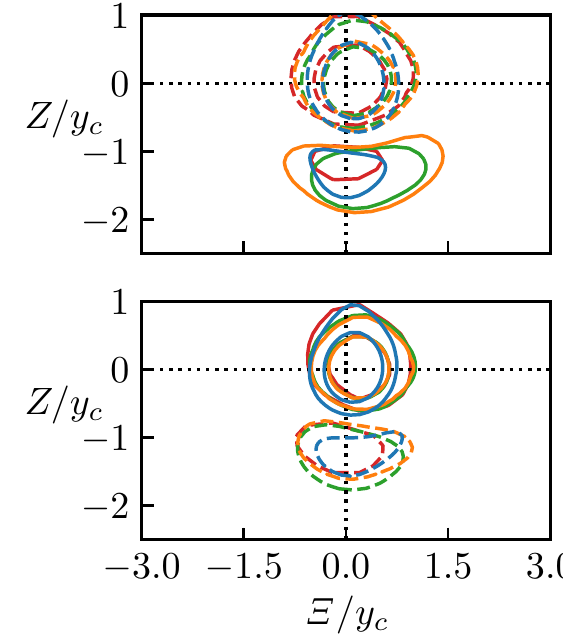}\mletter{5.4cm}{-11.0cm}{(d)}\mletter{5.4cm}{-1.4cm}{\raggedleft sweep}\mletter{2.76cm}{-1.55cm}{\raggedleft ejection}

\caption{ (a, b, d) Different sections of the conditional structure of $\widetilde{v}$ at the
peak of the $v$-burst, $\Sl T=0$, for the four filters. F2000. \solid, positive wall-normal
velocity; \dashed, negative. The contours are (-0.15, 0.15, .45) of the corresponding
extremum. In each figure the top panel is conditioned to the sweep, and the bottom one, to
the ejection. The lines are: blue, $\widetilde v_1$; orange, $\widetilde v_2$; green,
$\widetilde v_3$; red, $\widetilde v_4$. (c) Autocorrelation function of the wall normal
velocity, $C_{vv}(\rmDelta x, y, y^\prime)$ at $y^\prime=0.05h = 100 \nu/u_\tau$. The shaded contours
are the unconditional autocorrelation function; \textcolor{DodgerBlue}{\solid},
autocorrelation only for positive $v$ events; \textcolor{Red}{\dashed}, autocorrelation
only for negative $v$ events. Contours are (0.15, 0.3, 0.6) of the maximum correlation
in each case.
}\label{fig:cond4d_static}
\end{figure}

\begin{figure}
  \centering
  \setlength{\mywidth}{0.38\textwidth}

  \includegraphics[width=.85\textwidth]{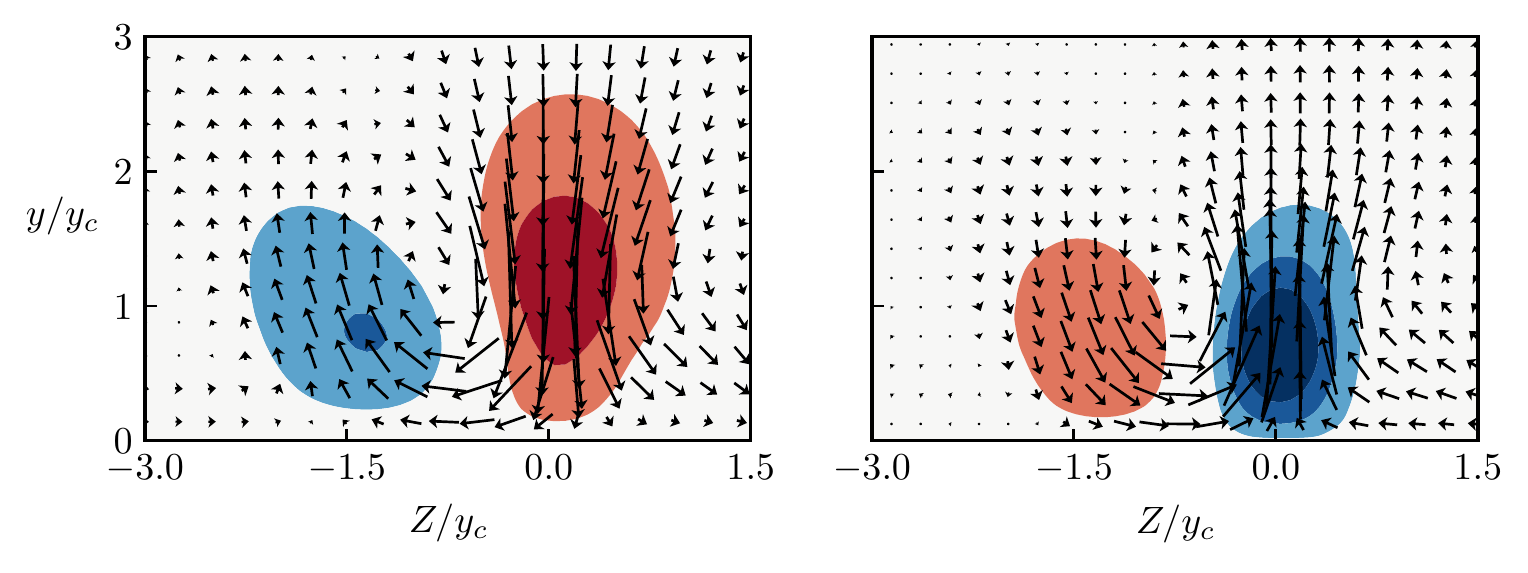}%
  \mletter{3.7cm}{-22cm}{(a)}%
  \mletter{3.7cm}{-10.6cm}{(b)}
  \caption{$(y, Z)-$section of the conditional mean structure of $\widetilde{\ub}_2$ at the peak of the $v$-burst, $\Sl T\!=\!0$.
  F2000. The shaded
  contours are streamwise velocity perturbations, $\widetilde u_2^+ = \pm(0.5, 0.9, 1.3)$. The
  arrows are the cross-plane velocities, with values in $(\widetilde v_2^2 +
  \widetilde w_2^2)^{1/2} = [0, u_\tau]$, represented by the arrow length. {\color{refcolor3} (a) Conditional sweep. (b) Conditional ejection.}
}\label{fig:cond4d_quiver}
\end{figure}

Figure \ref{fig:cond4d_static} provides three orthogonal sections of the three-dimensional
structure of the bursts at $\Sl T=0$, and shows that the scaling with $y_c$ is extremely good,
only interrupted by the channel centreline. The largest filters produce `cropped' bursts at
$y = h$, but they agree well below that height. The dotted lines in each figure show the
location of the other two sections. Individual ejections and sweeps are often accompanied by
a single strong structure with opposite wall-normal velocity \citep{loz:flo:jim:2012},
sitting at $\Xi\approx 0,\, Z/y_c \approx \pm 1.2$. Inspection of a representative sample of
individual bursts shows that only about $5\%$ of them have more than one companion, and are
approximately symmetric. As in \S\ref{ssec:cm}, the bursts in \r{eq:cond4d} are oriented to
retain as much as possible the asymmetry of this arrangement in the conditional mean. This
is done by placing on a negative $Z$ the strongest ejection found in the neighbourhood
\begin{equation}
  |\Xi|\le 0.2\lambdaf,\, |Z|\le \lambdaf.
\end{equation}
of the conditioning sweep (or vice versa), reflecting the velocity field as required. For
all the filters and conditions tested, the companion appears in the conditional evolution as
a single opposite-signed `partner' with half the intensity of the main structure.

Small differences can be seen between the conditional sweeps and the ejections. For example,
the ejections in the bottom panel of figure \ref{fig:cond4d_static}(b) have small upstream
$(\Xi<0)$ `tails' near the wall, while the sweeps in the upper panel have a more rounded
downstream `nose' farther from the wall. Interestingly, both features are also found in the
autocorrelation function of the wall-normal velocity, added to figure \ref{fig:cond4d_static}(b) as
a shaded area, suggesting that different regions of the autocorrelation function of the
wall-normal velocity are controlled by perturbations of different sign.

A similar effect was found by \citet{sil:jim:14} for the $(x, z)$ sections of the
autocorrelation of the spanwise velocity. They found them to be approximately square, and
showed that, when the correlation is conditioned to only positive or negative values of $w$
at the reference point, the correlation separates into two almost diagonal patterns, which
form the square when added together.
Here, sweeps and ejections contribute to the `nose' and `tail' of the unconditional correlation of $v$, respectively. This is confirmed in figure \ref{fig:cond4d_static}(c) which shows the autocorrelation functions of $v$ at $y/h \approx 0.05$: 
\begin{equation}
  C_{vv}(\rmDelta x, y, y'=0.05h) = \langle v(x + \rmDelta x, y)v(x, y'=0.05h) \rangle,
\end{equation}
where the average $\langle\cdot\rangle$ is performed in three different ways:
unconditionally (shaded in the figure), and conditioned to either negative (dashed lines) or positive (solid lines)
values of $v$ at the reference point $v(x, 0.05h)$.

Figure \ref{fig:cond4d_static}(d) shows that the wall-parallel sections of the conditioning
burst are roughly circular, whereas the conditional partners are bean-shaped, but we will
show in \S\ref{sec:spacetime} that the mismatch between the shapes of the bursts and of their
partners is almost surely an artefact of the conditional average, not of the instantaneous
flow fields.

Figure \ref{fig:cond4d_quiver} shows a cross-flow section of the conditional
$\widetilde\ub_2$ field. The figure is conditioned on $v$, and the shaded contours are
streamwise velocity perturbations. The sign of the primary $u$ structure (near $Z=0$) is
opposite to the one of $v$, forming a classical $(\mathrm{Q}_2,\, u<0, v>0)$ or
$(\mathrm{Q}_4,\, u>0, v<0)$ eddy. Both cases have a weaker partner of opposite polarity at
$Z/y_c\approx -1.2$, and the complete structure shares many features with the conditional
pairs of $uv$-$\mathrm{Q}$s in \citet{loz:flo:jim:2012} and \cite{don:loz:sek:jim:17}, which
were conditioned on their intensity, without regard to their temporal evolution. Those
authors found that strong $\mathrm{Q}$s can be classified as `attached', with roots that
extend very near the wall, or `detached', which do not. The attached $\mathrm{Q}$s are
self-similar, as in the present case, form $\mathrm{Q}_2$--$\mathrm{Q}_4$ pairs of size
comparable to the present ones and, when conditionally averaged by centring them on the
centre of gravity of the pair, they also contain a roller between streaks. They are
responsible for most of the tangential Reynolds stress in the flow. Similar rollers have been
identified as part of the self-sustaining cycle of the buffer layer, where they tend to be
almost parallel to the wall
\citep{jim:moi:91, ham:kim:wal:95, jim:pin:99, sch:hus:2002, jim:kaw:sim:nag:shi:05}, and are
also believed to be important in the logarithmic layer \citep{flo:jim:2010, gio:sun:hwa:17}.
In our case, the roller remains inclined in the $(\Xi, y)$ plane at approximately 15$^{\rm
o}$ during the evolution (not shown), sitting between the pair of streaks. This is similar
to the inclinations previously found from the autocorrelation of $u$ \citep[see][for a
review]{sil:jim:14}. The similarity between all these structures suggests that bursts, Qs
and quasi-streamwise vortices are different manifestations of the same phenomenon.

\begin{sidewaysfigure}[p]
  \vspace{380pt}
  \centering
  \includegraphics[width=.8\textwidth]{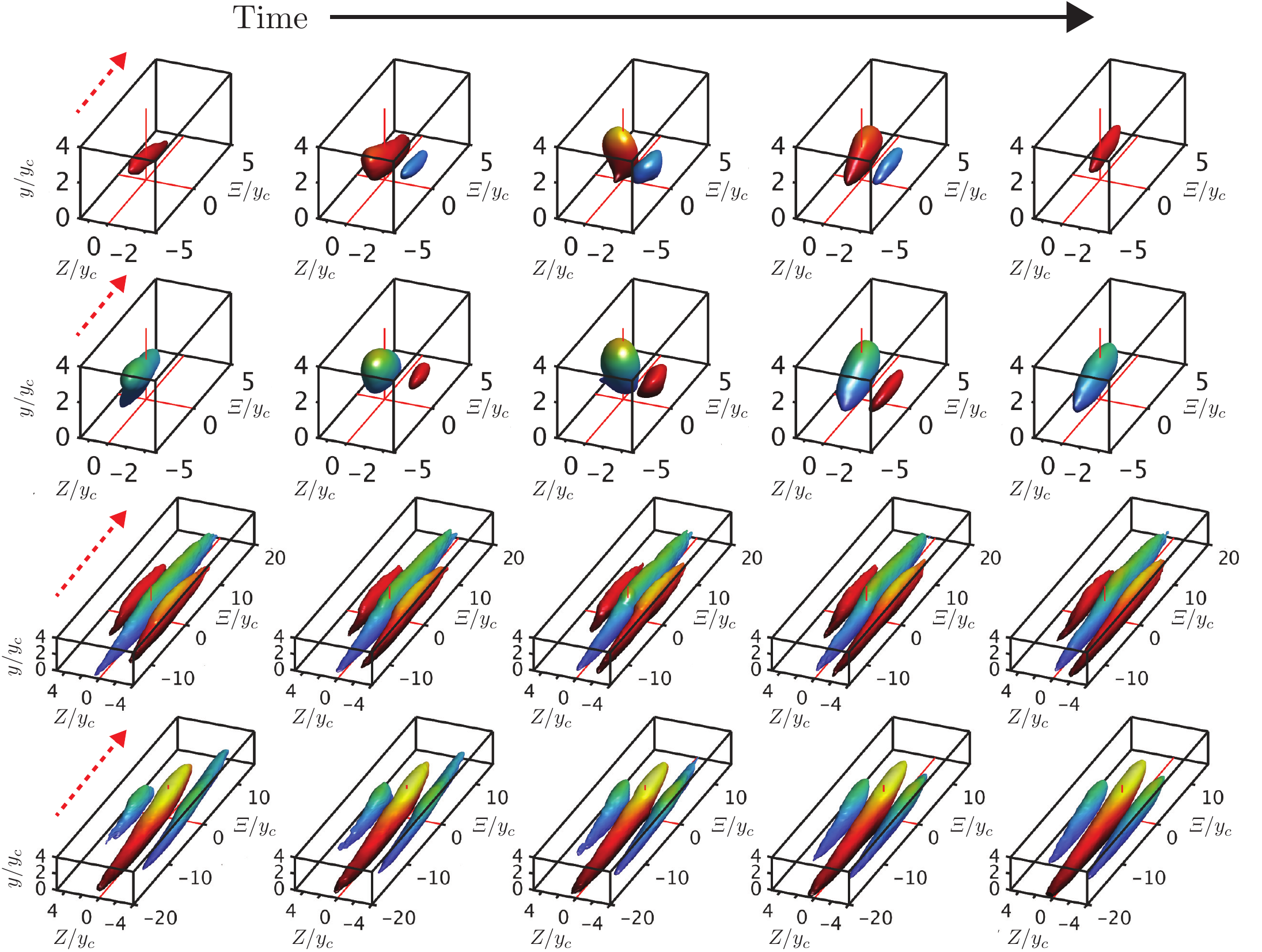}\mletter{12.2cm}{-33.8cm}{\large (a)}\mletter{9.2cm}{-33.8cm}{\large (b)}\mletter{5.6cm}{-33.8cm}{\large (c)}\mletter{2.6cm}{-33.8cm}{\large (d)}

  \caption{Mean evolution of $\widetilde{v}_2$ and $\widetilde{u}_2$ conditioned to the
  presence of a burst in F2000. The snapshots range from $\Sl T=-4(2)4$. The flow moves in the
  direction of the dashed arrow, an the mean advection between snapshots has been removed. (a,
  b) The blue surface corresponds to $\widetilde{v}_2^+ < -0.15$ and the red to
  $\widetilde{v}_2^+ > 0.15$. (a) Conditioned to a sweep. (b) Conditioned to an ejection.
  (c, d) As in (a, b), but for the evolution of $\widetilde{u}_2$; the surfaces are, blue,
  $\widetilde{u}_2^+ < -0.15$; red, $\widetilde{u}_2^+ > 0.15$.}\label{fig:cond4d_evo}
  
\end{sidewaysfigure}

\begin{figure}
  \includegraphics[width=\textwidth]{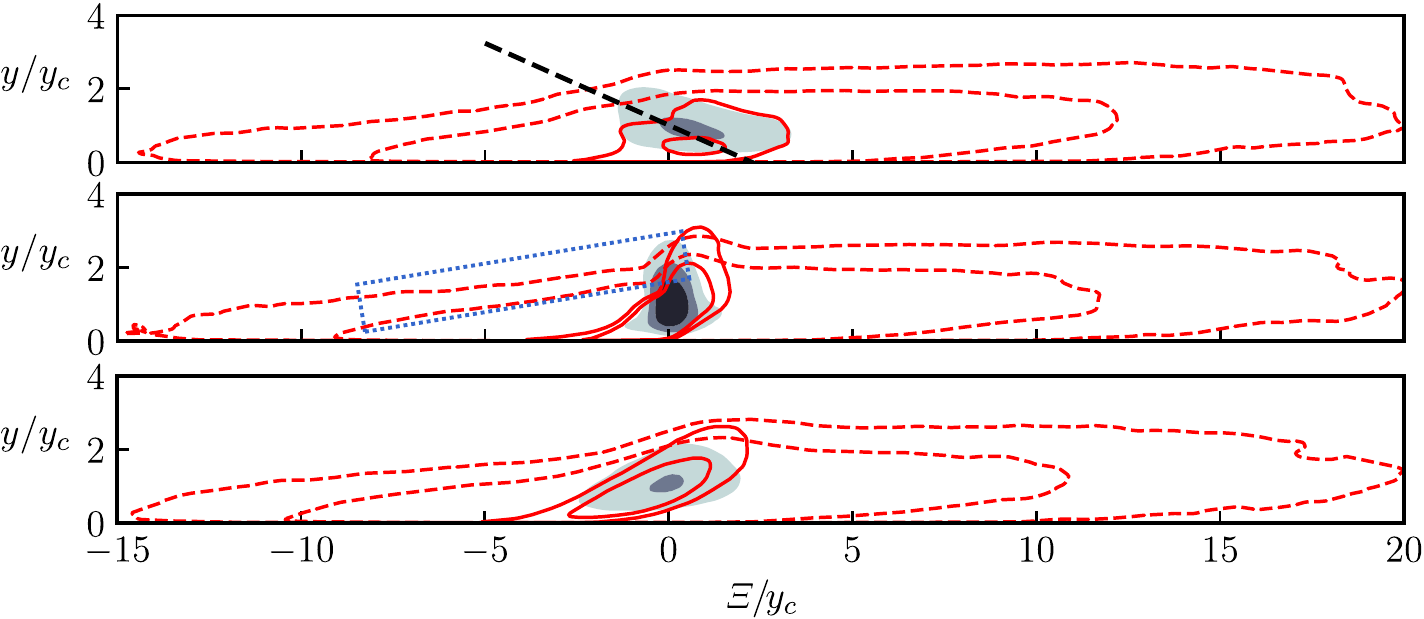}%
  \mletter{5.3cm}{-24.1cm}{\large (a)}%
  \mletter{3.6cm}{-24.1cm}{\large (b)}%
  \mletter{1.9cm}{-24.1cm}{\large (c)}%
  \vspace{20pt}
  \includegraphics[width=\textwidth]{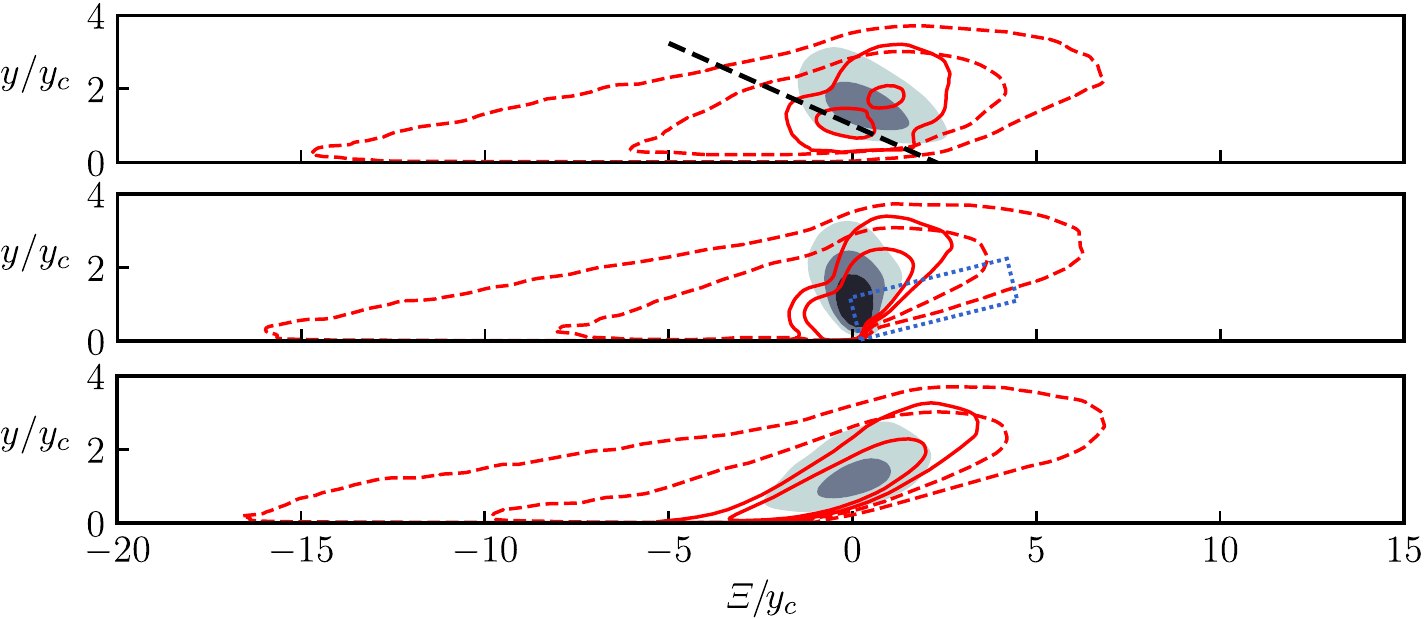}%
  \mletter{5.3cm}{-24.1cm}{\large (d)}%
  \mletter{3.6cm}{-24.1cm}{\large (e)}%
  \mletter{1.9cm}{-24.1cm}{\large (f)}%

  \caption{(a--c) $Z=0$ section of the low-passed-filtered low-speed streak (dashed lines,
  $|\tilde u_2^{+}| = [0.15, 0.3]$) and its associated  ejection (shaded contours,
  $|\tilde v_2^+| = [0.15, 0.3, 0.6]$). The solid lines are the shorter wavelengths of the
  filtered streamwise velocity, comparable to the band-pass-filtered velocity in
  \S\ref{sec:bandpass}, as explained in the text.
  (a) $\Sl T = -2$. (b) $\Sl T = 0$. (c) $\Sl T = 2$. (d--f) As in (a--c), but for the high-speed streak, 
  and the wall-normal velocity sweep. F2000.
  }\label{fig:cond4d_streakcut}
\end{figure}

Up to now, we have described bursts at the moment of their maximum intensity, but figures
\ref{fig:cond4d_evo}(a) and \ref{fig:cond4d_evo}(b) show the time evolution of a constant
isosurface of $\widetilde{v}_2$ for the conditional evolution of an ejection and a sweep,
respectively. The burst is always located at $Z/y_c \approx 0$, and its partner structure at
$Z/y_c \approx -1.2$, as in figure \ref{fig:cond4d_static}(d). The conditional evolution spans
a total of $\Sl T=8$, clearly showing an Orr-like evolution of the inclination angle and
amplitude of the burst in both cases. The partner structure is present during the full
evolution, and is amplified by approximately the same amount as the conditioning burst, but
with half its average intensity. Figure \ref{fig:cond4d_evo}(c, d) shows the evolution of the
streamwise velocity streaks during the same conditional event. They are considerably longer
($\sim 20y_c$) than the $v$-bursts ($\sim3y_c$) and, unlike the latter, which have a single
secondary lateral structure, they have a pair of comparable streaks of the opposite sign at
each side. However, although there is some amplification of the streak during the burst,
evidenced in the figure by the thickening and lengthening of their $u$ isosurface, it is
less clear than the amplification of $v$. In particular, the streaks are already present in
the conditioning volume when the $v$-burst begins to form, and remain in it when the burst
disappears. {\color{refcolor1} The alternating pattern of streaks contains a maximum in the magnitude of the spanwise variation of the streamwise velocity, $\partial \tilde u/\partial z$, located within the pair of bursts. This local extrema of the spanwise variation could be a marker of the sinous instability of the streamwise streaks, reminiscent of the dynamics of the buffer-layer streaks analysed by other authors \citep{ham:kim:wal:95, wal:97}. However, it should be considered that the conditional structure is strongly affected by the correlation of the wall-normal velocity with the rest of the flow, and thus the local maxima of $\partial \tilde u/\partial z$ observed in the conditional eddy could be an effect of the correlation of the spanwise derivative with $v$.}


Figure \ref{fig:cond4d_streakcut} shows longitudinal $Z = 0$ sections of the conditional
evolutions in figure \ref{fig:cond4d_evo}, at different moments during the burst. To be
able to compare it with the band-pass-filtered conditional evolution in \S\ref{ssec:cm}, we
also provide contours of the band-pass-filtered streamwise velocity, obtained by filtering the
conditional field (which already is an average of low-pass-filtered fields) with a high-pass
filter chosen so that the scales retained by the combined effect of the two filters are
comparable to the band-pass filter in \S\ref{sec:bandpass}. Filtering the streamwise
velocity in this way reveals an inner `core' of the streak. While the longer `body' of the
streak is always tilted forwards, the much shorter core tilts both backwards and forwards in
synchrony with the wall-normal velocity. This confirms the different nature of the short
(band-pass-filtered) and long (low-pass-filtered) perturbations of the streamwise velocity
discussed in \S\ref{sec:bandpass}. It is worth noting that the sweep is located towards the
front of the high-speed streak, whereas the ejection tends to be closer to the back of the
low-speed streak, recalling a similar arrangement of streaks and vortices in the buffer
layer \citep[see figure 9\emph{b} in][]{jim:ala:flo:04}.

Before interpreting these results, it is important to understand that figures
\ref{fig:cond4d_evo}(c, d) and \ref{fig:cond4d_streakcut} do not represent instantaneous
streaks, or even averaged ones, but the part of the streak conditioned to the presence of a
strong burst of $v$. As such, the intensity of the structures in those figures reflects both
the intensity of the fluctuations of $u$, and how is $u$ correlated to the burst. For
example, the isolines in figure \ref{fig:cond4d_streakcut} do not represent the intensity of
$u$, and can not be used to estimate that intensity except probably near the inner core. Any
attempt to derive from them whether the burst is a consequence of the presence of the
streak, or the other way around, is bound to be speculative. In the same way, the fact that
the streaks in figure \ref{fig:cond4d_evo}(c, d) are not seen to strengthen during the burst
does not mean that they do not do so (or vice versa).

With these restrictions in mind, the simplest interpretation of the positional bias in
figure \ref{fig:cond4d_streakcut} is that the streaks are `wakes' created by sweeps and
ejections advected by the mean flow at the wall-distance at which they originate
\citep{jim:ala:flo:04}. Sweeps, coming from above, move faster than the local mean speed and
bring high-speed fluid down, leaving a high-speed wake upstream. The effect of the ejections
is the opposite. However, it was argued by \cite{ala:jim:zan:mos:06} that this explanation
is unlikely, because the $v$-bursts do not live long enough. Thus, if we take their lifetime
to be the same as for Qs, $T^+\approx 2 y_c^+$ and the velocity difference to be $O(1.5
u_\tau)$ \citep{loz:jim:2014}, the maximum length of their wake would be $O(3 y_c)$. This is
the length of the inner core in figure \ref{fig:cond4d_streakcut}, but much shorter than the
length of the streak. A more likely possibility is that each streak is created by several
bursts, each of which contributes a small fraction to its length \citep{ala:jim:zan:mos:06}.
The fact that the conditional streaks weaken so little away from the peak of the burst in
figure \ref{fig:cond4d_evo}(c, d) also strongly suggests that the streaks are stable features
of the flow, while the burst grows and eventually disappears within them. This is supported
by the known streamwise distance between consecutive $\mathrm{Q}$s, $\rmDelta x \approx 7 y_c$
\citep{loz:flo:jim:2012, don:loz:sek:jim:17}, which is much shorter than the streak length.
These dimensions will be confirmed for the bursts in the next section.

\begin{figure}
  \centering
  \includegraphics[scale=1]{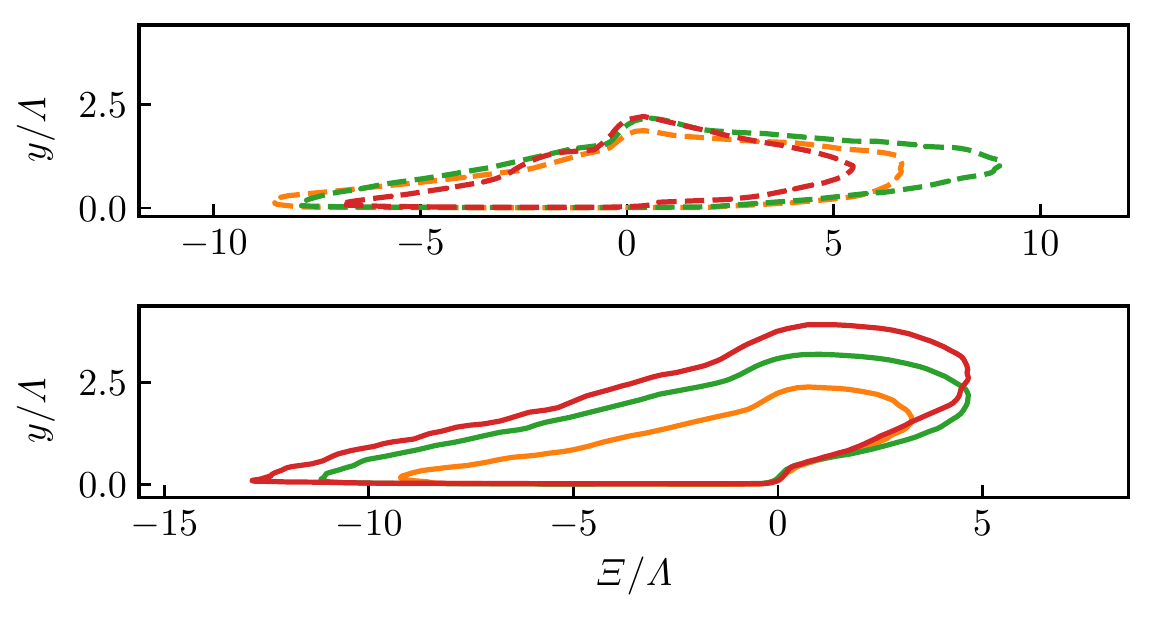}%
  \mletter{5.7cm}{-20.0cm}{\large (a)}%
  \mletter{2.8cm}{-20.0cm}{\large (b)}%
  {\color{refcolor2}  
  \caption{$Z=0$ section of the low-passed-filtered streamwise velocity streaks ($|\tilde u_2^{+}| = [0.3]$) for F2000. Lines are: (orange), $\Lambda_2$; (green) $\Lambda_3$; (red), $\Lambda_4$. (a) Low-speed streaks. (b) High-speed streaks.
  }}\label{fig:cond4d_streakcut_confirm}
\end{figure}

In fact, if we accept the argument above that very long wakes cannot be a consequence of
short-lived bursts, the scaling of figure \ref{fig:cond4d_streakcut} with the filter width
can be used to extract some causal information. We saw in discussing figure
\ref{fig:cond4d_static} that the size on the $v$-burst is proportional to
$y_c\propto\lambdaf$. The inner core of the streak in \ref{fig:cond4d_streakcut} also scales
with $\lambdaf$, which is not surprising because it is created by a pseudo-band-pass filter
of that size. The causal information is contained in the scaling of the rest of the streak,
which is obtained with a low-pass filter, and therefore contains all the scales larger than
$\lambdaf$. Although not shown in the figure to avoid clutter, most of the streak dimensions
scale with $h$, not with $\lambdaf$, suggesting that most of the streak it is not caused by
the burst, nor does directly causes it. The only regions that scale with $\lambdaf$ are the
{\em upper edge} of the leftward tail of the low-speed streak in figure
\ref{fig:cond4d_streakcut}(a--c), and the {\em underside} of the rightward nose of the
high-speed streak in figure \ref{fig:cond4d_streakcut}(d--f). Both regions are approximately indicated by a dotted rectangle in figure \ref{fig:cond4d_streakcut}(b, e){\color{refcolor2} , and shown for the filters $\Lambda_2$--$\Lambda_4$ in figure \ref{fig:cond4d_streakcut_confirm}, which scales the high-speed and low-speed streaks with $y_c$}.

We may now come back to the question of what causes the positional bias in figure \ref{fig:cond4d_streakcut}, which is part of the wider question of what causes sweeps and ejections to be organised along streaks. It is generally accepted that {\color{refcolor2} wall-normal velocity perturbations} create streaks by deforming the mean profile. {\color{refcolor2} However, the typical intense $v$-structure, i.e. an Orr burst, is much shorter than the streak, and it is clear from} the discussion in the previous paragraphs that it implies that something else organises the bursts so that the short streak segments join into longer objects. The possibility that streak instability is responsible for the bursts has been mentioned often, although the detailed mechanism is unclear \citep[see discussions in][]{sch:hus:2002, far:ioa:12}, as well as the possibility that long streaks are compound objects \citep{jim:18}. However, most of these analyses deal with infinite uniform streaks, and cannot explain a preferential longitudinal distribution of the bursts within them. Although the question of which are the original perturbations that give rise to the formation of bursts, is beyond the scope of the present paper, an obvious suggestion of the above observations is that bursts are preferentially created at the `active' end, {\color{refcolor2} i.e. where the burst is located. This is the same location where} the streak `collides' with the ambient flow, either where the upper surface of the back of the slow streaks, is overrun by the faster flow behind, or where the bottom of the `nose' of the faster streak overruns the slower flow ahead. {\color{refcolor2} This is justified if we consider that it is only these edges that scale with $y_c$, and thus can be affected by the bursts. The rear and upper parts of the high-speed streaks scale better with $h$, and so do the front parts of the low-speed streaks. The dynamics affecting this parts should in principle be independent of the bursts.} {\color{refcolor3} A somewhat similar distribution of active regions was observed by \citet{abe:kaw:cho:04}, who studied the relation of the large-scale structures in channels with the stresses at the wall. The sweeps were found to be related to the part of the high-speed streaks that had a footprint of intense spanwise shear at the wall, this being the `active' part.}

A slightly different interpretation of the same data is that the streak interaction does not take place at the end of a streak, but in the front of a meander in which high-speed flow pushes into a low-speed one. The conditional data in figure \ref{fig:cond4d_streakcut} are not enough to distinguish between those two possibilities, even when inspected in other flow sections or in three-dimensional views, and all that can probably be said is that bursts tend to be created at preexisting longitudinal inhomogeneities of the streaks.
\subsection{Space--time organisation of sweeps and ejections}\label{sec:spacetime}

It remains unclear from the previous discussion whether the secondary $v$ structure that
appears to one side of the conditioning burst in figures \ref{fig:cond4d_static} and
\ref{fig:cond4d_evo}(a, b) is an independent Orr burst, a true companion of the conditioning
burst, or another unrelated structure. However, it seems unlikely that $v$ `monopoles' of any given
sign, strong enough to be reflected in the conditional average, appear isolated in the flow,
because continuity requires that the mean wall-normal velocity over patches of dimensions
comparable to the distance from the wall should vanish. Our baseline hypothesis, motivated
in part by the similarity of the conditional averages in figure \ref{fig:cond4d_static} to
the conditional Qs in \cite{loz:flo:jim:2012} and \cite{loz:jim:2014}, is that bursts appear
in side-by-side pairs of opposite sign, and that the secondary structure in figures
\ref{fig:cond4d_static} and \ref{fig:cond4d_evo} is just a poorly centred reflection of the
opposite-signed partner of the burst being used to condition the evolution.

\begin{figure}
  \centering
  \scalebox{0.87}{%
  \raisebox{10mm}{\includegraphics{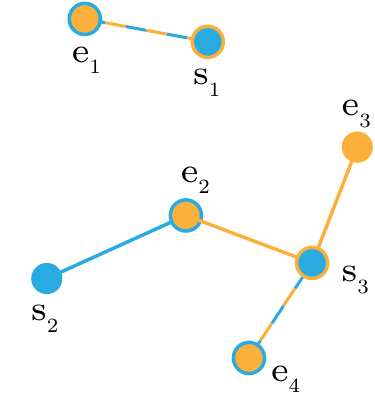}\hspace{6pt}}%
  \includegraphics{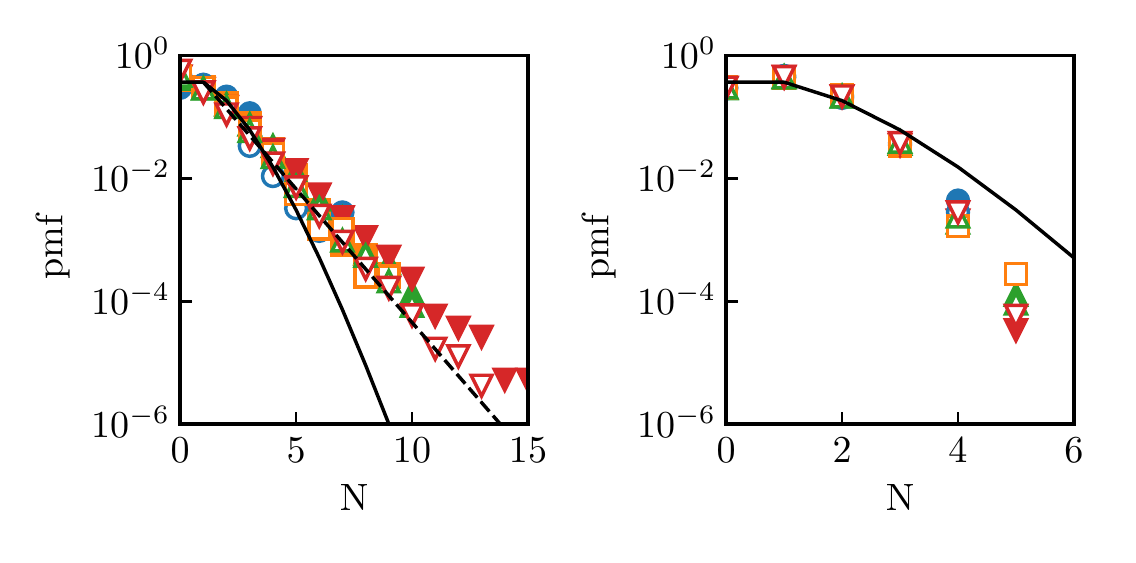}%
  }\mletter{0.5cm}{-26.2cm}{(a)}\mletter{0.5cm}{-17.2cm}{(b)}\mletter{0.5cm}{-7.6cm}{(c)}
  \caption{Probability mass functions (pmfs) of the number of partners of each structure.
  (a) Sketch of the definitions of pair, partner and number of partners (node degree). 
  (b) Pmf of the number of sweep-partners for ejections (open symbols) and ejection-partners 
  for sweeps (closed symbols). 
  (c) Pmf of the number of ejection-ejection partners (open symbols) and sweep-sweep partners
  (closed). (a--b) \fullcirc, $\lambdaf_1$; \fullsquare, $\lambdaf_2$; \fulltri, $\lambdaf_3$;
  \fulltridown, $\lambdaf_4$. The solid line is the corresponding distribution for a Poisson
  process.\label{fig:cond4d_pdf_objects}}
\end{figure}

To verify this, we compute the distance between the centres of each detected sweep and its
closest ejection, and vice versa, defined in the mean-convective frame of reference:
\begin{equation}
L_2 = \left[(\rmDelta tu_\tau)^2 + (\rmDelta z)^2 + (\rmDelta \xi)^2\right]^{1/2}.
\label{eq:L23D}
\end{equation}
\cite{loz:flo:jim:2012} and \cite{don:loz:sek:jim:17} give similar statistics for 
unconditional Qs, although without the inclusion of the time difference. In their case, they
require restrictions on the relative size of the eddies to avoid pairs with very different
components, but this is made unnecessary in our case by only considering structures obtained with a
particular filter. The definition of a pair is not straightforward, and figure
\ref{fig:cond4d_pdf_objects}(a) shows a two-dimensional sketch of possible complications.
Each sweep, such as s$_1$, chooses as partner its closest ejection, such as e$_1$, and may in
turn be chosen as partner by other ejections. Partnership is not symmetric. A given sweep
may be chosen as partner by several ejections, such as e$_2$--e$_4$ in the case of s$_3$,
and each of those may or may not be the ejection chosen by the sweep as its closest partner.
Couples which choose each other are defined as pairs, such as (e$_1$, s$_1$) and (s$_3$,
e$_4$) in figure \ref{fig:cond4d_pdf_objects}(a), and this relation is mutual and unique.
Each structure can be part of a pair, or of none at all, but being paired does not preclude
having been chosen as a partner by other structures. Examples of unpaired structures are
e$_2$, e$_3$ and s$_2$, which have not been chosen by their preferred partner, and a paired
structure with a large number of partners (or node degree, in graph notation) is s$_3$.
Approximately 60\% of the sweeps are in pairs, and 14\% of them are in exclusive pairs (see
table \ref{tab:nswejc}).

There are no obvious statistical differences between the components of exclusive and
non-exclusive pairs, but pairs are not random occurrences. The probability distribution of
the degree of the sweeps and ejections is shown in figure \ref{fig:cond4d_pdf_objects}(b),
where it is compared with the expected density function for Poisson-distributed structures,
represented by the solid line. It is clear that sweeps and ejections are clustered in the
sense of having a higher number of partners of the opposite type than the
Poisson model, obtained by randomising the position of the structures. On the other
hand, figure \ref{fig:cond4d_pdf_objects}(c), shows the distribution of the degree of
structures of the same type, which is considerably steeper than that for the Poisson process. These two
behaviours can be informally summarised as that bursts of opposite type `attract' each
other, while those of the same sign `repel', and can probably be explained by the
aforementioned effect of continuity, which requires that the net wall-normal mass flux over
large areas should tend to vanish.

It is interesting that the probability of having $N > 3$ partners of the opposite type in
figure \ref{fig:cond4d_pdf_objects}(b) follows quite closely a geometric distribution $P(N)
\sim q^N$ with $q=0.4$, represented by the dashed line. This is intriguing, because
it implies that the probability of being chosen as a partner is independent of how many
partners a structure already has, probably implying that some structures have to choose a
distant partner because they cannot find a closer one. This, in turn, suggests that bursts
that are not part of a local pair are relatively isolated from prospective partners of the
opposite sign. The same is not true for the same-sign partners in figure
\ref{fig:cond4d_pdf_objects}(c) which do not have an exponential tail or, if they do, it is one with
a much smaller parameter than in the previous case $(q\approx 0.04)$. The default model
proposed by \cite{loz:flo:jim:2012} from the analysis of the $uv$ Qs, is
that Qs tend to cluster into streamwise trains of tight pairs. The previous discussion
suggests that the organisation into trains is better developed than the one in pairs, or at
least that many pairs are asymmetric, with one of the partners too weak to be identified
as such by the analysis.

\begin{figure}
  \includegraphics[width=\textwidth]{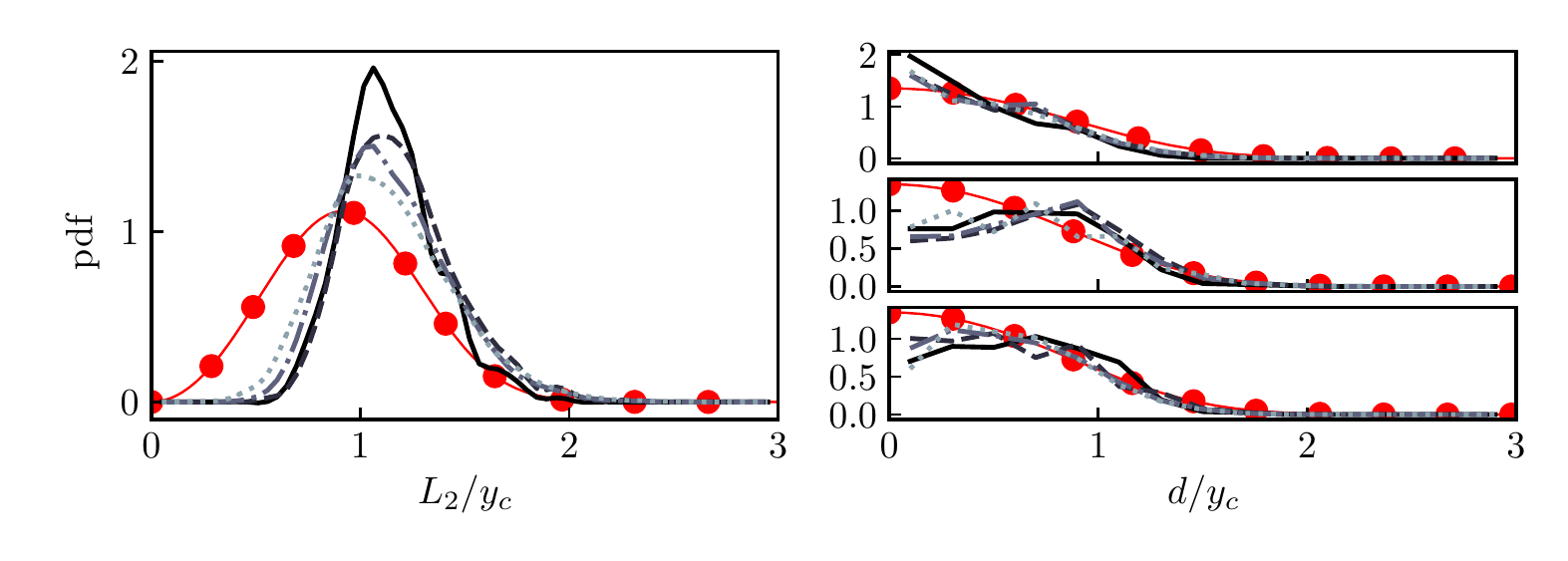}\mletter{4cm}{-26cm}{\large (a)}\mletter{3.8cm}{-1.7cm}{\large (b)}\mletter{2.7cm}{-1.7cm}{\large (c)}\mletter{1.6cm}{-1.7cm}{\large (d)}
  \includegraphics[width=\textwidth]{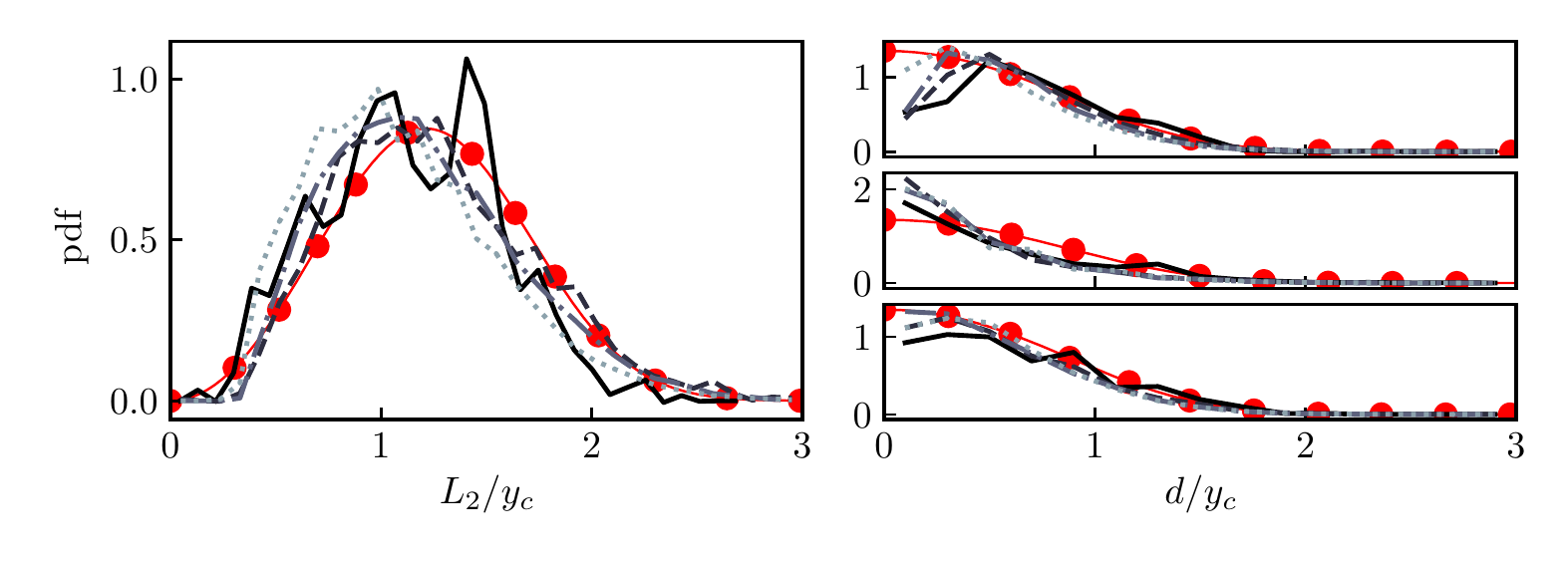}\mletter{4cm}{-26cm}{\large (e)}\mletter{3.8cm}{-1.7cm}{\large (f)}\mletter{2.7cm}{-1.7cm}{\large (g)}\mletter{1.6cm}{-1.7cm}{\large (h)}
  \caption{(a--d) Probability density functions of the distance between components of sweep-ejection pairs.
  (e--h) Same for sweep-sweep pairs.
  (a, e) p.d.f. of the space-time distance $L_2$, defined in \r{eq:L23D}.
  (b, f) p.d.f. of $|\rmDelta tu_\tau/y_c|$. (c, g) $|\rmDelta z/y_c|$. (d, h) $|\rmDelta \xi/y_c|$.
  In all cases,  lines are: \full, $\lambdaf_1$; \dashed, $\lambdaf_2$; \chain, $\lambdaf_3$;
  \dotted, $\lambdaf_4$; \linesolidcircle, p.d.f. of the intra-pair distance for an equivalent
  Poisson process. \label{fig:cond4d_dpdf}}
\end{figure}

Figure \ref{fig:cond4d_dpdf} shows the probability distribution of the distance between the
components of established pairs, compared with its Poisson counterpart. Figure
\ref{fig:cond4d_dpdf}(a--d) refers to sweep--ejection pairs, and figure
\ref{fig:cond4d_dpdf}(e--h) to pairs of the same type. More than 90\% of the normalised
distance between sweeps and ejections fall within $L_2/y_c\in (0.5,2)$, with its mode at
$L_2/y_c \approx 1.1$. Its p.d.f. is significantly different from the Poisson distribution,
shown in the figure as a line with filled circles. Figure \ref{fig:cond4d_dpdf}(b--d) shows the
contributions  to $L_2$ of its different components. The main one is from the spanwise
distance, which peaks at approximately $\rmDelta z/y_c \approx 1$, similar to the spanwise
offset between the main and secondary eddies in figures \ref{fig:cond4d_static} and
\ref{fig:cond4d_evo}. The streamwise component also plateaus for $\rmDelta \xi/y_c \lesssim
1$, but the temporal offset peaks for contemporaneous events $(\rmDelta t=0)$ and decays
quickly. As with the total $L_2$ distance, the Poisson model fails for all components.
Events tend to be spatially farther from each other than  those for the Poisson model, but closer in
the time of their peak development. The former is most probably due to the finite size of
the objects being considered. Figure \ref{fig:cond4d_static} shows that the $(x, z)$ diameter
of the sweeps and ejections is approximately $y_c$, and it is not surprising that there are
few pairs in figure \ref{fig:cond4d_dpdf} spatially closer than that distance. On the other
hand, this restriction does not apply to time, and figure \ref{fig:cond4d_dpdf}(b) shows
that pairs preferentially evolve in synchrony, as in figure \ref{fig:cond4d_evo}. 

Figure \ref{fig:cond4d_dpdf}(e--h) tells a slightly different story about homogeneous
pairs. In the first place, although the average distances are approximately the same, the
homogeneous pairs do not show the same probability deficit at short distances as in the
heterogeneous case. Sweeps are closer to each other than their size would appear to allow.
The reason can be understood from the individual p.d.f.s in figure \ref{fig:cond4d_dpdf}(f--h).
While the exclusion in dissimilar pairs takes place in space, similar pairs are
mainly spaced in time. The p.d.f. for $\rmDelta z$ peaks at the origin, in agreement with the
model of a train of aligned pairs, and $\rmDelta \xi$ is also only weakly constrained at short
distances, but these events do not happen simultaneously, and figure
\ref{fig:cond4d_dpdf}(f) reveals a clear deficit for $\rmDelta tu_\tau/y_c\lesssim 0.5$, or
$S\rmDelta t\lesssim 2$. This is of the order of the lifetime of the burst, suggesting that,
although consecutive pairs may align along streaks, they are created asynchronously, so that
neighbouring sweeps are typically at different stages of their life. Note that this would be
consistent with the observation in figure \ref{fig:cond4d_streakcut} that sweeps tend to
form at the `active' front of the streak, and suggests a model in which the trains of bursts
propagate incrementally while they extend the streak length. Although it has to be
emphasised that, as we have mentioned several times in this discussion, these bursts are
only very seldom symmetric hairpins, one cannot avoid noting the similarities between this
propagation model and the incremental formation of hairpin packets analysed by
\cite{zho:adr:bal:ken:99} in the absence of turbulent background activity. 

\section{Conclusions}\label{sec:conclusions}

Using wavelets as band-pass filters, we have extended the identification in \citet{jim:pof:15} of
Orr-like bursts in minimal channel flows, to computational boxes of arbitrary size. Wavelets
allow us to reduce the flow to a field of local single-scale wavetrains whose inclination
and intensity can be computed, but lack the connection with the surrounding flow and the
multiscale character of turbulence. We have complemented this analysis with low-pass
filters that retain the very large motions of the streamwise velocity. The evolution of both
filtered velocities is tracked in time. Regardless of the filter used, we find significant
statistical evidence of Orr-like evolution of localised structures with strong wall-normal
velocity.

The intense wall-normal velocity wavefronts are always vertical, and are consistently
preceded by backwards-leaning perturbations. The length and time scales of these wavetrains
are self-similar with respect to the filter width, especially in the logarithmic layer.
Increasing the Reynolds number in the range $Re_\tau=950$--4000 simply increases the range
of scales in the logarithmic region, resulting in a wider hierarchy of self-similar eddies.
When normalised with their filter width, the events have roughly constant duration-a-rea,
approximately filling 10\% of the time--area at a given height.

We also study the effect of the box size, using short boxes of dimensions $2\pi h \times 2h
\times \pi h$, long ones of $8\pi h \times 2h \times 3\pi h$, and comparisons with the
single-mode analysis in minimal boxes by \citet{jim:pof:15}. We find good agreement between
the three box sizes, suggesting that the dynamics of the Orr-like bursts is well captured
in minimal domains. {\color{refcolor3} The similarities with \citet{jim:pof:15} imply that further studies of the bursting process or related dynamics could be carried using smaller, more affordable simulations.} The evolution of the inclination angle and amplitude of the wavefronts of the
three velocity components is reasonably well predicted by linearised optimal transient
growth, corroborating that the evolution of these events is dominated by linear dynamics.

Because the statistical analysis of the wavetrains shows that the most intense events of the
wall-normal velocity are always vertical, Orr events can be identified solely from the
amplitude, allowing low-pass filters to identify Orr events while retaining the large-scale
motions and the sign of the velocity perturbations. The result is that sweeps and
ejections tend to form contemporaneous `pairs', i.e. the peaking of the sweep and of the
ejection happens at the same time. The components of the pairs are roughly aligned in the
spanwise direction, forming streamwise-oriented tilted rollers, and the rollers tend to be aligned in the
streamwise direction with respect to one another. However, rollers tend not to be contemporaneous, and
the temporal offset among contiguous rollers is comparable to their lifespan. Sweeps and
ejections are embedded in streamwise-velocity streaks of the opposite velocity sign, thus
generating negative tangential Reynolds stresses that reinforce the streaks in which they
reside. In fact, the evidence from the conditional
flow histories in
the space--time neighbourhood of strong bursts is that bursts tend to
form along preexisting
streaks, especially near streamwise streak inhomogeneities.

The characteristics of the Orr events, such as their size, time scale, shape and
spatio-temporal organisation, are consistent with the attached three-dimensional regions of
intense tangential Reynolds stress studied in
\citep{loz:flo:jim:2012, loz:jim:2014, don:loz:sek:jim:17}. These works provide strong
indications that the mean shear controls the dynamics of the attached momentum-carrying
structures. The new evidence corroborates these findings, suggesting that the `tall' attached
momentum-transfer structures are Orr bursts, and can thus be reasonably well described by
linear dynamics.


\section*{Acknowledgements}
This work was supported by the COTURB project of the European Research Council (ERC2014.AdG-669505). We are grateful to M. Lee for providing the data used in figure \ref{fig:speck}.

\bibliographystyle{jfm}
\bibliography{biblio}

\begin{thebibliography}{62}
\expandafter\ifx\csname natexlab\endcsname\relax\def\natexlab#1{#1}\fi
\def\au#1{#1} \def\ed#1{#1} \def\yr#1{#1}\def\at#1{#1}\def\jt#1{\textit{#1}}
  \def\bt#1{#1}\def\bvol#1{\textbf{#1}} \def\vol#1{#1} \def\pg#1{#1}
  \def\publ#1{#1}\def\arxiv#1{#1}\def\org#1{#1}\def\st#1{\textit{#1}}

\bibitem[Abe {\em et~al.\/}(2018)Abe, Antonia \& Toh]{abe:ant:18}
{\sc \au{Abe, H.}, \au{Antonia, R.~A.} \& \au{Toh, S.}} \yr{2018}
  \at{Large-scale structures in a turbulent channel flow with a minimal
  streamwise flow unit}.  \jt{J. Fluid Mech.}  \bvol{850},  \pg{733--768}.

\bibitem[Abe {\em et~al.\/}(2004)Abe, Kawamura \& Choi]{abe:kaw:cho:04}
{\sc \au{Abe, H.}, \au{Kawamura, H.} \& \au{Choi, H.}} \yr{2004}  \at{Very
  large-scale structures and their effects on the wall shear-stress
  fluctuations in a turbulent channel flow up to re $\tau$= 640}.  \jt{J.
  Fluids Eng.}  \bvol{126}~(5),  \pg{835--843}.

\bibitem[Adrian(2007)]{adr:2007}
{\sc \au{Adrian, R.~J.}} \yr{2007}  \at{{Hairpin vortex organization in wall
  turbulence}}.  \jt{Phys. Fluids}  \bvol{19}~(4),  \pg{041301}.

\bibitem[Adrian {\em et~al.\/}(2000)Adrian, Meinhart \&
  Tomkins]{adr:mei:tom:00}
{\sc \au{Adrian, R.~J.}, \au{Meinhart, C.~D.} \& \au{Tomkins, C.~D.}} \yr{2000}
   \at{{Vortex organization in the outer region of the turbulent boundary
  layer}}.  \jt{J. Fluid Mech.}  \bvol{422},  \pg{1--54}.

\bibitem[del {\'A}lamo \& Jim{\'e}nez(2006)]{ala:jim:06}
{\sc \au{del {\'A}lamo, J.~C.} \& \au{Jim{\'e}nez, J.}} \yr{2006}  \at{Linear
  energy amplification in turbulent channels}.  \jt{J. Fluid Mech.}
  \bvol{559},  \pg{205--213}.

\bibitem[del {\'A}lamo {\em et~al.\/}(2004)del {\'A}lamo, Jim{\'e}nez,
  Zandonade \& Moser]{ala:jim:zan:mos:04}
{\sc \au{del {\'A}lamo, J.~C.}, \au{Jim{\'e}nez, J.}, \au{Zandonade, P.} \&
  \au{Moser, R.~D.}} \yr{2004}  \at{{Scaling of the energy spectra in turbulent
  channels}}.  \jt{J. Fluid Mech.}  \bvol{500},  \pg{135--144}.

\bibitem[del {\'A}lamo {\em et~al.\/}(2006)del {\'A}lamo, Jim{\'e}nez,
  Zandonade \& Moser]{ala:jim:zan:mos:06}
{\sc \au{del {\'A}lamo, J.~C.}, \au{Jim{\'e}nez, J.}, \au{Zandonade, P.} \&
  \au{Moser, R.~D.}} \yr{2006}  \at{{Self-similar vortex clusters in the
  logarithmic region}}.  \jt{J. Fluid Mech.}  \bvol{561},  \pg{329--358}.

\bibitem[Brandt \& Henningson(2002)]{bra:hen:02}
{\sc \au{Brandt, L.} \& \au{Henningson, D.~S.}} \yr{2002}  \at{Transition of
  streamwise streaks in zero-pressure-gradient boundary layers}.  \jt{J. Fluid
  Mech.}  \bvol{472},  \pg{229--261}.

\bibitem[Brown \& Roshko(1974)]{bro:ros:74}
{\sc \au{Brown, G.~L.} \& \au{Roshko, A.}} \yr{1974}  \at{{On density effects
  and large structure in turbulent mixing layers}}.  \jt{J. Fluid Mech.}
  \bvol{64},  \pg{775--816}.

\bibitem[Butler \& Farrell(1992)]{but:92}
{\sc \au{Butler, K.~M.} \& \au{Farrell, B.~F.}} \yr{1992}
  \at{{Three-dimensional optimal perturbations in viscous shear flow}}.
  \jt{Phys. Fluids A: Fluid Dynamics}  \bvol{4}~(8),  \pg{1637--1650}.

\bibitem[Canuto {\em et~al.\/}(1988)Canuto, Hussaini, Quarteroni \&
  Zang]{can:hus:qua:zan:88}
{\sc \au{Canuto, C.}, \au{Hussaini, M.~Y.}, \au{Quarteroni, A.} \& \au{Zang,
  T.~A.}} \yr{1988} {\em {Spectral Methods in Fluid Dynamics}\/}.
  \publ{Springer-Verlag, Heidelberg}.

\bibitem[Cess(1958)]{ces:58}
{\sc \au{Cess, R.~D.}} \yr{1958}  \at{A survey of the literature on heat
  transfer in turbulent tube flow}.  \jt{Westinghouse Research Rep.}  \pg{pp.
  no. 8--0529--R24}.

\bibitem[Dong {\em et~al.\/}(2017)Dong, Lozano-Dur{\'a}n, Sekimoto \&
  Jim{\'e}nez]{don:loz:sek:jim:17}
{\sc \au{Dong, S.}, \au{Lozano-Dur{\'a}n, A.}, \au{Sekimoto, A.} \&
  \au{Jim{\'e}nez, J.}} \yr{2017}  \at{Coherent structures in statistically
  stationary homogeneous shear turbulence}.  \jt{J. Fluid Mech.}  \bvol{816},
  \pg{167--208}.

\bibitem[Farge(1992)]{farge:annrev}
{\sc \au{Farge, M.}} \yr{1992}  \at{Wavelet transforms and their applications
  to turbulence}.  \jt{Annu. rev. fluid mech.}  \bvol{24}~(1),  \pg{395--458}.

\bibitem[Farrell \& Ioannou(1993)]{far:93}
{\sc \au{Farrell, B.~F.} \& \au{Ioannou, P.~J.}} \yr{1993}  \at{{Optimal
  excitation of three-dimensional perturbations in viscous constant shear
  flow}}.  \jt{Phys. Fluids A: Fluid Dynamics}  \bvol{5}~(6),  \pg{1390--1400}.

\bibitem[Farrell \& Ioannou(2012)]{far:ioa:12}
{\sc \au{Farrell, B.~F.} \& \au{Ioannou, P.~J.}} \yr{2012}  \at{Dynamics of
  streamwise rolls and streaks in turbulent wall-bounded shear flow}.  \jt{J.
  Fluid Mech.}  \bvol{708},  \pg{149--196}.

\bibitem[Flores \& Jim\'enez(2006)]{flo:jim:06}
{\sc \au{Flores, O.} \& \au{Jim\'enez, J.}} \yr{2006}  \at{Effect of
  wall-boundary disturbances on turbulent channel flows}.  \jt{J. Fluid Mech.}
  \bvol{566},  \pg{357--376}.

\bibitem[Flores \& Jim{\'e}nez(2010)]{flo:jim:2010}
{\sc \au{Flores, O.} \& \au{Jim{\'e}nez, J.}} \yr{2010}  \at{{Hierarchy of
  minimal flow units in the logarithmic layer}}.  \jt{Phys. Fluids}
  \bvol{22}~(7),  \pg{071704}.

\bibitem[de~Giovanetti {\em et~al.\/}(2017)de~Giovanetti, Sung \&
  Hwang]{gio:sun:hwa:17}
{\sc \au{de~Giovanetti, M.}, \au{Sung, H.~J.} \& \au{Hwang, Y.}} \yr{2017}
  \at{Streak instability in turbulent channel flow: the seeding mechanism of
  large-scale motions}.  \jt{J. Fluid Mech.}  \bvol{832},  \pg{483–513}.

\bibitem[Hamilton {\em et~al.\/}(1995)Hamilton, Kim \& Waleffe]{ham:kim:wal:95}
{\sc \au{Hamilton, J.~M.}, \au{Kim, J.} \& \au{Waleffe, F.}} \yr{1995}
  \at{Regeneration mechanisms of near-wall turbulence structures}.  \jt{J.
  Fluid Mech.}  \bvol{287},  \pg{317--348}.

\bibitem[Hoyas \& Jim{\'e}nez(2006)]{hoy:jim:06}
{\sc \au{Hoyas, S.} \& \au{Jim{\'e}nez, J.}} \yr{2006}  \at{{Scaling of the
  velocity fluctuations in turbulent channels up to ${Re}_\tau= 2003$}}.
  \jt{Phys. Fluids}  \bvol{18}~(1),  \pg{011702}.

\bibitem[Hwang \& Cossu(2010)]{hwa:cos:10}
{\sc \au{Hwang, Y.} \& \au{Cossu, C.}} \yr{2010}  \at{Linear non-normal energy
  amplification of harmonic and stochastic forcing in the turbulent channel
  flow}.  \jt{J. Fluid Mech.}  \bvol{664},  \pg{51--73}.

\bibitem[Jim{\'e}nez(2013)]{jim:pof:13}
{\sc \au{Jim{\'e}nez, J.}} \yr{2013}  \at{How linear is wall-bounded
  turbulence?}  \jt{Phys. Fluids}  \bvol{25}~(11),  \pg{110814}.

\bibitem[Jim{\'e}nez(2015)]{jim:pof:15}
{\sc \au{Jim{\'e}nez, J.}} \yr{2015}  \at{Direct detection of linearized bursts
  in turbulence}.  \jt{Phys. Fluids}  \bvol{27}~(6),  \pg{065102}.

\bibitem[Jim{\'e}nez(2018)]{jim:18}
{\sc \au{Jim{\'e}nez, J.}} \yr{2018}  \at{Coherent structures in wall-bounded
  turbulence}.  \jt{J. Fluid Mech.}  \bvol{842},  \pg{1}.

\bibitem[Jim{\'e}nez {\em et~al.\/}(2004)Jim{\'e}nez, del {\'A}lamo \&
  Flores]{jim:ala:flo:04}
{\sc \au{Jim{\'e}nez, J.}, \au{del {\'A}lamo, J.~C.} \& \au{Flores, O.}}
  \yr{2004}  \at{{The large-scale dynamics of near-wall turbulence}}.  \jt{J.
  Fluid Mech.}  \bvol{505},  \pg{179--199}.

\bibitem[Jim{\'e}nez \& Hoyas(2008)]{jim:hoy:2008}
{\sc \au{Jim{\'e}nez, J.} \& \au{Hoyas, S.}} \yr{2008}  \at{{Turbulent
  fluctuations above the buffer layer of wall-bounded flows}}.  \jt{J. Fluid
  Mech.}  \bvol{611},  \pg{215--236}.

\bibitem[Jim{\'e}nez {\em et~al.\/}(2005)Jim{\'e}nez, Kawahara, Simens, Nagata
  \& Shiba]{jim:kaw:sim:nag:shi:05}
{\sc \au{Jim{\'e}nez, J.}, \au{Kawahara, G.}, \au{Simens, M.~P.}, \au{Nagata,
  M.} \& \au{Shiba, M.}} \yr{2005}  \at{{Characterization of near-wall
  turbulence in terms of equilibrium and bursting solutions}}.  \jt{Phys.
  Fluids}  \bvol{15}~(015105).

\bibitem[Jim{\'e}nez \& Moin(1991)]{jim:moi:91}
{\sc \au{Jim{\'e}nez, J.} \& \au{Moin, P.}} \yr{1991}  \at{{The minimal flow
  unit in near-wall turbulence}}.  \jt{J. Fluid Mech.}  \bvol{225},
  \pg{221--240}.

\bibitem[Jim{\'e}nez \& Pinelli(1999)]{jim:pin:99}
{\sc \au{Jim{\'e}nez, J.} \& \au{Pinelli, A.}} \yr{1999}  \at{{The autonomous
  cycle of near-wall turbulence}}.  \jt{J. Fluid Mech.}  \bvol{389},
  \pg{335--359}.

\bibitem[Kawahara \& Kida(2001)]{kaw:kid:01}
{\sc \au{Kawahara, G.} \& \au{Kida, S.}} \yr{2001}  \at{Periodic motion
  embedded in plane couette turbulence: regeneration cycle and burst}.  \jt{J.
  Fluid Mech.}  \bvol{449},  \pg{291--300}.

\bibitem[Kim {\em et~al.\/}(1971)Kim, Kline \& Reynolds]{kim:kli:rey:71}
{\sc \au{Kim, H.~T.}, \au{Kline, S.~J.} \& \au{Reynolds, W.~C.}} \yr{1971}
  \at{{The production of turbulence near a smooth wall in a turbulent boundary
  layer}}.  \jt{J. Fluid Mech.}  \bvol{47},  \pg{133--160}.

\bibitem[Kim(1989)]{kim:89}
{\sc \au{Kim, J.}} \yr{1989}  \at{On the structure of pressure fluctuations in
  simulated turbulent channel flow}.  \jt{J. Fluid Mech.}  \bvol{205},
  \pg{421--451}.

\bibitem[Kim {\em et~al.\/}(1987)Kim, Moin \& Moser]{kim:moi:mos:87}
{\sc \au{Kim, J.}, \au{Moin, P.} \& \au{Moser, R.~D}} \yr{1987}
  \at{{Turbulence statistics in fully developed channel flow at low {R}eynolds
  number}}.  \jt{J. Fluid Mech.}  \bvol{177},  \pg{133--166}.

\bibitem[Kline {\em et~al.\/}(1967)Kline, Reynolds, Schraub \&
  Runstadler]{kli:rey:sch:run:67}
{\sc \au{Kline, S.~J.}, \au{Reynolds, W.~C.}, \au{Schraub, F.~A.} \&
  \au{Runstadler, P.~W.}} \yr{1967}  \at{{The structure of turbulent boundary
  layers}}.  \jt{J. Fluid Mech.}  \bvol{30},  \pg{741--773}.

\bibitem[Kolmogorov(1941)]{kol:41b}
{\sc \au{Kolmogorov, A.~N.}} \yr{1941}  \at{{Local structure of turbulence in
  an incompressible fluid at very high {R}eynolds numbers}}.  \jt{Dokl. Akad.
  Nauk. SSSR}  \bvol{30},  \pg{9--13}.

\bibitem[Lee \& Moser(2015)]{lee:mos:15}
{\sc \au{Lee, M.} \& \au{Moser, R.~D.}} \yr{2015}  \at{Direct numerical
  simulation of turbulent channel flow up to ${R}e_\tau \sim 5200$}.  \jt{J.
  Fluid Mech.}  \bvol{774},  \pg{395--415}.

\bibitem[Lele(1992)]{lel:92}
{\sc \au{Lele, S.~K.}} \yr{1992}  \at{{Compact finite difference schemes with
  spectral-like resolution}}.  \jt{J. Comput. Phys.}  \bvol{103}~(1),
  \pg{16--42}.

\bibitem[Leung {\em et~al.\/}(2012)Leung, Swaminathan \&
  Davidson]{leu:swa:dav:12}
{\sc \au{Leung, T.}, \au{Swaminathan, N.} \& \au{Davidson, P.A.}} \yr{2012}
  \at{Geometry and interaction of structures in homogeneous isotropic
  turbulence}.  \jt{J. Fluid Mech.}  \bvol{710},  \pg{453--481}.

\bibitem[Lozano-Dur{\'a}n {\em et~al.\/}(2012)Lozano-Dur{\'a}n, Flores \&
  Jim{\'e}nez]{loz:flo:jim:2012}
{\sc \au{Lozano-Dur{\'a}n, A.}, \au{Flores, O.} \& \au{Jim{\'e}nez, J.}}
  \yr{2012}  \at{{The three-dimensional structure of momentum transfer in
  turbulent channels}}.  \jt{J. Fluid Mech.}  \bvol{694},  \pg{100--130}.

\bibitem[Lozano-Dur{\'a}n \& Jim{\'e}nez(2014)]{loz:jim:2014}
{\sc \au{Lozano-Dur{\'a}n, A.} \& \au{Jim{\'e}nez, J.}} \yr{2014}
  \at{Time-resolved evolution of coherent structures in turbulent channels:
  characterization of eddies and cascades}.  \jt{J. Fluid Mech.}  \bvol{759},
  \pg{432–471}.

\bibitem[Lu \& Willmarth(1973)]{lu:wil:73}
{\sc \au{Lu, S.~S.} \& \au{Willmarth, W.~W.}} \yr{1973}  \at{{Measurements of
  the structure of the {R}eynolds stress in a turbulent boundary layer}}.
  \jt{J. Fluid Mech.}  \bvol{60},  \pg{481--511}.

\bibitem[Malkus(1956)]{mal:56}
{\sc \au{Malkus, W. V.~R.}} \yr{1956}  \at{Outline of a theory of turbulent
  shear flow}.  \jt{J. Fluid Mech.}  \bvol{1}~(5),  \pg{521--539}.

\bibitem[Meneveau(1991)]{men:91}
{\sc \au{Meneveau, C.}} \yr{1991}  \at{Analysis of turbulence in the
  orthonormal wavelet representation}.  \jt{J. Fluid Mech.}  \bvol{232},
  \pg{469--520}.

\bibitem[Moin \& Moser(1989)]{moi:mos:1989}
{\sc \au{Moin, P.} \& \au{Moser, R.~D.}} \yr{1989}  \at{{Characteristic-eddy
  decomposition of turbulence in a channel}}.  \jt{J. Fluid Mech.}  \bvol{200},
   \pg{471--509}.

\bibitem[Orr(1907)]{orr:1907}
{\sc \au{Orr, W.~M.}} \yr{1907} The stability or instability of the steady
  motions of a perfect liquid and of a viscous liquid. {P}art {I}: A perfect
  liquid.  \bt{In {\em Proc. R. Ir. Acad.\/}}, ,  \vol{vol. A27},  \pg{pp.
  9--68}.

\bibitem[Pujals {\em et~al.\/}(2009)Pujals, Garc{\'\i}a-Villalba, Cossu \&
  Depardon]{puj:gar:09}
{\sc \au{Pujals, G.}, \au{Garc{\'\i}a-Villalba, M.}, \au{Cossu, C.} \&
  \au{Depardon, S.}} \yr{2009}  \at{A note on optimal transient growth in
  turbulent channel flows}.  \jt{Phys. Fluids}  \bvol{21}~(1),  \pg{015109}.

\bibitem[Reynolds \& Tiederman(1967)]{rey:tie:67}
{\sc \au{Reynolds, W.~C.} \& \au{Tiederman, W.~G.}} \yr{1967}  \at{Stability of
  turbulent channel flow, with application to {M}alkus' theory}.  \jt{J. Fluid
  Mech.}  \bvol{27},  \pg{253--272}.

\bibitem[Ruelle \& Takens(1971)]{rue:71}
{\sc \au{Ruelle, D.} \& \au{Takens, F.}} \yr{1971}  \at{On the nature of
  turbulence}.  \jt{Commun. math. phys.}  \bvol{20}~(3),  \pg{167--192}.

\bibitem[Schmid \& Henningson(2001)]{sch:hen:01}
{\sc \au{Schmid, P.~J.} \& \au{Henningson, D.~S.}} \yr{2001} {\em {Stability
  and transition in shear flows}\/}.  \publ{Springer, New York}.

\bibitem[Schoppa \& Hussain(2002)]{sch:hus:2002}
{\sc \au{Schoppa, W.} \& \au{Hussain, F.}} \yr{2002}  \at{{Coherent structure
  generation in near-wall turbulence}}.  \jt{J. Fluid Mech.}  \bvol{453},
  \pg{57--108}.

\bibitem[Sillero {\em et~al.\/}(2014)Sillero, Jim\'enez \& Moser]{sil:jim:14}
{\sc \au{Sillero, J.}, \au{Jim\'enez, J.} \& \au{Moser, R.~D.}} \yr{2014}
  \at{Two-point statistics for turbulent boundary layers and channels at
  reynolds numbers up to $\delta^+ \sim 2000$}.  \jt{Phys. Fluids}
  \bvol{26}~(10),  \pg{105--109}.

\bibitem[Squire(1933)]{squ:33}
{\sc \au{Squire, H.~B.}} \yr{1933}  \at{On the stability for three-dimensional
  disturbances of viscous fluid flow between parallel walls}.  \jt{Proc. R.
  Soc. Lond. A}  \bvol{142}~(847),  \pg{621--628}.

\bibitem[Sreenivasan(1985)]{sre:85}
{\sc \au{Sreenivasan, K.~R.}} \yr{1985}  \at{On the fine-scale intermittency of
  turbulence}.  \jt{J. Fluid Mech.}  \bvol{151},  \pg{81--103}.

\bibitem[Stretch(1990)]{stretch90}
{\sc \au{Stretch, D.~D.}} \yr{1990} Automated pattern eduction from turbulent
  flow diagnostics.  \bt{In {\em CTR Annu. Res. Briefs\/}},  \pg{pp. 145--157}.
  Stanford Univ.

\bibitem[Tennekes \& Lumley(1972)]{ten:lum:72}
{\sc \au{Tennekes, H.} \& \au{Lumley, J.~L.}} \yr{1972} {\em {A first course on
  turbulence}\/}.  \publ{MIT Press}.

\bibitem[Townsend(1976)]{Tow:76}
{\sc \au{Townsend, A.~A.}} \yr{1976} {\em {The structure of turbulent shear
  flows}\/}, 2nd edn.  \publ{Cambridge U. Press}.

\bibitem[Vela-Mart\'in {\em et~al.\/}(2018)Vela-Mart\'in, Encinar,
  Garc\'ia-Guti\'errez \& Jim\'enez]{encinar18}
{\sc \au{Vela-Mart\'in, A.}, \au{Encinar, M.~P.}, \au{Garc\'ia-Guti\'errez, A.}
  \& \au{Jim\'enez, J.}} \yr{2018}  \at{A second-order consistent, low-storage
  method for time-resolved channel flow simulations} ,  \arxiv{arXiv:
  https://doi.org/1808.06461}.

\bibitem[Waleffe(1997)]{wal:97}
{\sc \au{Waleffe, F.}} \yr{1997}  \at{On a self-sustaining process in shear
  flows}.  \jt{Phys. Fluids}  \bvol{9}~(4),  \pg{883--900}.

\bibitem[Wallace {\em et~al.\/}(1972)Wallace, Eckelmann \&
  Brodkey]{wal:eck:bro:72}
{\sc \au{Wallace, J.~M.}, \au{Eckelmann, H.} \& \au{Brodkey, R.~S.}} \yr{1972}
  \at{{The wall region in turbulent shear flow}}.  \jt{J. Fluid Mech.}
  \bvol{54},  \pg{39--48}.

\bibitem[Willmarth \& Lu(1972)]{wil:lu:72}
{\sc \au{Willmarth, W.~W.} \& \au{Lu, S.~S.}} \yr{1972}  \at{{Structure of the
  {R}eynolds stress near the wall}}.  \jt{J. Fluid Mech.}  \bvol{55},
  \pg{65--92}.

\bibitem[Zhou {\em et~al.\/}(1999)Zhou, Adrian, Balachandar \&
  Kendall]{zho:adr:bal:ken:99}
{\sc \au{Zhou, J.}, \au{Adrian, R.~J.}, \au{Balachandar, S.} \& \au{Kendall,
  T.~M.}} \yr{1999}  \at{{Mechanisms for generating coherent packets of hairpin
  vortices in channel flow}}.  \jt{J. Fluid Mech.}  \bvol{387},
  \pg{{353--396}}.

\end{thebibliography}

\end{document}